\documentclass[twocolumn,prd,aps,superscriptaddress,preprintnumbers,tightenlines,showpacs,nofootinbib,eqsecnum,amsfonts,amsmath, floatfix]{revtex4-1}

\usepackage{graphicx} 
\usepackage[usenames,dvipsnames]{xcolor}
\usepackage{xspace} 
\usepackage{bm} 
\usepackage{enumerate}
\usepackage[utf8]{inputenc}
\usepackage[normalem]{ulem}
\usepackage{graphicx}
\usepackage{lineno}
\usepackage{mathrsfs,amsmath,amsfonts,amsthm,bm}

\xdefinecolor{mylinkcolor}{rgb}{0,0,0.5}
\usepackage[
	bookmarksnumbered, bookmarksopen, bookmarksopenlevel=2,
	breaklinks=true,
	colorlinks=true, filecolor=mylinkcolor, citecolor=mylinkcolor,
	linkcolor=mylinkcolor, urlcolor=mylinkcolor, menucolor=mylinkcolor,
]{hyperref}

\def\vct#1{{\bm{#1}}}

\def\nl{\\ & \quad}

\newcommand{\be}{\begin{equation}}
\newcommand{\ee}{\end{equation}}
\newcommand{\mr}{\mathrm}
\newcommand{\ms}{\mathsf}

\newcommand{\nnm}{\nonumber}
\newcommand{\doe}{\partial}
\newcommand{\bse}{\begin{subequations}}
	\newcommand{\ese}{\end{subequations}}

\newcommand{\tr}{\textrm}
\newcommand{\mc}{\mathcal}

\newcommand{\bs}{\boldsymbol}
\newcommand{\ve}{\varepsilon}

\newcommand{\W}{W}

\newcommand{\bpm}{\begin{pmatrix}}
	\newcommand{\epm}{\end{pmatrix}}

\DeclareMathOperator{\Order}{\mathcal{O}}

\allowdisplaybreaks

\newcommand{\AEI}{\affiliation{Max Planck Institute for Gravitational Physics (Albert Einstein Institute), Am M\"uhlenberg 1, Potsdam 14476, Germany}}
\newcommand{\Maryland}{\affiliation{Department of Physics, University of Maryland, College Park, MD 20742, USA}}

\newcommand{\NNNLOSO}{N$^3$LO-PN SO}

\newcommand{\spinonespintwo}{spin$_1$-spin$_2$}
\newcommand{\sonestwo}{S$_1$S$_2$}

\begin{document}

\title{
Gravitational 
spin-orbit and aligned spin$_1$-spin$_2$
couplings through third-subleading post-Newtonian orders
}

\author{Andrea Antonelli}\AEI
\author{Chris Kavanagh}\AEI
\author{Mohammed Khalil}\AEI\Maryland
\author{Jan Steinhoff}\AEI
\author{Justin Vines}\AEI

\date{\today}

\begin{abstract}

The study of scattering encounters continues to provide new insights into the general relativistic two-body problem. The local-in-time conservative dynamics of an aligned-spin binary, for both unbound and bound orbits, is fully encoded in the gauge-invariant scattering-angle function, which is most naturally expressed in a post-Minkowskian (PM) expansion, and which exhibits a remarkably simple dependence on the masses of the two bodies (in terms of appropriate geometric variables). This dependence links the PM and small-mass-ratio approximations, allowing gravitational self-force results to determine new post-Newtonian (PN) information to all orders in the mass ratio. In this paper, we exploit this interplay between relativistic scattering and self-force theory to obtain the third-subleading (4.5PN) spin-orbit dynamics for generic spins, and the third-subleading (5PN) spin$_1$-spin$_2$ dynamics for aligned spins. We further implement these novel PN results in an effective-one-body framework, and demonstrate the improvement in accuracy by comparing against numerical-relativity simulations.

\end{abstract}

\maketitle

\section{Introduction}

The burgeoning field of gravitational-wave (GW) astronomy has already shown its potential to revolutionize our understanding of our universe \cite{Abbott:2019yzh}, gravity \cite{LIGOScientific:2019fpa}, and the nature of compact objects \cite{LIGOScientific:2018jsj,LIGOScientific:2018mvr}, such as black holes (BHs) and neutron stars. The detection of compact-binary GW sources and the accurate inference of their parameters is contingent on having accurate theoretical predictions for their coalescence. As a result of this, a variety of techniques, both analytical and numerical, have been developed to understand the coalescence of binary compact objects, with the final goal of providing faithful waveform models that can be used in GW data analysis.

\emph{Post-Newtonian} (PN) theory, the best known of the analytical techniques, has provided the foundation for the analytical studies of the two-body problem in general relativity which are most directly useful for gravitational-wave astronomy \cite{Blanchet:2013haa,Schafer:2018kuf,Rothstein:2014sra,Goldberger:2007hy,Futamase:2007zz,Pati:2000vt,Porto:2016pyg,Levi:2018nxp}. In this approximation, most applicable to bound systems, one simultaneously assumes weak gravitational potential and small velocities, i.e., $GM/rc^2 \sim v^2/c^2 \ll 1$. The PN expansion is thus a powerful tool for describing the early inspiral of the binaries observed by LIGO and Virgo~\cite{Abbott:2016blz,LIGOScientific:2018mvr}.
PN studies have been carried out at high orders both in the nonspinning~\cite{Damour:2014jta,Damour:2015isa,Bernard:2016wrg,Bernard:2017ktp,Foffa:2019hrb,Foffa:2019rdf,Foffa:2019yfl,Blumlein:2019zku,Blumlein:2020pog,Bini:2019nra,Bini:2020wpo, Bini:2020hmy,Bini:2020nsb,Bini:2020uiq} and in the spinning sectors, including spin-orbit (SO)~\cite{Hartung:2011te,Hartung:2013dza,Marsat:2012fn, Bohe:2012mr,Levi:2015uxa,Levi:2020kvb}, bilinear-in-spin ({\spinonespintwo}, {\sonestwo})~\cite{Hartung:2011ea,Levi:2011eq,Levi:2014sba,Levi:2020uwu} and spin-squared (S$^2$)~\cite{Levi:2015ixa,Levi:2015msa,Levi:2016ofk,Levi:2020uwu} couplings, as well as cubic and higher-in-spin corrections~\cite{Levi:2019kgk,Levi:2014gsa,Levi:2020lfn,Vines:2016qwa,Siemonsen:2019dsu}.
PN information on the spin dynamics has also been included in effective-one-body (EOB) waveform models~\cite{Nagar:2011fx,Barausse:2011ys,Bohe:2016gbl,Babak:2016tgq,Cotesta:2018fcv,Ossokine:2020kjp,Nagar:2018plt,Nagar:2018zoe,Khalil:2020mmr}.

In parallel to PN formalisms, the small-mass-ratio approximation, based on gravitational \emph{self-force} (GSF) theory, has also seen rapid development (see Ref.~\cite{Barack:2018yvs} and references therein for a review). As suggested by the name, the expansion parameter in this limit is the mass ratio of the two bodies $q=m_1/m_2 \ll 1$. The leading order in this approximation is given by the geodesic motion of a test-body in a Schwarzschild or Kerr background. Successive corrections,  which can be interpreted as a force moving the body away from geodesic motion, are due to the perturbation of the background sourced by the small body's nonzero stress-energy tensor.
This self-force effect on the motion of a nonspinning body has currently been numerically calculated to first order in $q$ for generic orbits in Kerr spacetime~\cite{vandeMeent:2017bcc}. In a recent breakthrough~\cite{Pound:2019lzj}, the second-order-in-$q$ binding energy in a Schwarzschild background has been calculated and compared to predictions from the first law of binary black-hole mechanics~\cite{LeTiec:2011ab}.
Meanwhile, much activity has led to the analytic calculation at very high PN orders (but at first order in $q$) of gauge-invariant quantities, such as the Detweiler redshift~\cite{Detweiler:2008ft,Bini:2013rfa,Kavanagh:2015lva,Johnson-McDaniel:2015vva,Hopper:2015icj,Bini:2015bfb,Kavanagh:2016idg,Bini:2018zde,Bini:2019lcd,Bini:2020zqy} and the precession frequency~\cite{Dolan:2013roa,Bini:2014ica,Akcay:2016dku,Kavanagh:2017wot,Akcay:2017azq,Bini:2018ylh,Bini:2019lcd}, including effects of the smaller body's spin. This has quite naturally led to related activity in confronting and validating the PN and GSF approximations~\cite{LeTiec:2011ab,Blanchet:2011aha,Akcay:2015pza} in the domain which both are valid, i.e., for large orbital separations and small mass ratios, as well as in constructing EOB models based on both approximations~\cite{Barausse:2011dq,Akcay:2012ea,Akcay:2015pjz,Antonelli:2019fmq}.

Recently, there has also been rapid advance in understanding and employing \emph{post-Minkowskian} (PM) techniques, using a weak-field approximation $GM/rc^2\ll 1$ in a background Minkowski spacetime, with no restriction on the relative velocity of the two bodies \cite{Damour:2016gwp,Damour:2017zjx,Bjerrum-Bohr:2018xdl,Cheung:2018wkq,Bern:2019nnu,Bern:2019crd}. This approximation most naturally applies to the weak-field scattering of compact objects, in which possibly relativistic velocities can be reached. Recent advances in PM gravity and in our understanding of the scattering of compact objects have been spearheaded by modern on-shell scattering-amplitude techniques, developed originally in the context of quantum particle physics (see, e.g., Ref.~\cite{Bern:2019crd} and references therein).

Scattering amplitudes were used in Ref.~\cite{Bjerrum-Bohr:2018xdl} to calculate the nonspinning 2PM ($\mathcal{O}(G^2)$, one-loop) scattering angle, reproducing with astonishing efficiency the decades-old results of Westpfhal~\cite{Westpfahl:1980mk,Westpfahl:1979gu} obtained by classical methods; an equivalent canonical Hamiltonian at 2PM order was derived from amplitudes in Ref.~\cite{Cheung:2018wkq}. The scattering angle plays a key role in PM gravity: it encodes the complete local-in-time conservative dynamics of the system (at least in a perturbative sense) and it can be used to specify a Hamiltonian in a given unique gauge~\cite{Damour:2017zjx}, which can in turn be used for unbound as well as \emph{bound} systems (with potential relevance for improving waveform models~\cite{Antonelli:2019ytb}); see in particular Refs.~\cite{Kalin:2019rwq,Kalin:2019inp}.
In Refs.~\cite{Bern:2019nnu,Bern:2019crd}, the scattering angle and a corresponding Hamiltonian have been obtained at 3PM (two-loop) order for nonspinning systems, and the results have been confirmed and expounded upon in  Refs.~\cite{Cheung:2020gyp,Blumlein:2020znm,Bini:2020wpo,Kalin:2020fhe}.

The PM approximation for two-\emph{spinning}-body systems was first tackled only very recently, with the SO dynamics at the 1PM and 2PM levels first derived by classical means in Refs.~\cite{Bini:2017xzy,Bini:2018ywr}.  These results have since been confirmed by amplitudes methods in Ref.~\cite{Bern:2020buy}, which also gave the 1PM and 2PM dynamics for the \sonestwo~sector, rounding out the current state of the art for generic-spin PM results beyond tree level.  Several other works have also considered amplitudes methods in relation to spinning two-body systems, also beyond the SO and \sonestwo~sectors (beyond the dipole level in the bodies' multipole expansions), in particular for special cases such as bodies with black-hole-like spin-induced multipole structure and/or for the aligned-spin configuration (in which the bodies' spins are [anti-]parallel to the orbital angular momentum); see, e.g., \cite{Siemonsen:2019dsu,Aoude:2020onz,Chung:2020rrz} and references reviewed therein.

These works demonstrate that the study of gravitational scattering continues to provide novel results and useful insights on the relativistic two-body problem, with implications for precision gravitational-wave astronomy yet to be explored.  A particularly powerful example of such an insight concerns the nontrivially simple dependence of the scattering-angle function on the masses \cite{Damour:2019lcq} (see also \cite{Vines:2018gqi,Bern:2019crd,Kalin:2019rwq}). This was exploited in Refs.~\cite{Bini:2019nra,Bini:2020wpo} to obtain almost all the 5PN dynamics (with the exception of 2 out of 36 coefficients in the EOB Hamiltonian; see also Refs.~\cite{Foffa:2019hrb,Blumlein:2019zku}) from first-order self-force calculations (while appropriately dealing with nonlocal-in-time tail terms). This approach has also been used in Ref.~\cite{Bini:2020nsb, Bini:2020uiq} to obtain most of the 6PN dynamics.  An extension of this approach to spinning systems was used by the current authors in Ref.~\cite{Antonelli:2020aeb} to obtain the next-to-next-to-next-to-leading order (N$^3$LO) SO PN dynamics.

In this paper, we provide details for the calculation of the {\NNNLOSO} dynamics presented in Ref.~\cite{Antonelli:2020aeb}, which completes the PN knowledge at 4.5PN order together with the NLO S$^3$ dynamics from Ref.~\cite{Levi:2019kgk} (see also \cite{Siemonsen:2019dsu}).
Furthermore, we extend our analysis to include a derivation of the N$^3$LO {\sonestwo} effects, contributing at 5PN order, for the case of spins aligned with the orbital angular momentum. We note that partial results of the {\NNNLOSO} and N$^3$LO {\sonestwo} dynamics have previously been presented in Refs.~\cite{Levi:2020kvb,Levi:2020uwu}, where all terms at $G^4$ were calculated within the powerful effective field theory framework using Feynman integral calculus. The latter of these references gives further results for \emph{all} quadratic-in-spin terms at  N$^3$LO.

Our derivations are organized in the following procedures.
\begin{enumerate}
\item
We argue that the scattering angle for an aligned-spin binary has a simple dependence on the masses (when expressed in terms of appropriate geometrical variables), which extends the result of Ref.~\cite{Damour:2019lcq} for nonspinning binaries. 
This mass dependence implies that the 4PM part of the scattering angle, which encodes the N$^3$LO PN dynamics, is determined by terms up to linear order in the mass ratio.
We use analytic results for the test-spin scattering angle to fix all terms at zeroth-order in the mass ratio, leaving the linear terms to be fixed by first-order GSF results.
\item
Assuming the existence of a PN Hamiltonian at the desired 4.5PN SO and 5PN {\sonestwo} orders, and making use of its associated mass-shell constraint with undetermined coefficients, we calculate the scattering angle and match it to the constrained form from step 1. This procedure fixes its lower orders in velocity at 3PM and 4PM orders, leaving but half of the linear-in-mass-ratio coefficients to be determined by GSF calculations. We construct the bound-orbit radial action from the scattering angle (via the Hamiltonian dynamics), noting its simple dependence on the bodies' masses.
\item
From the radial action, we calculate the redshift and spin-precession invariants and compare them with GSF results available in the literature to determine the remaining coefficients of the scattering angle. Vital to this step is the first law of spinning binary mechanics~\cite{LeTiec:2011ab,Blanchet:2012at,Tiec:2015cxa}, which is used to relate the radial action to the redshift and precession frequency, and for which we herein discuss an extension to arbitrary-mass-ratio aligned-spin eccentric orbits.
\end{enumerate}
(Although we work with aligned spins throughout, we note that the aligned SO result actually fixes the SO Hamiltonian also for precessing spins \cite{Antonelli:2020aeb}.)

The paper is organized as follows. Sections~\ref{sec:massdep},~\ref{Irsec} and~\ref{sec:N3LOSO} discuss points 1, 2 and 3, respectively.
In Sec.~\ref{sec:NRcomp}, we implement the new PN results in the scattering angle in an EOB model, and use it to compare our results against NR simulations.
We conclude in Sec.~\ref{sec:conc} with a discussion of results and potential future work.
Finally, Appendix A contains expressions for tail terms in the radial action, while Appendix B contains explicit expressions for a certain mapping between variables used to connect redshift and precession-invariant results from the radial action to GSF results in the literature, which have been previously erroneously (yet innocuously) reported in the literature.

\subsection*{Notation}

We use the metric signature $(-,+,+,+)$, and use units in which the speed of light is $c = 1$.
For a binary of compact objects with masses $m_1$ and $m_2$, we use the following combinations of the masses
\begin{gather}
M= m_1 + m_2, \quad \mu = \frac{m_1m_2}{M}, \quad \nu = \frac{\mu}{M}, \nonumber\\
q = \frac{m_1}{m_2}, \quad \delta =\frac{m_2 - m_1}{M}, \label{massmap}
\end{gather}
with $m_1<m_2$.
We often make use of the rescaled versions of the canonical spins $\bm{S}_1$ and $\bm{S}_2$, i.e.,
\begin{gather}
\bm{a}_1 = \frac{\bm{S}_1}{m_1}, \qquad 
\bm{a}_2 = \frac{\bm{S}_2}{m_2},
\end{gather}
and define the following combinations of spins
\begin{gather}
\bm{S}=\bm{S}_1+\bm{S}_2, \quad \bm{S}_* = \frac{m_2}{m_1}\bm{S}_1+\frac{m_1}{m_2}\bm{S}_2, \nonumber\\
\bm{a}_\mr b =\frac{\bm{S}}{M}, \quad \bm{a}_\mr t =\frac{\bm{S}_*}{M}.
\end{gather}
The relative position and momentum 3-vectors are denoted by $\vct{r}$ and $\vct{p}$, respectively.
Using an implicit Euclidean background, it holds that
\begin{equation}
\vct p^2 = p_r^2 + \frac{L^2}{r^2}, \quad
p_r= \bm{n}\cdot\bm{p}, \quad
\bm{L}=\bm{r}\times\bm{p},
\end{equation}
where $\bm{n}=\bm{r}/r$ with $r=|\vct r|$, and $\vct L$ is the orbital angular momentum with magnitude $L$.

\section{The mass dependence of the scattering angle}
\label{sec:massdep}
Here we argue that the structure of the PM expansion, applied to the conservative orbital dynamics of a two-massive-body system, leads to simple constraints on the dependence of the scattering-angle function on the bodies' masses, at fixed geometric quantities characterizing the incoming state.  We closely follow the arguments given in Sec.~II of Ref.~\cite{Damour:2019lcq} for the nonspinning case, considering only the local-in-time, conservative part of the dynamics, while generalizing to the case of spinning bodies, finally, in the aligned-spin configuration.  

The motion of a two-point-mass system (the nonspinning case) is effectively governed by the coupled system of (i) geodesic equations for the worldlines of the two point masses, using the full two-body spacetime metric (with a suitable regularization or renormalization procedure), and (ii) Einstein's equations for the metric, sourced by effective point-mass energy-momentum tensors.  In the case of spinning bodies, to dipolar order in the bodies' multipole expansions, the geodesic equations are replaced by the pole-dipole Mathisson-Papapetrou-Dixon (MPD) equations~\cite{Mathisson:1937zz,Papapetrou:1951pa,Dixon:1970zza},
\bse\label{MPD}
\begin{alignat}{3}\label{MPDp}
\frac{\mr Dp_{\mr i\mu}}{\mr d\tau_\mr i}&=-\frac{1}{2}R_{\mu\nu\rho\sigma}\dot x_\mr i^\nu S_\mr i^{\rho\sigma},
\\\label{MPDS}
\frac{\mr DS_\mr i^{\mu\nu}}{\mr d\tau_\mr i}&=2p_\mr i^{[\mu}\dot x_\mr i^{\nu]},
\\\label{TDSSC}
0&=p_{\mr i\mu}S_\mr i^{\mu\nu},
\end{alignat}
\ese
where, for the $\mr i$th body ($\mr i=1,2$), $p_\mr i^\mu(\tau_\mr i)$ is the linear momentum vector, $S_\mr i^{\mu\nu}(\tau_\mr i)$ is the antisymmetric spin (intrinsic angular momentum) tensor, and $\dot x_\mr i^\mu(\tau_\mr i)$ is the tangent to the body's worldline $x_\mr i(\tau_\mr i)$.   The constraint (\ref{TDSSC}), the ``covariant'' or Tulczyjew-Dixon  spin supplementary condition~\cite{Dixon:1979,Steinhoff:2014kwa,Tulczyjew:1959,fokker1929relativiteitstheorie}, combined with (\ref{MPDp}) and (\ref{MPDS}), uniquely determines a first-order equation of motion for the worldline, $\dot x_\mr i^\mu=\dot x_\mr i^\mu(x_\mr i,p_\mr i,S_\mr i)[g]$.  The corresponding effective energy-momentum tensor,
\begin{alignat}{3}\label{Tmunu}
\begin{aligned}
T^{\mu\nu}(x)&=\sum_\mr i\int d\tau_\mr i\bigg[ p_\mr i^{(\mu}\dot x_\mr i^{\nu)}\frac{\delta^4(x-x_\mr i)}{\sqrt{-g}}
\\
&\qquad+\nabla_\lambda\bigg(S_\mr i^{\lambda(\mu}\dot x_\mr i^{\nu)}\frac{\delta^4(x-x_\mr i)}{\sqrt{-g}}\bigg)\bigg],
\end{aligned}
\end{alignat}
sources Einstein's equations,
\be\label{Einstein}
R_{\mu\nu}-\frac{1}{2}Rg_{\mu\nu}=8\pi GT_{\mu\nu}.
\ee
In the PM scheme, an iterative solution to these equations is obtained as an expansion in $G$ of the worldlines, momenta and spins,
\begin{alignat}{3}\label{expand_xpS}
x_\mr i^\mu(\tau_\mr i)&=x_{\mr i0}^\mu(\tau_\mr i)+G x_{\mr i1}^\mu(\tau_\mr i)+G^2 x_{\mr i2}^\mu(\tau_\mr i)+\cdots,
\nnm\\
p_\mr i^\mu(\tau_\mr i)&=p_{\mr i0}^\mu(\tau_\mr i)+G p_{\mr i1}^\mu(\tau_\mr i)+G^2 p_{\mr i2}^\mu(\tau_\mr i)+\cdots,
\\\nnm
S_\mr i^{\mu\nu}(\tau_\mr i)&=S_{\mr i0}^{\mu\nu}(\tau_\mr i)+G S_{\mr i1}^{\mu\nu}(\tau_\mr i)+G^2 S_{\mr i2}^{\mu\nu}(\tau_\mr i)+\cdots,
\end{alignat}
and of the metric,
\be\label{expand_g}
g_{\mu\nu}(x)=\eta_{\mu\nu}+G h_1{}_{\mu\nu}(x)+G^2 h_2{}_{\mu\nu}(x)+\cdots,
\ee
where $\eta_{\mu\nu}$ is the Minkowski metric, which we henceforth use instead of the full metric $g_{\mu\nu}$ for all 4-vector manipulations (index raising and lowering, dot products and squares of vectors, etc.).

At the leading orders in (\ref{expand_xpS}), given by the solutions to (\ref{MPD}) with $g=\eta$, each body moves inertially in flat spacetime,
\begin{alignat}{3}\label{xpS0}
\begin{aligned}
x_{\mr i0}^\mu(\tau_\mr i)&=y_{\mr i}^\mu+u_{\mr i}^\mu\tau_\mr i,
\\
p_{\mr i0}^\mu(\tau_\mr i)&=m_\mr i u_{\mr i}^\mu,
\\
S_{\mr i0}^{\mu\nu}(\tau_\mr i)&=m_\mr i\epsilon^{\mu\nu}{}_{\rho\sigma}u_{\mr i}^\rho a_{\mr i}^\sigma.
\end{aligned}
\end{alignat}
Here, $y_\mr i^\mu$ are constant displacements from the origin at $\tau_\mr i=0$, and $u_\mr i^\mu$ are constant 4-velocities, with $u_\mr i^2=-1$, so that $\tau_\mr i$ are the (Minkowski) proper times, and $p_\mr i^2=-m_\mr i^2$ where $m_\mr i$ are the constant rest masses.  The zeroth-order spin tensors $S_{\mr i0}^{\mu\nu}$ are also constant, and, being orthogonal to $u_{\mr i\mu}$, have been parametrized in terms of a constant mass-rescaled (Pauli-Lubanski, ``covariant'') spin vector,
\be
a_\mr i^\mu=-\frac{1}{2m_\mr i}\epsilon^\mu{}_{\nu\rho\sigma}u_{\mr i}^\nu S_{\mr i0}^{\rho\sigma},
\ee
with dimensions of length, the magnitude of which would measure the radius of the ring singularity of a corresponding (linearized) Kerr black hole.  We identify the zeroth-order geometric (mass-independent) quantities, $y_\mr i^\mu$, $u_\mr i^\mu$ and $a_\mr i^\mu$, with those characterizing the asymptotic incoming state, along with the masses $m_1$ and $m_2$.

Inserting (\ref{xpS0}) into (\ref{Tmunu}) (with $g=\eta$) yields the zeroth-order stress-energy tensor, which serves as a source for the first-order metric perturbation $h_{1\mu\nu}$ in the linearization of (\ref{Einstein}).  The solution for the trace-reversed $\bar h_1^{\mu\nu}=h_1^{\mu\nu}-\frac{1}{2}\eta^{\mu\nu}h_{1\rho}{}^\rho$, in harmonic gauge ($\doe_\mu \bar h_1^{\mu\nu}=0$), reads
\begin{alignat}{3}\label{h1PM}
\bar h_1^{\mu\nu}(x)&=4 \sum_{\mr i}m_\mr i \Big(u_\mr i^\mu u_\mr i^\nu+u_\mr i^{(\mu}\epsilon^{\nu)}{}_{\rho\sigma\lambda}u_\mr i^\rho a_{\mr i}^\sigma\doe^\lambda\Big)\frac{1}{r_\mr i},
\end{alignat}
where $r_\mr i=\{(x-y_\mr i)^2+[u_\mr i\cdot(x-y_\mr i)]^2\}^{1/2}$ is the (Minkowski) distance of the field point $x$ from the (zeroth-order, flat geodesic) worldline $x_{\mr i0}=y_\mr i+u_\mr i \tau_\mr i$ in its rest frame, and $\doe_\mu$ is the flat covariant derivative. (Note that the result for the first-order field (\ref{h1PM}) would be the same whether we used the physical retarded Green's function or the time-symmetric Green's function, given the nature of the zeroth-order source, constant momentum and spin along a flat-spacetime geodesic.)
A key property to be noted here is that $h_1$ is linear in the masses $m_\mr i$, while having a more intricate dependence on the geometric quantities $y_\mr i^\mu$, $u_\mr i^\mu$ and $a_\mr i^\mu$.  (It is linear in the spins $a_\mr i^\mu$ here only because we are working to linear order in the spins, to dipolar order in the multipole expansions.)

In the next step of the iterative scheme, one uses $g=\eta+h_1$ in the bodies' equations of motion (\ref{MPD}) to solve for the first-order perturbations in (\ref{expand_xpS}) [for which it is sufficient to integrate the RHSs of (\ref{MPDp}) and (\ref{MPDS}) along the zeroth-order motion (\ref{xpS0}), and to regularize by simply dropping the divergent self-field contribution].
Importantly, one finds that $x_{\mr i1}^\mu$, $p_{\mr i1}^\mu/m_\mr i$ and $S_{\mr i1}^{\mu\nu}/m_\mr i$ are each linear functionals of $h_{1\mu\nu}(x)$, and are thus linear in the masses.  From Poincar\'e symmetry, it follows that these results can depend on the positions $y_\mr i$ only through the vectorial impact parameter $b^\mu=y_1^\mu-y_2^\mu$, where the $y_\mr i^\mu$ here are chosen along the two zeroth-order worldlines by the conditions $u_1\cdot b=u_2\cdot b=0$ (at mutual closest approach).  For example, the impulse (net change in momentum) for body 1, $\Delta p_1^\mu=Gp_{11}^\mu(\tau_1\to\infty)+\mc O(G^2)$, is given by\footnote{Results equivalent to the first two lines of Eq.~(\ref{Deltap1PM}) were first derived in Ref.~\cite{Bini:2017xzy}, and the last line results from an expansion in spins of the all-orders-in-spin results for black holes from Ref.~\cite{Vines:2017hyw}, both references having worked from purely classical considerations; see also \cite{Maybee:2019jus,Guevara:2019fsj} for derivations from quantum scattering amplitudes.}
\begin{alignat}{3}\label{Deltap1PM}
\Delta p_1^\mu&=\frac{2Gm_1m_2}{\sqrt{\gamma^2-1}}\bigg[{-}(2\gamma^2-1)\frac{b^\mu}{b^2}
\\\nnm
&\quad+\frac{2\gamma}{b^4}(2b^\mu b^\nu-b^2\eta^{\mu\nu})\epsilon_{\nu\rho\sigma\lambda}u_1^\rho u_2^\sigma (a_1^\lambda+a_2^\lambda)
\\\nnm
&+2\frac{2\gamma^2-1}{b^6}(4b^\mu b^\nu b^\rho-3b^2 b^{(\mu}\Pi^{\nu\rho)})a_{1\nu}a_{2\rho}\bigg]
+\mc O(G^2),
\end{alignat}
where
\be\label{gamma}
\gamma=-u_1\cdot u_2
\ee
is the asymptotic relative Lorentz factor, and $\Pi^\mu{}_\nu=\epsilon^{\mu\rho\alpha\beta}\epsilon_{\nu\rho\gamma\delta}u_{1\alpha}u_{2\beta}u_1^\gamma u_2^\delta/(\gamma^2-1)$ is the projector into the plane orthogonal to both $u_1$ and $u_2$.  Here, as below, we work to linear order in each spin, $a_1$ and $a_2$, keeping the cross term.  We note again in (\ref{Deltap1PM}) the simple dependence on the masses, with an overall factor of $m_1m_2$, at fixed geometric quantities $b^\mu$, $u_\mr i^\mu$ and $a_\mr i^\mu$.  

In continuing the iterative PM solution, the $\mc O(G^n)$ terms in the bodies' degrees of freedom (\ref{expand_xpS}) correct the source (\ref{Tmunu}) for the field equation (\ref{Einstein}), determining the $\mc O(G^{n+1})$ metric perturbation in (\ref{expand_g}); the latter, via the bodies' equations of motion (\ref{MPD}), determines the $\mc O(G^{n+1})$ corrections in (\ref{expand_xpS}).  As in Ref.~\cite{Damour:2019lcq} we are assuming here a systematic use of the time-symmetric Green's function, to pick out the conservative sector of the dynamics.  It becomes evident from the structure of these expansions that the $\mc O(G^n)$ metric perturbation $h_n^{\mu\nu}$ in (\ref{expand_g}) can be expressed as a homogeneous polynomial of degree $n$ in the masses,
\begin{alignat}{3}\label{h1h2}
h_1^{\mu\nu}(x)&=m_1h_{m_1}^{\mu\nu}(x)+m_2h_{m_2}^{\mu\nu}(x),
\nnm\\
h_2^{\mu\nu}(x)&=m_1^2h_{m_1^2}^{\mu\nu}(x)+m_2^2h_{m_2^2}^{\mu\nu}(x)+m_1m_2h_{m_1m_2}^{\mu\nu}(x),
\nnm\\
&\cdots
\end{alignat}
where the $h_{\cdots}^{\mu\nu}$ on the RHSs are functions only of the (asymptotic incoming) geometric quantities $(y_\mr i^\mu,u_\mr i^\mu,a_\mr i^\mu)$ and the field point $x$.  The first line of (\ref{h1h2}) matches (\ref{h1PM}). Similarly, the $\mc O(G^n)$ corrections $x_{\mr in}^\mu$, $p_{\mr in}^\mu/m_\mr i$, $S_{\mr in}^{\mu\nu}/m_\mr i$ for the body degrees of freedom (\ref{expand_xpS}) will be homogeneous polynomials of degree $n$ in the masses; this is the crucial point for the following analysis (and for its conceivable extensions beyond the aligned-spin case).  The zeroth-order quantities $x_{\mr i0}^\mu=y_\mr i^\mu+u_\mr i^\mu\tau_\mr i$, $p_{\mr i0}^\mu/m_\mr i=u_\mr i^\mu$ and $S_{\mr i0}^{\mu\nu}/m_\mr i=\epsilon^{\mu\nu}{}_{\rho\sigma}u_\mr i^\rho a_\mr i^\sigma$ from (\ref{xpS0}) are (taken to be) independent of the masses, as is the zeroth-order metric $h_0=\eta$; they, along with the masses, both (i) fully parametrize the asymptotic incoming state and (ii) can be used to parametrize all the higher-order corrections.

Let us now specialize to the case of aligned spins, in which both spin vectors $a_\mr i^\mu$ are (anti-)parallel to the orbital angular momentum, all of which remain constant throughout the scattering, while the orbital motion is confined to the fixed plane orthogonal to the angular momenta (just as for the nonspinning case).  This entails $u_1\cdot a_\mr i=u_2\cdot a_\mr i=0$ and $b\cdot a_\mr i=0$.  Choosing $\hat z^\mu$ (with $\hat z^2=1$) to be the direction of the orbital angular momentum ($\propto-\epsilon_{\mu\nu\rho\sigma}u_1^\nu u_2^\rho b^\sigma$), let us write $a_\mr i^\mu=a_\mr i\hat z^\mu$ for the constant rescaled spin vectors (equal to their incoming values), where the scalars $a_\mr i$ are positive for spins aligned with $\hat z^\mu$ and negative for anti-aligned.  Crucially, in this case, the only nontrivial independent Lorentz-invariant scalars that can be constructed from the vectors $u_\mr i^\mu$, $a_\mr i^\mu$ and $b^\mu$ are the magnitude $b=(b^2)^{1/2}$ of the impact parameter and the two spin lengths $a_1$ and $a_2$, all three with dimensions of length, and the dimensionless Lorentz factor $\gamma=-u_1\cdot u_2$.

Now consider the extension to higher orders in $G$ of the impulse $\Delta p_1^\mu$ (\ref{Deltap1PM}), which equals $-\Delta p_2^\mu$ (under the conservative dynamics) as the total momentum $p_1^\mu+p_2^\mu$ is conserved.  Its magnitude $\ms Q:=(\Delta p_{1\mu}\Delta p_1^\mu)^{1/2}$ must be a Lorentz-invariant scalar.  In the aligned-spin case, given the previous discussion, and due to Poincar\'e symmetry and dimensional analysis, it must be a function only of the dimensionless scalar $\gamma$ and the dimension-length scalars $b$, $a_1$, $a_2$, $Gm_1$ and $Gm_2$.  Given also the conclusion from above that, in (\ref{expand_xpS}) with $\mr i=1$, $p_{1n}^\mu/m_1$ is a homogeneous polynomial of degree $n$ in the masses, with the leading $n=1$ result seen in (\ref{Deltap1PM}), it follows that the magnitude $\ms Q$ of the impulse must take the following form through fourth order in $G$ (through 4PM order),
\bse\label{QQ}
\begin{alignat}{3}\label{masterQ}
\ms Q&=\frac{2Gm_1m_2}{b}\bigg[\ms Q^\mr{1PM}
\\\nnm
&\quad+\frac{G}{b}\bigg(m_1\ms Q^\mr{2PM}_{m_1}+m_2\ms Q^\mr{2PM}_{m_2}\bigg)
\\\nnm
&\quad+\frac{G^2}{b^2}\bigg(m_1^2\ms Q^\mr{3PM}_{m_1^2}+m_2^2\ms Q^\mr{3PM}_{m_2^2}+m_1m_2 \ms Q_{m_1m_2}^\mr{3PM}\bigg)
\\\nnm
&\quad+\frac{G^3}{b^3}\bigg(m_1^3\ms Q^\mr{4PM}_{m_1^3}+m_2^3\ms Q^\mr{4PM}_{m_2^3}
\\\nnm
&\qquad\qquad+m_1^2m_2 \ms Q_{m_1^2m_2}^\mr{4PM}+m_1m_2^2 \ms Q_{m_1m_2^2}^\mr{4PM}\bigg)\bigg]
\\\nnm
&\quad+\mc O(G^5),
\end{alignat}
where the $\ms Q$'s on the RHS are functions of the dimensionless scalars $\gamma$, $a_1/b$ and $a_2/b$,
\begin{alignat}{3}\label{Qcoeffs}
\ms Q^{n\mr{PM}}_{m_1^im_2^j}&=\ms Q^{n\mr{PM}}_{m_1^im_2^j}(\gamma,\frac{a_1}{b},\frac{a_2}{b})
\\\nnm
&=\ms Q^{n\mr{PM}}_{m_1^im_2^ja^0}(\gamma)
\\\nnm
&\quad+\frac{a_1}{b}\ms Q^{n\mr{PM}}_{m_1^im_2^ja_1}(\gamma)
+\frac{a_2}{b}\ms Q^{n\mr{PM}}_{m_1^im_2^ja_2}(\gamma)
\\\nnm
&\quad+\frac{a_1a_2}{b^2}\ms Q^{n\mr{PM}}_{m_1^im_2^ja_1a_2}(\gamma)
\end{alignat}
\ese
(with $i+j=n-1$).  In the second equality, we have expanded to linear order in each spin (assuming regular limits as the spins go to zero), and we are finally left with a set of undetermined functions depending only on the Lorentz factor $\gamma$.

Furthermore, $\ms Q$ must be invariant under an exchange of the two bodies' identities, $(m_1,a_1)\leftrightarrow(m_2,a_2)$.  At 1PM order, this tells us that $\ms Q^\mr{1PM}(\gamma,a_1/b,a_2/b)$ is symmetric under $a_1\leftrightarrow a_2$, and thus $\ms Q^{\mr{1PM}}_{a_1}=\ms Q^{\mr{1PM}}_{a_2}$, so that the third line of (\ref{Qcoeffs}) in this case is proportional to $a_1+a_2$.  Indeed, the explicit expression for $\ms Q^\mr{1PM}$ is given by the magnitude of the aligned-spin specialization of (\ref{Deltap1PM}) (divided by $2Gm_1m_2/b$),\footnote{Note that this is the expansion to linear order in the spins of the result (80) from~\cite{Vines:2017hyw} for a two-black-hole system,
\be
\ms Q^\mr{1PM}=\bigg(\frac{2\gamma^2-1}{\sqrt{\gamma^2-1}}-2\gamma\frac{a_1+a_2}{b}\bigg)\bigg(1-\frac{(a_1+a_2)^2}{b^2}\bigg)^{-1},
\ee
to all orders in the spin-multipole expansion at 1PM order.}
\be\label{Q1PM}
\ms Q^\mr{1PM}=\frac{2\gamma^2-1}{\sqrt{\gamma^2-1}}\bigg(1+2\frac{a_1a_2}{b^2}\bigg)-2\gamma\frac{a_1+a_2}{b}.
\ee
At 2PM order, the $1\leftrightarrow2$ symmetry tells us that each of the two functions in the second line of (\ref{masterQ}) determines the other,
\be\label{Q2PMsymm}
\ms Q^\mr{2PM}_{m_1}(\gamma,\frac{a_1}{b},\frac{a_2}{b})
=
\ms Q^\mr{2PM}_{m_2}(\gamma,\frac{a_2}{b},\frac{a_1}{b}).
\ee
This function, like $\ms Q^\mr{1PM}$, is in fact fully determined by the (extended) test-body limit of $\ms Q/(m_1m_2)$ --- the limit where one of the masses, say, $m_1$, goes to zero, while keeping fixed $m_2$, $a_2$ and $a_1$ (and $\gamma$ and $b$).  The result for $\ms Q/m_1$ in this limit can be consistently determined by solving the pole-dipole MPD equations (\ref{MPD}) for a spinning test body in a stationary Kerr background;
we will present explicit results from this procedure below in terms of the scattering-angle function.  This test-body limit, with $m_1\to0$, determines all of the functions $\ms Q^{n\mr{PM}}_{m_2^{n-1}}$ with no powers of $m_1$, for all $n$, and the $1\leftrightarrow2$ symmetry also tells us that
\be\label{Qmnm1}
\ms Q^{n\mr{PM}}_{m_1^{n-1}}(\gamma,\frac{a_1}{b},\frac{a_2}{b})
=
\ms Q^{n\mr{PM}}_{m_2^{n-1}}(\gamma,\frac{a_2}{b},\frac{a_1}{b}).
\ee
The only remaining functions in (\ref{masterQ}), those not determined by the test-body limit and exchange symmetry, are $\ms Q^\mr{3PM}_{m_1m_2}$, $\ms Q^\mr{4PM}_{m_1^2m_2}$ and $\ms Q^\mr{4PM}_{m_1m_2^2}$.  They are however still constrained by the exchange symmetry as follows. Firstly,
\be\label{Q3PMsymm}
\ms Q^\mr{3PM}_{m_1m_2}(\gamma,\frac{a_1}{b},\frac{a_2}{b})
=
\ms Q^\mr{3PM}_{m_1m_2}(\gamma,\frac{a_2}{b},\frac{a_1}{b}),
\ee
which implies that the third line of (\ref{Qcoeffs}) for $\ms Q^\mr{3PM}_{m_1m_2}$ (like for $\ms Q^\mr{1PM}$ above) is proportional to $a_1+a_2$.  Secondly,
\be
\ms Q^\mr{4PM}_{m_1^2m_2}(\gamma,\frac{a_1}{b},\frac{a_2}{b})
=
\ms Q^\mr{4PM}_{m_1m_2^2}(\gamma,\frac{a_2}{b},\frac{a_1}{b}),
\ee
so that one of these two functions determines the other.

Taking all of these constraints from exchange symmetry, we can eliminate all of the $\ms Q$'s with more $m_1$'s in the subscript for those with more $m_2$'s, while those with the same number of $m_1$'s and $m_2$'s must be symmetric under $a_1\leftrightarrow a_2$.  First focusing on the nonspinning ($a^0$) part of (\ref{masterQ}), this becomes
\begin{alignat}{3}
&\ms Q_{a^0}=\frac{2Gm_1m_2}{b}\bigg[\ms Q^\mr{1PM}_{a^0}
+\frac{G}{b}(m_1+m_2)\ms Q^\mr{2PM}_{m_2a^0}
\nnm\\
&\quad+\frac{G^2}{b^2}\bigg((m_1^2+m_2^2)\ms Q^\mr{3PM}_{m_2^2a^0}+m_1m_2 \ms Q_{m_1m_2a^0}^\mr{3PM}\bigg)
\\\nnm
&+\frac{G^3}{b^3}\bigg((m_1^3+m_2^3)\ms Q^\mr{4PM}_{m_2^3a^0}
+m_1m_2(m_1+m_2) \ms Q_{m_1m_2^2a^0}^\mr{4PM}\bigg)\bigg],
\end{alignat}
recalling that all the $\ms Q$'s on the right-hand side are functions only of $\gamma$ [henceforth dropping $+\mc O(G^5)$].
Introducing the total rest mass $M=m_1+m_2$ and the symmetric mass ratio $\nu=m_1m_2/M^2=\mu/M$ as in (\ref{massmap}), and noting
\begin{alignat}{3}
m_1+m_2&=M,
\nnm\\
m_1^2+m_2^2&=M^2(1-2\nu),
\\\nnm
m_1^3+m_2^3&=M^3(1-3\nu),
\end{alignat}
this becomes 
\begin{alignat}{3}\label{Qa0}
\ms Q_{a^0}&=\frac{2Gm_1m_2}{b}\bigg[\ms Q^\mr{1PM}_{a^0}
+\frac{GM}{b}\ms Q^\mr{2PM}_{m_2a^0}
\nnm\\
&\quad+\Big(\frac{GM}{b}\Big)^2\bigg(\ms Q^\mr{3PM}_{m_2^2a^0}+\nu \tilde{\ms Q}_{m_1m_2a^0}^\mr{3PM}\bigg)
\\\nnm
&\quad+\Big(\frac{GM}{b}\Big)^3\bigg(\ms Q^\mr{4PM}_{m_2^3a^0}+\nu \tilde{\ms Q}_{m_1m_2^2a^0}^\mr{4PM}\bigg)\bigg],
\end{alignat}
where we defined $\tilde{\ms Q}_{m_1m_2a^0}^\mr{3PM}:={\ms Q}_{m_1m_2a^0}^\mr{3PM}-2\ms Q^\mr{3PM}_{m_2^2a^0}$ and $\tilde{\ms Q}_{m_1m_2^2a^0}^\mr{4PM}:={\ms Q}_{m_1m_2^2a^0}^\mr{4PM}-3\ms Q^\mr{4PM}_{m_2^3a^0}$, still functions only of $\gamma$.  Remarkably, through 4PM order, this is just linear in the mass ratio $\nu$ at fixed $M$.  Precisely the same manipulations go through for the $a_1a_2$ terms, replacing $a^0$ with $a_1a_2$ in all the subscripts and with an overall factor of $a_1a_2/b^2$ on the right-hand side.

Next consider just the 1PM and 2PM terms of the SO ($a^1$) part of (\ref{masterQ}), after accounting for the exchange symmetry in the same way as in the previous paragraph (with $\ms Q^\mr{1PM}_{a_1}=\ms Q^\mr{1PM}_{a_2}$, $\ms Q^\mr{2PM}_{m_1a_1}=\ms Q^\mr{2PM}_{m_2a_2}$ and $\ms Q^\mr{2PM}_{m_1a_2}=\ms Q^\mr{2PM}_{m_2a_1}$); we find
\begin{alignat}{3}\label{Qa12}
&\ms Q_{a^1}+\mc O(G^3)=\frac{2Gm_1m_2}{b}\bigg[\frac{a_1+a_2}{b}\ms Q^\mr{1PM}_{a_2}
\\\nnm
&+\frac{G}{b}\bigg(
\frac{m_1a_1+m_2a_2}{b}\ms Q^\mr{2PM}_{m_2a_2}+\frac{m_2a_1+m_1a_2}{b}\ms Q^\mr{2PM}_{m_2a_1}
\bigg)\bigg].
\end{alignat}
We recognize in the second line the following spin combinations often used in the PN and EOB literature,
\begin{alignat}{3}
\begin{aligned}
S&:=m_1a_1+m_2a_2=S_1+S_2,
\\
S_*&:=m_2a_1+m_1a_2=\frac{m_2}{m_1}S_1+\frac{m_1}{m_2}S_2.
\end{aligned}
\end{alignat}
We will find it convenient to rescale each of these by the total rest mass $M$, defining
\begin{alignat}{3}\label{abat}
\begin{aligned}
a_\mr b&:=\frac{S}{M}=\frac{m_1a_1+m_2a_2}{m_1+m_2},
\\
a_\mr t&:=\frac{S_*}{M}=\frac{m_2a_1+m_1a_2}{m_1+m_2},
\end{aligned}
\end{alignat}
where b stands for background (or big) and t stands for test (or tiny).  The (first) reason for these labels is that, in the extended test-body limit [$m_1\to0$ at fixed $m_2$ (or $M$) and fixed $a_1$ and $a_2$], we see that $a_\mr b\to a_2$ becomes the spin-per-mass of the big background object with mass $M=m_2$, and $a_\mr t\to a_1$ becomes the spin-per-mass of the tiny spinning test body with negligible mass (with a further reason explained below).  Note that $a_\mr b+a_\mr t=a_1+a_2$.  Now extending (\ref{Qa12}) to 4PM order, from (\ref{masterQ}) accounting for exchange symmetry, using our new notation, we find
\begin{alignat}{3}\label{Qa1234}
&\ms Q_{a^1}=\frac{2Gm_1m_2}{b^2}\bigg[\ms Q^\mr{1PM}_{a_2}(a_\mr b+a_\mr t)
\\\nnm
&\quad+\frac{GM}{b}\bigg(
\ms Q^\mr{2PM}_{m_2a_2}a_\mr b+\ms Q^\mr{2PM}_{m_2a_1}a_\mr t
\bigg)
\\\nnm
&+\Big(\frac{GM}{b}\Big)^2\bigg(
\ms Q^\mr{3PM}_{m_2^2a_2}a_\mr b+\ms Q^\mr{3PM}_{m_2^2a_1}a_\mr t+\nu \tilde{\ms Q}^\mr{3PM}_{m_1m_2a_2}(a_\mr b+a_\mr t)
\bigg)
\\\nnm
&+\Big(\frac{GM}{b}\Big)^3\bigg(
\ms Q^\mr{4PM}_{m_2^3a_2}a_\mr b+\ms Q^\mr{4PM}_{m_2^3a_1}a_\mr t
\\\nnm
&\qquad\quad+\nu\Big[ \tilde{\ms Q}^\mr{4PM}_{m_1m_2^2a_2}a_\mr b+\tilde{\ms Q}^\mr{4PM}_{m_1m_2^2a_2}a_\mr t\Big]
\bigg)\bigg],
\end{alignat}
where we defined $\tilde{\ms Q}^\mr{3PM}_{m_1m_2a_2}={\ms Q}^\mr{3PM}_{m_1m_2a_2}-{\ms Q}^\mr{3PM}_{m_2^2a_2}-{\ms Q}^\mr{3PM}_{m_2^2a_1}$, $\tilde{\ms Q}^\mr{4PM}_{m_1m_2^2a_2}:={\ms Q}^\mr{4PM}_{m_1m_2^2a_2}-2\ms Q^\mr{4PM}_{m_2^3a_2}-\ms Q^\mr{4PM}_{m_2^3a_1}$ and $\tilde{\ms Q}^\mr{4PM}_{m_1m_2^2a_2}:={\ms Q}^\mr{4PM}_{m_1m_2^2a_2}-\ms Q^\mr{4PM}_{m_2^3a_2}-2\ms Q^\mr{4PM}_{m_2^3a_1}$, all still functions only of $\gamma$.  We see that (\ref{Qa1234}), like (\ref{Qa0}), is linear in the symmetric mass ratio $\nu$ (at fixed $M$, $a_\mr b$ and $a_\mr t$).

Now, just as in Eq.~(2.14) of \cite{Damour:2019lcq} --- following from conservation of the total momentum $p_1^\mu+p_2^\mu$ and simple geometry and kinematics (which is identical for the nonspinning and aligned-spin cases) --- the scattering angle $\chi$, by which both bodies are deflected in the system's center-of-mass (cm) frame, is related to the magnitude $\ms Q$ of the impulse by
\be\label{chiQ}
\sin\frac{\chi}{2}=\frac{\ms Q}{2p_\infty},
\ee
where $p_\infty$ (called ``$P_\mr{c.m.}$'' by Damour) is the magnitude of the bodies' equal and opposite spatial momenta in the cm frame, at infinity,
\be\label{pcm}
p_\infty=\frac{m_1m_2}{E}\sqrt{\gamma^2-1}.
\ee
Here, $E$ is the total energy in the cm frame,
\begin{alignat}{3}\label{Etotal}
E^2&=m_1^2+m_2^2+2m_1m_2\gamma
\nnm\\
&=M^2(1+2\nu(\gamma-1)),
\end{alignat}
determined by the asymptotic Lorentz factor $\gamma$ and the rest masses.  Note also the definition of the asymptotic relative velocity $v$ as used e.g.\ in \cite{Vines:2018gqi,Siemonsen:2019dsu,Antonelli:2020aeb},
\be\label{gammav}
v=\frac{\sqrt{\gamma^2-1}}{\gamma}\quad\Leftrightarrow\quad\gamma=\frac{1}{\sqrt{1-v^2}}.
\ee
We will find it convenient to define yet another variable equivalent to $\gamma$ or $v$, namely
\be
\ve:=\gamma^2-1=\gamma^2v^2=\Big(\frac{p_\infty E}{m_1m_2}\Big)^2,
\ee
which, like $v^2$, can serve as a PN expansion parameter, and unlike $v$, is real for both unbound and bound orbits,
\begin{alignat}{3}
\begin{aligned}
\tr{unbound:}\quad E>M\quad\Leftrightarrow\quad\ve&>0, 
\\
\tr{bound:}\quad E<M\quad\Leftrightarrow\quad\ve&<0, 
\end{aligned}
\end{alignat}
noting that $v=i\sqrt{1-\gamma^2}/\gamma$ and $p_\infty$ are imaginary for bound orbits.  (Note that our $\ve=\gamma^2v^2$ is Damour's ``$p_\infty^2=p_\mr{eob}^2$'' [the squared momentum per mass of the effective test body], while our $p_\infty$ is Damour's ``$P_\mr{c.m.}$''.)
We will also find it convenient to define a notation for the dimensionless ratio $\Gamma$ (Damour's ``$h$'') between the total energy and the total rest mass,
\be\label{defGamma}
\Gamma:=\frac{E}{M}=\sqrt{1+2\nu(\gamma-1)},
\ee
with $\Gamma>1$ ($\gamma>1$) for unbound orbits, and $\Gamma<1$ ($\gamma<1$) for bound orbits.  Then $p_\infty=\mu\gamma v/\Gamma=\mu\sqrt{\ve}/\Gamma$.

With this notation in order, we can take our simplified result for the impulse magnitude $\ms Q$ (\ref{masterQ}) [namely the sum of (\ref{Qa0}), its analogous $a_1a_2$ version, and the SO part (\ref{Qa1234})], insert it into (\ref{chiQ}), and solve for the aligned-spin scattering angle $\chi$.  After this process, $\chi/\Gamma$ turns out to be linear in $\nu$ in the same way that $\ms Q$ is, thanks to the facts that the sine function is odd in its argument and that $\Gamma^2$ is linear in $\nu$.  The result can be expressed as follows,
\bse\label{masterchi}
\begin{alignat}{3}\label{chiform}
\frac{\chi}{\Gamma} &= \frac{GM}{b\sqrt{\ve}}\ms X_{G^1}^{\nu^0}
\\\nnm
&\quad+\Big(\frac{GM}{b\sqrt{\ve}}\Big)^2\ms X_{G^2}^{\nu^0}
\\\nnm
&\quad +\Big(\frac{GM}{b\sqrt{\ve}}\Big)^3\Big[\ms X_{G^3}^{\nu^0}+\nu \ms X_{G^3}^{\nu^1}\Big]
\\\nnm
&\quad +\Big(\frac{GM}{b\sqrt{\ve}}\Big)^4\Big[\ms X_{G^4}^{\nu^0}+\nu \ms X_{G^4}^{\nu^1}\Big]+\mc O\Big(\frac{GM}{b}\Big)^5,
\end{alignat}
where each $\ms X_{G^k}^{\nu^m}$ takes the form
\begin{alignat}{3}\label{Xkm}
\ms X_{G^k}^{\nu^m} 
&= \ms X_{k}^{m}(\ve)
\\\nnm
&\quad+\frac{a_\mr b}{b\sqrt{\ve}}\ms X_{k}^{m\mr b}(\ve)+\frac{a_\mr t}{b\sqrt{\ve}}\ms X_{k}^{m\mr t}(\ve)
\\\nnm
&\quad+\frac{a_1a_2}{b^2\ve}\ms X_{k}^{m\times}(\ve),
\end{alignat}
\ese
with $\times$ standing for the ``cross term'' $a_1a_2$, and with the special constraints
\be\label{speccons}
\ms X^{0\mr b}_{1}=\ms X^{0\mr t}_{1},\qquad \ms X^{1\mr b}_{3}=\ms X^{1\mr t}_{3},
\ee
recalling from (\ref{abat}) that $Ma_\mr b=m_1a_1+m_2a_2$ and $M a_\mr t=m_2a_1+m_1a_2$.\footnote{In Ref.~\cite{Antonelli:2020aeb}, the expression of the result (\ref{masterchi}) for the mass dependence of the scattering angle differed in that (i) we did not pull a factor of $1/\sqrt{\ve}$ out of the $X$'s for every factor of $1/b$, (ii) we used $v$ instead of $\ve$, and (iii) we used $a_+$ and $\delta\,a_-$ in place of $a_\mr b$ and $a_\mr t$, with $a_\pm:=a_2\pm a_1$ and $\delta:=(m_2-m_1)/M$; the equivalence of the two expressions is apparent since
\begin{alignat}{3}
\begin{aligned}
a_++\delta\,a_-&=2a_\mr b,
\\
a_+-\delta\,a_-&=2a_\mr t.
\end{aligned}
\end{alignat}}
All the $\ms X$'s on the right-hand side of (\ref{Xkm}) are dimensionless and are functions \emph{only} of the dimensionless $\ve=\gamma^2-1$; they can be expressed in terms of the above $\ms Q(\gamma)$'s alone.

We see that the 1PM and 2PM terms in (\ref{masterchi}) are independent of the symmetric mass ratio $\nu$ and are thus fully preserved in the (\emph{extended}) test-body limit $\nu\to0$ (at fixed $M$, or equivalently $m_1\to 0$ at fixed $M$, \emph{and} at fixed $a_1$, $a_2$, $b$ and $\gamma$), while the 3PM and 4PM terms are linear in $\nu$.  This allows us to deduce the complete 1PM and 2PM results for $\chi/\Gamma$ from its test-body limit, and the complete 3PM and 4PM results from first-order self-force (linear-in-mass-ratio) calculations.

The special constraints (\ref{speccons}) are consequences of the $1\leftrightarrow2$ symmetry, as seen in the $G^1\nu^0$ and $G^3\nu^1$ SO terms in (\ref{Qa1234}).  This is a prediction of the above arguments which our considerations below will be able to test, rather than to rely on.  For the case of the $G^3\nu^1$ SO terms, which we will determine (in a PN expansion) below from matching to first-order self-force calculations, we will allow $\ms X^1_{3\mr b}$ and $\ms X^1_{3\mr t}$ to be independent --- in fact, $\ms X^1_{3\mr b}$ will be determined by the redshift invariant in a Kerr background and $\ms X^1_{3\mr t}$ by the spin-precession invariant in a Schwarzschild background --- and we will find from the matching procedure that they are indeed equal through the considered PN orders.
The fact that the complete content of Eqs.~(\ref{masterchi}) holds through N$^2$LO in the PN expansion can be seen in Eqs.~(4.32) of Ref.~\cite{Vines:2018gqi}.

The $\nu^0$ terms in (\ref{masterchi}) can be determined by solving the MPD equations of motion (\ref{MPD}) for a spinning (pole-dipole) test body in a stationary background Kerr spacetime.  An integrand for the test-spin-in-Kerr aligned-spin scattering angle function, to all PM orders, was derived in Ref.~\cite{Bini:2017pee}; see, e.g., their Eq.~(66) (which also includes pole-dipole-quadrupole terms for a test black hole).  The results of the integration are as follows, to all orders in $\ve$ (to all PN orders at each PM order), extending Eq.~(5.5) of Ref.~\cite{Vines:2018gqi} to 4PM order in the spin-orbit and bilinear-in-spin terms.  The nonspinning parts are
\bse\label{X0s}
\begin{alignat}{3}\label{X0a0}
\ms X_{1}^{0}&=2\frac{1+2\ve}{\sqrt{\ve}}=2\frac{2\gamma^2-1}{\sqrt{\gamma^2-1}}=2\frac{1+v^2}{v\sqrt{1-v^2}},
\\\nnm
\ms X_{2}^{0}&=\frac{3\pi}{4}(4+5\ve)=\frac{3\pi}{4}(5\gamma^2-1),
\\\nnm
\ms X_{3}^{0}&=2\frac{-1+12\ve+72\ve^2+64\ve^3}{3\ve^{3/2}},
\\\nnm
\ms X_{4}^{0}&=\frac{105\pi}{64}(16+48\ve+33\ve^2),
\end{alignat}
the SO parts are
\begin{alignat}{3}\label{X0a1}
\ms X_{1}^{0\mr b}a_\mr b+\ms X_{1}^{0\mr t}a_\mr t&=-4\gamma\sqrt{\ve}(a_\mr b+a_\mr t),
\\\nnm
\ms X_{2}^{0\mr b}a_\mr b+\ms X_{2}^{0\mr t}a_\mr t&=-\frac{\pi}{2}\gamma(2+5\ve)(4a_\mr b+3a_\mr t),
\\\nnm
\ms X_{3}^{0\mr b}a_\mr b+\ms X_{3}^{0\mr t}a_\mr t&=-4\gamma\frac{1+12\ve+16\ve^2}{\sqrt{\ve}}(3a_\mr b+2a_\mr t),
\\\nnm
\ms X_{4}^{0\mr b}a_\mr b+\ms X_{4}^{0\mr t}a_\mr t&=-\frac{21\pi}{16}\gamma(8+36\ve+33\ve^2)(8a_\mr b+5\mr a_t),
\end{alignat}
and the bilinear-in-spin parts are
\begin{alignat}{3}\label{X0a1a2}
\ms X_{1}^{0\times}&=4\sqrt{\ve}(1+2\ve),
\\\nnm
\ms X_{2}^{0\times}&=\frac{3\pi}{2}(2+19\ve+20\ve^2),
\\\nnm
\ms X_{3}^{0\times}&=8\frac{1+38\ve+128\ve^2+96\ve^3}{\sqrt{\ve}},
\\\nnm
\ms X_{4}^{0\times}&=\frac{105\pi}{16} (24+212\ve+447\ve^2+264\ve^3)
\end{alignat}
\ese
with $\gamma=\sqrt{1+\ve}$.\footnote{Note that, through 2PM order and up through the SO terms, the first two lines of the right-hand side of (\ref{chiform}), with (\ref{X0a0}) and (\ref{X0a1}) plugged into the first two lines of (\ref{Xkm}), correctly give either (i) the aligned-spin scattering angle for a spinning test body with rescaled spin $a_\mr t$ in a Kerr background with mass $M$ and rescaled spin $a_\mr b$, or (ii) the rescaled aligned-spin scattering angle $\chi/\Gamma$ for the arbitrary-mass two-spinning-body system, using the ``spin maps'' (\ref{abat}); this is a further reason for the labels $a_\mr t$ and $a_\mr b$.  This gives a different ``EOB scattering-angle mapping,'' an alternative to Eq.~(3.16) of \cite{Vines:2018gqi}, which produces the 1PM and 2PM SO terms in the two-body scattering angle from its extended test-body limit.  [Note however that this different mapping fails at quadratic order in the spins, while Eq.~(3.16) of \cite{Vines:2018gqi} still holds, according to all known results.]}

The $\nu^1$ terms  in (\ref{masterchi}), at 3PM and 4PM orders, can be determined in a PN expansion (here, an expansion in $\ve$) from first-order self-force results (as well as from consistency with lower orders), as we will explicitly demonstrate below for the spin parts.  We will use the known nonspinning coefficients through 4PM-3PN order \cite{Vines:2018gqi},
\bse\label{X1s}
\begin{alignat}{3}\label{X1a0}
\ms X_{3}^{1}&=-\frac{8+94\ve+313\ve^2+\mc O(\ve^3)}{12\sqrt{\ve}},
\\\nnm
\ms X_{4}^{1}&=\pi \bigg[{-}\frac{15}{2}+\bigg(\frac{123}{128}\pi^2-\frac{557}{8}\bigg)\ve+\mc O(\ve^2)\bigg],
\end{alignat}
noting the transcendental $\zeta(2)$ contribution in the last term (the 4PM-3PN term).  We will parametrize 
the SO coefficients as
\begin{alignat}{3}\label{SOchipar}
\ms X_{3}^{1\mr i}&=\frac{\gamma}{\sqrt{\ve}}\Big(\ms X^{1\mr i}_{30}+\ms X^{1\mr i}_{31}\ve+\ms X^{1\mr i}_{32}\ve^2+\ms X^{1\mr i}_{33}\ve^3+\mc O(\ve^4)\Big),
\nnm\\
\ms X_{4}^{1\mr i}&=\pi\gamma\Big(\ms X^{1\mr i}_{41}+\ms X^{1\mr i}_{42}\ve+\ms X^{1\mr i}_{43}\ve^2+\mc O(\ve^3)\Big),
\end{alignat}
with $\mr i=\mr b, \mr t$, and the bilinear-in-spin coefficients as
\begin{alignat}{3}\label{S1S2chipar}
\ms X_{3}^{1\times}&=\frac{1}{\sqrt{\ve}}\Big(\ms X^{1\times}_{30}+\ms X^{1\times}_{31}\ve+\ms X^{1\times}_{32}\ve^2+\ms X^{1\times}_{33}\ve^3+\mc O(\ve^4)\Big),
\nnm\\
\ms X_{4}^{1\times}&=\pi\Big(\ms X^{1\times}_{41}+\ms X^{1\times}_{42}\ve+\ms X^{1\times}_{43}\ve^2+\mc O(\ve^3)\Big).
\end{alignat}
\ese
 We have included all the same powers of $\ve$ present in the $\nu^0$ coefficients (\ref{X0s}), up to the orders in $\ve$ which will contribute at the N$^3$LO PN level.  (We have also factored out $\gamma=\sqrt{1+\ve}$ in the SO terms and $\pi$ in the 4PM terms, following the patterns at $\nu^0$.) For these $\ms X^{1\cdots}_{kn}$, which are all pure numbers, $k$ gives the PM order, and $n$ gives the maximum PN order (N$^n$LO) which determines that coefficient.  This labeling and the consistency and sufficiency of this ansatz for the scattering angle will become evident in the matching between the scattering angle and a canonical Hamiltonian described in the following section.

Finally, it is important to note that the impact parameter $b$ appearing everywhere in this section is the distance orthogonally separating the two spinning bodies' asymptotic incoming worldlines as defined by the ``covariant'' or Tulczyjew-Dixon condition~\cite{Dixon:1979,Steinhoff:2014kwa,Tulczyjew:1959,fokker1929relativiteitstheorie}, Eq.~(\ref{TDSSC}) above, for each body---the so-called ``proper'' or ``covariant'' impact parameter $b\equiv b_\mr{cov}$ \cite{Vines:2018gqi,Guevara:2018wpp,Siemonsen:2019dsu}.  This is crucial to the above argument because only with the covariant condition (\ref{TDSSC}) (or something equivalent to it at 0PM order) does it hold that the first-order field (\ref{h1PM}) is linear in the masses.  Below, we will also work with the canonical orbital angular momentum $L\equiv L_\mr{can}=p_\infty b_\mr{can}$, where $b_\mr{can}$ is the impact parameter orthogonally separating the asymptotic incoming worldlines defined by cm-frame Newton-Wigner conditions~\cite{pryce1948mass,newton1949localized} for each body.  This coincides with the conserved canonical orbital angular momentum $L$ appearing in a canonical Hamiltonian formulation of aligned-spin two-body dynamics~\cite{Barausse:2009aa,Vines:2016unv}.  [Note that, for the aligned-spin case, the covariant/Pauli-Lubanski spin vectors $m_\mr ia_\mr i^\mu$ used above coincide with the canonical spin vectors $S_\mr i^\mu$ (spatial vectors in the cm frame) which would be associated with the cm-frame Newton-Wigner conditions, and thus so do the aligned-spin (signed) magnitudes, $S_\mr i=m_\mr ia_\mr i$.]  As shown in \cite{Vines:2017hyw,Vines:2018gqi}, the canonical $L=:L_\mr{can}$ is related to the covariant $b$ by
\begin{alignat}{3}\label{DeltaL}
L&=L_\mr{cov}+\Delta L,
\\\nnm
L_\mr{cov}&=p_\infty b=\frac{\mu}{\Gamma}\gamma v b=\frac{\mu}{\Gamma}\sqrt{\ve} b,
\\\nnm
\Delta L&=\Big(\sqrt{m_1^2+p_\infty^2}-m_1\Big)a_1+\Big(\sqrt{m_2^2+p_\infty^2}-m_2\Big)a_2
\\\nnm
&=M\frac{\Gamma-1}{2}\bigg(a_\mr b+a_\mr t-\frac{a_\mr b-a_\mr t}{\Gamma}\bigg).
\end{alignat}
Solving this for $b$, inserting the result into (\ref{masterchi}) [or (\ref{masterchi2})], and re-expanding to bilinear order in the (mass-rescaled) spins $a_1$ and $a_2$, one obtains the final parametrized form for the aligned-spin scattering angle function $\chi(E,L;m_\mr i,a_\mr i)$ used in the following matching calculations.

Let us finally rewrite the scattering angle to include both the $\nu^0$ and $\nu^1$ terms in single coefficients (or which could allow mass-dependence differing from that deduced above), and which would accommodate general quadratic-in-spin terms, with sums over i and j implied,
\be\label{masterchi2}
\frac{\chi}{\Gamma}=\sum_{k\ge1}\Big(\frac{GM}{b\sqrt{\ve}}\Big)^k\bigg[\ms X_k(\ve,\nu)+\frac{a_\mr i}{b\sqrt{\ve}}\ms X_k{}^{\mr i}(\ve,\nu)+\frac{a_\mr i a_\mr j}{b^2{\ve}}\ms X_k{}^{\mr i\mr j}(\ve,\nu)\bigg]
\ee
$+\mc O(a^3)$, with 
\begin{alignat}{3}
a_\mr i \ms X_k{}^{\mr i}&=a_\mr b \ms X_k{}^{\mr b}+a_\mr t \ms X_k{}^{\mr t},
\\\nnm
a_\mr i a_\mr j\ms X_k{}^{\mr i\mr j}&=a_1 a_2 \ms X_k{}^{\!\times}+\mc O(a_1^2,a_2^2).
\end{alignat} 
 Our prediction for the mass-ratio dependence of the $k$PM coefficients ${\ms X}_k{}^{\mr A}=\{{\ms X}_k,{\ms X}_k{}^{\mr b},{\ms X}_k{}^{\mr t},{\ms X}_k{}^{\!\times}\}$ is that 
\be\label{joinXs}
{\ms X}_k{}^{\mr A}(\ve,\nu)=\left\{
\begin{array}{cc}
{\ms X}_k^{0\mr A}(\ve),\quad &k=1,2 
\\
{\ms X}_k^{0\mr A}(\ve)+\nu {\ms X}_k^{1\mr A}(\ve),\quad &k=3,4
\end{array}\right..
\ee
The $\nu^0$ coefficients $\ms X^{0\mr A}_k(\ve)$ from the extended test-body limit are given explicitly in (\ref{X0s}), and the $\nu^1$ coefficients $\ms X^{1\mr A}_k(\ve)$ which we will determine from self-force results are parametrized in a PN expansion in (\ref{X1s}).  Note that we will also be able to use the self-force results to test the fact that there are no $\nu^1$ terms  at 1PM and 2PM orders in this parametrization of the scattering angle.  The fact that there are no $\nu^2$ or higher terms through 4PM order cannot be probed with first-order self-force results, but has already been confirmed by arbitrary-mass PN results through N$^2$LO.  Our prediction for the mass-dependence will yield new arbitrary-mass results at the N$^3$LO PN level once we have fixed the PN expansions of the coefficients $\ms X^{1\mr A}_k$ from first-order self-force calculations.

\section{From the unbound scattering angle to the bound radial action via canonical Hamiltonian dynamics}
\label{Irsec}

Besides the mass dependence of the scattering angle function established in the previous section, and the inputs of test-body results (discussed above) and first-order self-force results (discussed below), the other central ingredient in our derivation is the assumption of the existence of a (local-in-time) canonical Hamiltonian governing the aligned-spin conservative dynamics in the cm frame, for generic (\emph{both} bound and unbound) orbits, with the Hamiltonian having well-defined (regular, polynomial) PN and PM expansions. Through the desired 4.5PN order in the SO sector and 5PN {\sonestwo} one, we can safely ignore nonlocal-in-time (tail) contributions in the final dynamics/scattering angle. While these do appear at the 4PN level in the nonspinning sector~\cite{Damour:2014jta} (see e.g., Ref.~\cite{Bini:2017wfr} for a translation into a nonlocal-in-time scattering angle), they only start appearing at 5.5PN order in the spinning one. This can most easily be seen in the first line of Eq.(68a) in Ref.~\cite{Siemonsen:2017yux}, where the linear-in-spin tails are a relative 1.5PN order from the leading quadrupolar contributions to the tail. [As mentioned at the very end of this section, we find it necessary to include tail terms at 4PN order in the nonspinning sector to make contact with available results in the GSF literature.]

Our ultimate goal in this section is to take the gauge-invariant scattering-angle function $\chi$ for unbound orbits, parametrized in the previous section, and derive from it a parametrized expression for the gauge-invariant radial-action function $I_r$ which characterizes bound orbits, from which we can derive all the bound-orbit gauge invariants to be compared with self-force results in Sec.~\ref{sec:smallq} below.  

We do this by passing through the gauge-dependent canonical Hamiltonian dynamics.  It is to some extent true that this process (as we implement it here) can be bypassed by using relationships between gauge invariants for unbound and bound orbits found in \cite{Kalin:2019inp}, but not entirely.  Those relationships yield $I_r$ through $\mc O(G^4)$ from $\chi$ through $\mc O(G^4)$, but the complete PN expansion of $I_r$ through N$^3$LO extends to $\mc O(G^8)$ (for the spin terms).  The extra terms in $I_r$ are obtained here via the canonical Hamiltonian dynamics, which determines them from (the PN re-expansion of) $\chi$ through $\mc O(G^4)$.  Note that $\chi$ through $\mc O(G^4)$ does not contain the complete PN expansion of $\chi$ through N$^3$LO, nor through LO, since even the Newtonian scattering angle has contributions at all orders in $G$.  But the PN expansion of the 4PM scattering angle, $\chi$ through $\mc O(G^4)$, does contain the complete information of the N$^3$LO PN Hamiltonian (contained in its $O(G^4)$ truncation), which determines the N$^3$LO PN radial action $I_r$ (contained in its $O(G^8)$ truncation).

We begin in Sec.~\ref{sec:massshell} by discussing canonical Hamiltonians for aligned-spin binaries, the resultant equations of motion, and their gauge freedom under canonical transformations, in a PM-PN expansion.  We fix a unique gauge by imposing simplifying conditions not on the Hamiltonian function $H$ itself, but on its corresponding ``mass-shell constraint'' (or ``impetus formula''\cite{Kalin:2019rwq}), which is simply a rearrangement of the expression of the Hamiltonian, in which the squared momentum is given as a function of the Hamiltonian $H$ (of the energy $E=H$).  In Sec.~\ref{sec:angle}, we describe how the scattering-angle function can be derived from the canonical mass-shell constraint, or vice versa (with our gauge-fixing for the mass shell), and derive the explicit relationships between the scattering-angle coefficients and the mass-shell coefficients.  Finally, in Sec.~\ref{sec:radialaction}, we compute the radial action $I_r$, and point out a hidden simplicity in its dependence on the mass ratio, when expressed in terms of appropriate (covariant rather than canonical) variables, which is a simple consequence of the mass dependence of the scattering angle $\chi$ and the relationship between $\chi$ and $I_r$ discovered in \cite{Kalin:2019inp}.

\subsection{The canonical Hamiltonian and/or the mass-shell constraint}\label{sec:massshell}

For an aligned-spin binary canonical Hamiltonian,
\begin{alignat}{3}
&H(r,\phi,p_r,L;m_\mr i,a_\mr i)
\nnm\\
&=H(r,p_r,L;m_\mr i,a_\mr i)
\end{alignat}
the dynamical variables (depending on a time parameter $t$) are polar coordinates $(r,\phi)$ in the orbital plane, with $r$ being the orbital separation, and their conjugate momenta $(p_r,p_\phi\equiv L)$.  The Hamiltonian does not depend on the angular coordinate $\phi$ due to the system's axial symmetry, and it otherwise depends only on the constant masses and spins $(m_\mr i,a_\mr i)=(m_1,m_2,a_1,a_2)$.  The Hamiltonian equations of motions read
\begin{alignat}{3}\label{Hameqs}
\dot r&=\frac{\doe H}{\doe p_r},\qquad & \dot p_r&=-\frac{\doe H}{\doe r},
\\\nnm
\dot \phi&=\frac{\doe H}{\doe L},\qquad & \dot L&=-\frac{\doe H}{\doe\phi}=0,
\end{alignat}
where we note that the canonical orbital angular momentum $L$ is a constant of motion.

Such a Hamiltonian is not unique, but is subject to a type of gauge freedom, namely under canonical transformations: diffeomorphisms of the phase space which preserve the canonical form (\ref{Hameqs}) of the equations of motion.  In a quite general gauge (one which encompasses all gauges encountered in previous PN or PM aligned-spin Hamiltonians), the Hamiltonian takes the following form through quadratic order in the spins, through 4PM order,
\begin{alignat}{3}\label{Hgeneral}
&H=H_0(\bs p^2;m_\mr i)+\sum_{k=1}^4\frac{G^k}{r^k}\bigg[c_k(\bs p^2,\frac{L^2}{r^2};m_\mr i)
\\\nnm
&\quad+\frac{La_\mr i}{r^2}c_k^\mr i(\bs p^2,\frac{L^2}{r^2};m_\mr j)+\frac{a_\mr ia_\mr j}{r^2}c_k^{\mr i\mr j}(\bs p^2,\frac{L^2}{r^2};m_\mr k)\bigg]+\mc O(G^5),
\end{alignat}
where
\be\label{psqprsq}
\bs p^2=p_r^2+\frac{L^2}{r^2},
\ee
is the total squared canonical linear momentum.  Here, $H_0$ is the 0PM (free) Hamiltonian, and the functions $c_k$, $c_k^\mr i$ and $c_k^{\mr i\mr j}$ encode respectively the nonspinning, spin-orbit, and quadratic-in-spin gravitational couplings at the $k$PM orders.
The $c$'s are assumed to have regular Taylor series around $L^2=0$ and $\bs p^2=0$.  We will work here with the standard (gauge) choice for the free Hamiltonian in the cm frame,
\be\label{H0PM}
H_0=\sqrt{m_1^2+\bs p^2}+\sqrt{m_2^2+\bs p^2},
\ee
such that, as $r\to\infty$, the magnitude $\sqrt{\bs p^2}$ of the canonical linear momentum corresponds to the two bodies' physical equal and opposite spatial momenta in the cm frame.

The expression (\ref{Hgeneral}) of the Hamiltonian can be solved, working perturbatively in $G$, for $\bs p^2(r,E,L; m_\mr i,a_\mr i)$, where $E\equiv H(r,p_r,L; m_\mr i,a_\mr i)$ is the total energy; one finds
\begin{alignat}{3}\label{psqgeneral}
\bs p^2&=p_\infty^2(E;m_\mr i)+\sum_{k\ge1}\frac{G^k}{r^k}\bigg[f_k(E,\frac{L^2}{r^2};m_\mr i)
\\\nnm
&\qquad+\frac{La_\mr i}{r^2}f_k^\mr i(E,\frac{L^2}{r^2};m_\mr j)+\frac{a_\mr ia_\mr j}{r^2}f_k^{\mr i\mr j}(E,\frac{L^2}{r^2};m_\mr k)\bigg],
\end{alignat}
where the 0PM part $p_\infty^2$ is found by (exactly) inverting (\ref{H0PM}), \mbox{$H_0(\bs p^2)=E$ $\;\Leftrightarrow\;$ $p_\infty^2(E)=\bs p^2$},
\be
p_\infty^2=\frac{(E^2-m_1^2-m_2^2)^2-4m_1^2m_2^2}{4E^2}=\mu^2\frac{\gamma^2-1}{\Gamma^2},
\ee
which we recognize as the same $p_\infty$ from (\ref{pcm}). The functions $f_k$, $f^\mr i_k$ and $f^{\mr i\mr j}_k$ are determined by (and carry all of the information of) the $c_k^{\cdots}$ coefficients in the Hamiltonian (\ref{Hgeneral}).  Importantly, the $f^{\cdots}_k$ functions will have regular limits as $\gamma^2-1=\ve\to0$ (as $p_\infty\to0$) and as $L^2\to0$, given our assumption that the $c^{\cdots}_k$ functions were regular as $\bs p^2\to 0$ and $L^2\to 0$.  The quantities $\gamma$, $\ve$ and $\Gamma$ are all defined in terms of the energy $E$ and the rest masses just as in the previous section.

As discussed in Ref.~\cite{Vines:2018gqi} (through N$^2$LO in the PN expansion, and as we have explicitly verified through N$^3$LO), it is possible to find a perturbative canonical transformation which brings the Hamiltonian (\ref{Hgeneral}) into a ``quasi-isotropic'' form, i.e., a form in which the $c$'s depend only $\bs p^2$ and not on $L^2/r^2$.
Furthermore, the freedom in canonical transformations [among Hamiltonians of the form (\ref{Hgeneral})] is completely fixed once one imposes this quasi-isotropic-Hamiltonian condition \emph{and} uniquely specifies a 0PM Hamiltonian $H_0$, as we have done in (\ref{H0PM}).  For such a quasi-isotropic Hamiltonian, one finds that the corresponding ``mass shell constraint,'' the expression for $\bs p^2$ (\ref{psqgeneral}), has nonspinning and SO coefficients $f_k$  $f_k^\mr i$ which are independent of $L^2/r^2$, but its quadratic-in-spin coefficients $f_k^{\mr i\mr j}$ have terms at zeroth and first orders in $L^2/r^2$.  However, there also exists a different (non-quasi-isotropic) gauge for the Hamiltonian (\ref{Hgeneral}) (one with $L^2/r^2$ terms in $c_k^{\mr i\mr j}$) such that its mass shell constraint (\ref{psqgeneral}) is quasi-isotropic, with the $f_k$, $f_k^\mr i$ and $f^{\mr i\mr j}_k$ \emph{all} depending only on $E$ (and the masses) and not on $L^2/r^2$.
Because both the scattering angle and the radial action are more directly related to the $f$ coefficients in the mass shell, we will find it convenient to adopt this quasi-isotropic-mass-shell gauge (which is also unique with a given choice for $H_0$), specializing (\ref{psqgeneral}) to the form
\begin{alignat}{3}\label{psqiso}
\bs p^2&=p_\infty^2(E;m_\mr i)+\sum_{k\ge1}\frac{G^k}{r^k}\bigg[f_k(E;m_\mr i)
\\\nnm
&\qquad+\frac{La_\mr i}{r^2}f_k^\mr i(E;m_\mr j)+\frac{{a_\mr ia_\mr j}}{r^2}f_k^{\mr i\mr j}(E;m_\mr k)\bigg].
\end{alignat}
Regrouping in terms of powers of $r$ instead of powers of $G$, we have
\be\label{massshellsimple}
p_r^2+\frac{L^2}{r^2}=\bs p^2=p_\infty^2+\sum_{k\ge1}\frac{G^k}{r^k}\tilde f_k,
\ee
where we define
\be\label{fktotal}
\tilde f_{k}=f_k+\frac{La_\mr i}{G^2}f_{k-2}^\mr i+\frac{{a_\mr ia_\mr j}}{G^2}f_{k-2}^{\mr i\mr j},
\ee
with $f^{\cdots}_{-1}=f^{\cdots}_{0}=0$, and we need to extend the sum to $k=6$ (while dropping the nonspinning $f_5$ and $f_6$).  Our starting point for the following calculations will be this ansatz for the mass shell constraint, which is fully equivalent to an ansatz for a Hamiltonian of the form (\ref{Hgeneral}) modulo gauge freedom.  Our fundamental assumption is the existence of such a canonical Hamiltonian.  We will find that the coefficients $f_k^{\cdots}(E;m_\mr i)$ are uniquely determined by the expansion of the scattering-angle function to $k$PM order.

\subsection{The scattering angle}\label{sec:angle}

As shown in \cite{Damour:2017zjx}, the scattering angle $\chi(E,L;m_\mr i,a_\mr i)$ for an unbound orbit can be found directly from the canonical mass-shell constraint as follows.  The constraint (\ref{massshellsimple}) can be solved for the radial momentum $p_r(r,E,L;m_\mr i,a_\mr i)$, and then the scattering angle is given by the integral
\begin{alignat}{3}\label{chidef}
\pi+\chi(E,L)&=-\int_\infty^\infty\mr d r\, \frac{\doe}{\doe L}p_r(r,E,L)
\\\nnm
&=-2\int_{r_\mr{min}}^\infty\mr d r\, \frac{\doe}{\doe L}\sqrt{p_\infty^2-\frac{L^2}{r^2}+\sum_{k\ge1}\frac{G^k}{r^k}\tilde f_k},
\end{alignat}
where $r_\mr{min}$ is the largest real root of $p_r=0$.
In the direct evaluation of this integral, it would matter that the $\tilde f_k$ in (\ref{fktotal}) depend on $L$ (in the SO terms).  But let us define an antiderivative of $\pi+\chi$ with respect to $L$ to be ``the unbound radial action,'' 
\bse\label{defW}
\be\label{WdinvL}
\W=-\frac{1}{2\pi}\Big(\frac{\doe}{\doe L}\Big)^{-1}(\pi+\chi),
\ee
which is essentially a \emph{partie finie} of the radial action integral for unbound orbits,
\be
\W(E,L)=\frac{1}{2\pi}\mr{Pf}\int_\infty^\infty \mr d r\, p_r(r,E,L).
\ee
\ese
The eikonal phase~\cite{Bjerrum-Bohr:2018xdl,Kabat:1992tb,Akhoury:2013yua,Bern:2020buy} is $\W/\hbar$ (up to a constant).
For the expression of $\W$ in terms of the $\tilde f_k$, it does not matter that the $\tilde f_k$ depend on $L$.  That expression will be identical to the $L$-antiderivative of the nonspinning scattering angle expressed in terms of the nonspinning $f_k$, with $f_k\to\tilde f_k$, so this reduces the evaluation of the integral for the spinning case to the nonspinning problem, using the coefficient mapping (\ref{fktotal}).  The results of the nonspinning integral (for $\chi$, from which constructing $\W$ is trivial) have been tabulated at high orders, e.g., in \cite{Bjerrum-Bohr:2019kec}.  One finds
\be\label{WPM}
2\pi \W=-\pi L-\frac{G\ln L}{p_\infty}\tilde \chi_1+\sum_{k\ge 2}\frac{G^k}{p_\infty^k L^{k-1}}\frac{\tilde \chi_k}{k-1},
\ee
where $\tilde \chi_k$ are the entries of Table 1 in \cite{Bjerrum-Bohr:2019kec} with $f_k\to\tilde f_k$; the first few read
\begin{alignat}{3}\label{tildechis}
\tilde \chi_1&=\tilde f_1,\phantom{\bigg|}
\\\nnm
\tilde \chi_2&=\frac{\pi}{2}p_\infty^2\tilde f_2,
\\\nnm
\tilde \chi_3&=2p_\infty^4\tilde f_3+p_\infty^2\tilde f_1\tilde f_2-\frac{\tilde f_1^3}{12},
\\\nnm
\tilde \chi_4&=\frac{3\pi}{8}p_\infty^4(2p_\infty^2 \tilde f_4+\tilde f_2^2+2\tilde f_1\tilde f_3),
\\\nnm
& \cdots
\end{alignat}
The scattering angle $\chi$ is then given by
\be
\pi+\chi=-2\pi\frac{\doe\W}{\doe L},
\ee
with the $L$-derivative acting also inside the $\tilde f_k$ in (\ref{fktotal}).  
To obtain $\W$ or $\chi$ through quadratic order in spins and through 4PM order, $\mc O(G^4)$, counting both the $G^k$ in (\ref{WPM}) and the $1/G^2$ in (\ref{fktotal}), we must include parts of the contributions up to $\tilde f_6$ and up to $\tilde\chi_8$.  The resultant explicit expression of the scattering angle $\chi$ in terms of the $f_k^{\cdots}$ coefficients up to 4PM order and quadratic order in spins is 
\begin{widetext}
\begin{alignat}{3}\label{chifs}
\chi&=\frac{G}{p_\infty L}f_1+\frac{\pi G^2}{2L^2} f_2
+\frac{G^3}{ p_\infty^3L^3}\bigg[{-}\frac{1}{12}f_1^3+p_\infty^2 f_1f_2+2p_\infty^4 f_3\bigg]
+\frac{3\pi G^4}{8L^4}\bigg[f_2^2+2f_1f_3+2p_\infty^2f_4\bigg]
\\\nnm
&\quad+a_\mr i\Bigg\{\frac{Gp_\infty}{L^2}f_1^\mr i+\frac{\pi G^2}{2L^3}\bigg[f_1f_1^\mr i+p_\infty^2 f_2^\mr i\bigg]
+\frac{G^3}{p_\infty L^4}\bigg[\frac{3}{4}f_1^2 f_1^\mr i+3 p_\infty^2\Big(f_2 f_1^\mr i+f_1 f_2^\mr i\Big)+2p_\infty^4 f_3^\mr i\bigg]
\\\nnm
&\qquad\qquad+\frac{3\pi G^4}{4 L^5}\bigg[2 f_1 f_2 f_1^\mr i+f_1^2f_2^\mr i+2p_\infty^2\Big(f_3f_1^\mr i+f_2 f_2^\mr i+f_1 f_3^\mr i\Big)+p_\infty^4f_4^\mr i\bigg]\Bigg\}
\\\nnm
&\quad+a_\mr ia_\mr j\Bigg\{\frac{2Gp_\infty}{L^3}f_1^{\mr i\mr j}+\frac{3\pi G^2}{16L^4}\bigg[4f_1f_1^{\mr i\mr j}+p_\infty^2\Big(3 f_1^\mr if_1^\mr j+4f_2^{\mr i\mr j}\Big)\bigg]
+\frac{G^3}{p_\infty L^5}\bigg[f_1^2f_1^{\mr i\mr j}+4p_\infty^2\Big(f_2f_1^{\mr i\mr j}+f_1 f_1^\mr if_1^\mr j+f_1f_2^{\mr i\mr j}\Big)
\\\nnm
&\qquad\qquad
+\frac{8}{3}p_\infty^4\Big(2f_1^\mr if_2^\mr j+f_3^{\mr i\mr j}\Big)\bigg]
+\frac{15\pi G^4}{64 L^6}\bigg[
8f_1f_2f_1^{\mr i\mr j}+5f_1^2f_1^\mr if_1^\mr j+4f_1^2f_2^{\mr i\mr j}
\\\nnm
&\qquad\qquad
+2p_\infty^2\Big(4f_3 f_1^{\mr i\mr j}+5 f_2f_1^\mr if_1^\mr j+4 f_2f_2^{\mr i\mr j}+10f_1f_1^\mr if_2^\mr j+4f_1 f_3^{\mr i\mr j}\Big)
+p_\infty^4\Big(5f_2^\mr if_2^\mr j+10 f_1^\mr if_3^\mr j+4f_4^{\mr i\mr j}\Big)\bigg]\Bigg\}+\mc O(a^3)+\mc O(G^5).
\end{alignat}
\end{widetext}
We see that the $k$PM coefficients $f_k^{\cdots}$ first enter in the $G^k$ terms; however, they do not enter those terms at the leading orders in $p_\infty$ (in the PN expansion of each PM coefficient).  Recalling that all of the $f$'s are finite as $p_\infty\to0$ ($\ve\to 0$), we see that, within each set of square brackets multiplying $G^k$, the lowest orders in $p_\infty$ do not depend on $f_k^{\cdots}$, rather only on the lower-PM-order $f$'s (with some exceptions at $G^1$ and $G^2$).  Similarly, for the scattering-angle coefficients at even higher orders in $G$ (some of which will be relevant below), the lower orders in their PN expansions will be determined by coefficients from lower orders in $G$ already appearing here.

This gives the scattering angle $\chi$ in terms of the mass-shell coefficients $f_k$, $f_k^\mr i$, $f_k^{\mr i\mr j}$, as an expansion in the canonical orbital angular momentum $L$.  Equating that expression to a parametrization of $\chi$ of the form (\ref{masterchi2}) in terms of the covariant impact parameter $b$, using the translation (\ref{DeltaL}) while re-expanding in spins, one can solve for the $f$ coefficients in the mass shell in terms of the $\ms X$ coefficients in the scattering angle (or vice versa), order by order in the PM expansion.  Recall $p_\infty=\mu\sqrt{\ve}/\Gamma$.  Rewriting $\Delta L=L-p_\infty b$ from (\ref{DeltaL}) as a sum over (effective) spins,
\bse\label{dLxi}
\be
\Delta L=\frac{\mu}{\Gamma}\xi^\mr ia_\mr i=\frac{\mu}{\Gamma}\Big(\xi^\mr ba_\mr b+\xi^\mr ta_\mr t\Big),
\ee
with
\begin{alignat}{7}
\xi^\mr b&=\frac{(\Gamma-1)^2}{2\nu}&&=2\nu\Big(\frac{\gamma-1}{\Gamma+1}\Big)^2&&=\frac{\nu\ve^2}{8}+\mc O(\ve^3),
\nnm\\
\xi^\mr t&=\frac{\Gamma^2-1}{2\nu}&&=\gamma-1&&=\frac{\ve}{2}+\mc O(\ve^2),
\end{alignat}
\ese
the results for the $f$'s through 2PM order are as follows: nonspinning,
\bse\label{f12}
\begin{alignat}{3}\label{f012}
f_1&=\mu^2 M \frac{\sqrt{\ve}}{\Gamma}\ms X_1,
\\\nnm
f_2&=\frac{2\mu^2 M^2}{\pi\Gamma}\ms X_2,
\end{alignat}
spin-orbit,
\begin{alignat}{3}
f_1^\mr i&= \frac{\mu M}{\sqrt{\ve}}\Big(\ms X_1{}^{\mr i} + \ms X_1\xi^\mr i\Big),
\\\nnm
f_2^\mr i&=\frac{\mu M^2}{\ve}\bigg[\frac{2}{\pi}\ms X_2{}^{\mr i}-\Gamma \ms X_1 \ms X_1{}^{\mr i}+\Big(\frac{4}{\pi}\ms X_2-\Gamma (\ms X_1)^2\Big)\xi^\mr i\bigg]
\end{alignat}
and quadratic in spin,
\begin{alignat}{3}
f_1^{\mr i\mr j}&= \frac{\mu^2 M}{2\Gamma\sqrt{\ve}}\Big( \ms X_1{}^{\mr i\mr j} +2 \ms X_1\ms X_1{}^{\mr i}\xi{}^{{\mr j}}+ \ms X_1\xi^\mr i\xi^\mr j\Big),
\\\nnm
f_2^{\mr i\mr j}&=\frac{\mu^2 M^2}{\Gamma\ve}\bigg[\frac{4}{3\pi}\ms X_2{}^{\mr i\mr j}-\frac{1}{2}\Gamma\ms X_1\ms X_1{}^{\mr i\mr j}-\frac{3}{4}\Gamma\ms X_1{}^\mr i\ms X_1{}^\mr j
\\\nnm
&\quad+\Big(\frac{4}{\pi}\ms X_2{}^\mr i-\frac{5}{2}\Gamma\ms X_1\ms X_1{}^\mr i\Big)\xi^\mr j+\Big(\frac{4}{\pi}\ms X_2-\frac{5}{2}\Gamma (\ms X_1)^2\Big)\xi^\mr i\xi^\mr j\bigg],
\end{alignat}
\ese
with symmetrization over i and j understood.  These 1PM and 2PM results are exact (to all orders in $\ve$).  With our predicted mass-ratio dependence from the previous section, we have, for $k=1,2$, $\ms X_k(\ve,\nu)=\ms X_k^0(\ve)$, $a_\mr i\ms X_k{}^\mr i(\ve,\nu) =a_\mr b\ms X_k^{0\mr b}(\ve)+a_\mr t\ms X_k^{0\mr t}(\ve)$, and $a_\mr ia_\mr j\ms X_k{}^{\mr i\mr j}(\ve,\nu)=a_1a_2\ms X_k^{0{\times}}(\ve)+\mc O(a_1^2,a_2^2)$, all independent of $\nu$, and the $\ms X_k^{0{\cdots}}(\ve)$ from the extended test-body limit are given explicitly by (\ref{X0s}).  Though it is not immediately obvious here, each of these $f$'s has a finite limit as $\ve\to0$, as is required by our Hamiltonian ansatz.  We will need the expansions of the $f_1^{\cdots}$ up to $\mc O(\ve^3)$, and of the $f_2^{\cdots}$ up to $\mc O(\ve^2)$.  Along with $f_3^{\cdots}$ up to $\mc O(\ve^1)$ and $f_4^{\cdots}$ at $\mc O(\ve^0)$, we will then have a complete mass-shell constraint (\ref{psqiso}) up to N$^3$LO in the PN expansion, which could be solved for the corresponding canonical Hamiltonian (\ref{Hgeneral}).

At 3PM and 4PM orders, one can also solve for the $f$'s in terms of the $\ms X$'s, obtaining exact expressions analogous to the above.  But we will now work in a PN expansion, an expansion in $\ve$, while enforcing our predicted mass-ratio dependence [which (\ref{f12}) did not].  For the nonspinning coefficients, using the known results (\ref{X0a0}) and (\ref{X1a0}) for the $\ms X$'s, we find
\begin{alignat}{3}
\frac{f_3}{\mu^2 M^3}&=\frac{17-10\nu}{2}+\frac{36-91\nu+13\nu^2}{4}\ve+\mc O(\ve^2),
\nnm\\
\frac{f_4}{\mu^2M^4}&=8+\Big(\frac{41}{32}\pi^2-\frac{160}{3}\Big)\nu+\frac{7}{2}\nu^2+\mc O(\ve),
\end{alignat}
through the orders that contribute to the N$^3$LO PN level.  Here again we note the finite limits as $\ve\to0$.  For the spinning contributions, we must enforce that all the $f$'s have finite limits as $\ve\to0$, which will fix some of the unknown coefficients in our parametrization (\ref{X1s}) of the $\nu^1$ parts of the scattering angle, or relationships between them, from consistency with the lower-order $f$'s and $\ms X$'s [recall the discussion following (\ref{chifs})].  At the SO level, this determines or constrains the lower-PN-order scattering-angle coefficients,
\begin{alignat}{3}\label{lowPNchiSO}
\ms X_{30}^{1\mr i}a_\mr i&=0,\phantom{\bigg|}
\\\nnm
\ms X_{31}^{1\mr i}a_\mr i&=10(a_\mr b+a_\mr t),
\\\nnm
\ms X_{41}^{1\mr i}a_\mr i&=\frac{21}{2}a_\mr b+9a_\mr t,
\\\nnm
\ms X_{42}^{1\mr i}a_\mr i&=\frac{3}{4}\Big(68a_\mr b+49 a_\mr t+2 \ms X^{1\mr i}_{32}a_\mr i\Big),
\end{alignat}
and expressions for $f_3^\mr i$ and $f_4^\mr i$ which are explicitly regular as $\ve\to0$ and depend on the remaining unknowns $\ms X^{1\mr i}_{32}$, $\ms X^{1\mr i}_{33}$, and $\ms X^{1\mr i}_{43}$, with $\mr i=\mr b,\mr t$,
\bse\label{f34SO}
\begin{alignat}{3}
&\frac{f_3^\mr ia_\mr i}{\mu M^3}=\frac{-6+4\nu-5\nu^2}{2}a_\mr b+\frac{-3-31\nu-9\nu^2}{4}a_\mr t+\frac{\nu}{2}\ms X_{32}^{1\mr i}a_\mr i
\nnm\\
&\qquad\quad+\bigg[\frac{-24+172\nu-276\nu^2+21\nu^3}{16}a_\mr b
\\\nnm
&+\frac{-166\nu-90\nu^2+9\nu^3}{8}a_\mr t+\frac{\nu}{4} \Big(\ms X_{32}^{1\mr i}+2\ms X_{33}^{1\mr i}\Big)a_\mr i\bigg]\ve+\mc O(\ve^2),
\end{alignat}
and
\begin{alignat}{3}
\frac{f_4^\mr ia_\mr i}{\mu M^4}&=\bigg({-}2-\frac{811}{8}\nu-4\nu^2+\frac{13}{8}\nu^3\bigg)a_\mr b
\\\nnm
&\quad+\bigg(\frac{1}{8}-\frac{1577}{12}\nu+\frac{41}{16}\pi^2\nu+\frac{35}{4}\nu^2+\frac{3}{2}\nu^3\bigg)a_\mr t
\\\nnm
&\quad+\nu\Big[(4+\nu)\ms X_{32}^{1\mr i}-2\ms X^{1\mr i}_{33}+\frac{4}{3}\ms X^{1\mr i}_{43}\Big]a_\mr i+\mc O(\ve).
\end{alignat}
\ese
Similarly, for the bilinear-in-spin coefficients, we find
\begin{alignat}{3}\label{lowPNchiS1S2}
\ms X^{1{\times}}_{30}&=0,
\qquad 
\ms X^{1{\times}}_{31}=8,
\qquad
\ms X^{1{\times}}_{41}=\frac{15}{2},
\nnm\\
\ms X^{1{\times}}_{42}&=\frac{45}{32}\Big({-}22+\ms X^{1{\times}}_{32}\Big),
\end{alignat}
while $f_{k}^{\mr i\mr j}a_\mr ia_\mr j=f_k^\times a_1a_2+\mc O(a_1^2,a_2^2)$ with $f_3^\times$ and $f_4^\times$ given in terms of the remaining unknowns $\ms X^{1{\times}}_{32}$, $\ms X^{1{\times}}_{33}$, and $\ms X^{1{\times}}_{43}$ (and remaining unknowns from the SO level) by
\bse\label{f34S1S2}
\begin{alignat}{3}
\frac{f_3^\times}{\mu^2M^3}&=\frac{5}{2}+\Big(\frac{9}{2}+\frac{3}{8}\ms X_{32}^{1{\times}}\Big)\nu-\nu^2
\\\nnm
&\quad+\frac{3}{8}\bigg[4-15\nu-16\nu^2+4\nu^3+2\nu(1-2\nu)\ms X_{32}^{1\mr b}
\\\nnm
&\qquad\quad+4\nu^2\ms X_{32}^{1\mr t}-\frac{\nu^2}{2}\ms X_{32}^{1{\times}}+\nu \ms X_{33}^{1{\times}}\bigg]\ve+\mc O(\ve^2),
\end{alignat}
and
\begin{alignat}{3}
\frac{f_4^\times}{\mu^2M^4}&=2+\frac{187}{4}\nu-21\nu^2+\frac{13}{8}\nu^3
\\\nnm
&\quad+\frac{\nu}{4}(19+2\nu)\ms X_{32}^{1\mr b}+\frac{\nu}{2}(10-\nu)\ms X_{32}^{1\mr t}
\\\nnm
&\quad-\frac{3}{4}\nu(4+\nu)\ms X_{32}^{1{\times}}+\frac{3}{2}\nu\ms X_{33}^{1{\times}}+\frac{16}{15}\nu \ms X_{43}^{1{\times}}+\mc O(\ve).
\end{alignat}
\ese
We now have a complete expression of the mass-shell constraint (\ref{psqiso}) through N$^3$LO in the PN expansion and through bilinear order in spins, which could be solved for the corresponding canonical Hamiltonian. It depends on the remaining unknown (dimensionless, numerical) coefficients $\ms X^{1{\mr A}}_{32}$, $\ms X^{1{\mr A}}_{33}$, and $\ms X^{1{\mr A}}_{43}$ with $\mr A=\{\mr b,\mr t,{\times}\}$, from (\ref{X1s}).  Recall, for $\ms X^{1{\mr A}}_{kn}$, $k$ is the PM order, and $n$ is the relative PN order.

\subsection{The radial action}\label{sec:radialaction}

For a bound orbit ($\gamma^2-1=\gamma^2v^2=\ve<0$), the same canonical mass-shell constraint (\ref{psqiso}) governs the motion.  The (gauge-dependent) radial momentum function $p_r(r,E,L;m_\mr i,a_\mr i)$ is still given by
\be
p_r=\pm\sqrt{p_{\infty}^2-\frac{L^2}{r^2}+\sum_k\frac{G^k}{r^k}\bigg[f_k+\frac{L a_\mr i}{r^2}f_k^\mr i+\frac{a_\mr ia_\mr j}{r^2}f_k^{\mr i\mr j}\bigg]},
\ee
but now $p_\infty^2=(\mu/\Gamma)^2\ve$ is negative.  As a result, $p_r^2(r)$ has two positive real roots $r=r_\pm$ between which $p_r^2$ is positive, with $r_+$ being the largest real root, and the trajectory oscillates between these radial turning points $r_\pm$.  The \emph{canonical radial action} function $I_r(E,L,m_\mr i,a_\mr i)$ is defined as the integral of $p_r\mr dr$ over one period of the radial motion,
\begin{alignat}{3}\label{defIr}
2\pi I_r:=\oint \mr dr\,p_r&=\int_{r_-}^{r_+}\mr dr \Big({+}\sqrt{p_r^2}\Big)+\int_{r_+}^{r_-}\mr dr \Big({-}\sqrt{p_r^2}\Big)
\nnm\\
&=2\int_{r_-}^{r_+}\mr dr \sqrt{p_r^2},
\end{alignat}
and it is a gauge-invariant function, from which one can derive several other gauge-invariant functions physically characterizing bound orbits~\cite{Kalin:2019inp,Kalin:2019rwq}.  Like the ``unbound radial action'' $\W$ (the $L$-antiderivative of the scattering-angle $\chi$) (\ref{defW}), the bound radial action $I_r(E,L,m_\mr i,a_\mr i)$ encodes the complete gauge-invariant information content of the canonical Hamiltonian (governing both unbound and bound orbits) (at least up to the N$^3$LO PN level) --- though in a subtly different way, concerning orders in the PM-PN expansion of $I_r$ versus that of $\W$.

It was shown in \cite{Kalin:2019inp} that the periastron-advance angle, 
$
\Phi=2\pi+\Delta\Phi=-2\pi\doe I_r/\doe L
$,
the angle swept out by a bound orbit during one period of the radial motion, is related to the scattering angle, $\pi+\chi=-2\pi \doe \W/\doe L$, by
\begin{alignat}{3}
\Phi(E,L,m_\mr i,a_\mr i)=2\pi&+ \chi(E,L,m_\mr i,a_\mr i)
\\\nnm
&+ \chi(E,-L,m_\mr i,-a_\mr i),
\end{alignat}
where the right-hand side requires an analytic continuation from $E>M$ (unbound, for which $\chi$ is real) to $E<M$ (bound, for which $\chi$ is complex), as detailed below.
It follows from a straightforward extension of their argument that a particular $L$-antiderivative of this relation holds, giving the bound radial action $I_r$ in terms of the unbound radial action $\W$,
\begin{alignat}{3}\label{WtoIr}
I_r(E,L,m_\mr i,a_\mr i)&=\W(E,L,m_\mr i,a_\mr i)
\\\nnm
&\quad-\W(E,-L,m_\mr i,-a_\mr i),
\end{alignat}
as can also be verified by explicit calculation.

Consider the unbound radial action in the form (\ref{WPM}), after replacing $\tilde\chi_1$ using (\ref{tildechis}) and (\ref{f012}),
\be\label{Wto}
\W=-\frac{ L}{2}-GM\mu\frac{1+2\ve}{\sqrt{\ve}}\frac{\ln L}{\pi }+\frac{1}{2\pi }\sum_{k\ge 2}\frac{G^k}{p_\infty^k L^{k-1}}\frac{\tilde \chi_k}{k-1}.
\ee
In continuing this from the unbound case, $\ve>0$, $p_\infty^2>0$, to the bound case, $\ve<0$, $p_\infty^2<0$, the second term with $1/\sqrt{\ve}$ becomes imaginary, as do all of the terms in the sum with $k$ odd, having odd powers of $p_\infty=(\mu/\Gamma)\sqrt{\ve}$.  Note, from (\ref{tildechis}) and (\ref{fktotal}), and from the fact that all of the $f$'s have regular Taylor series in $\ve$ about $\ve=0$, that all of the $\tilde\chi_k$ are still real for the bound case, and that the $\tilde\chi_k$ are unchanged by $(L,a_\mr i)\to(-L,-a_\mr i)$.  Thus, plugging the continuation of (\ref{Wto}), with $\sqrt{\ve}=i\sqrt{-\ve}$, into (\ref{WtoIr}), we see that all of the odd-$k$ terms are canceled; after using $\ln L-\ln(-L)=\ln(-1)=-i\pi$ (choosing the branch which yields the physically sensible result), we are left with
\be\label{Irtildechi}
I_r=-L+GM\mu\frac{1+2\ve}{\sqrt{-\ve}}+\frac{1}{\pi }\sum_{l\ge 1}\frac{G^{2l}}{p_\infty^{2l} L^{2l-1}}\frac{\tilde \chi_{2l}}{2l-1},
\ee
which is real for bound orbits.  Only the $\tilde \chi_k$ with $k$ even ($k=2l$) remain, and those with $k$ odd are gone (except for $\tilde\chi_1$).  This may make it seem as though we have lost information in passing from $\W$ to $I_r$, but in fact we have not, as long as we are sure to keep all terms in the consistent PN expansion of $I_r$ (at least up to the N$^3$LO PN level); this is due to relationships between the $\tilde \chi_k$ as discussed below (\ref{chifs}).  

As we will make clearer below, the complete PN expansion of $I_r$ up to N$^3$LO is contained in its PM expansion up to $\mc O(G^6)$ for the nonspinning terms and up to $\mc O(G^8)$ for the spin-orbit and quadratic-in-spin terms.   This can be computed directly from (\ref{Irtildechi}), recalling that the $\tilde\chi_k$ are the entries of Table 1 of \cite{Bjerrum-Bohr:2019kec} with $f_k\to\tilde f_k$, as in (\ref{chifs}) above, with the $\tilde f_k$ given by (\ref{fktotal}).  We need again the contributions from $f_k$, $f_k^\mr i$, $f_k^{\mr i\mr j}$ up to $k=4$, contained in the $\tilde f_k=f_k+f_{k-2}^\mr iLa_\mr i/G^2+f_{k-2}^{\mr i\mr j}a_\mr ia_\mr j/G^2$ up to $k=6$.  To reach the all the $G^8$ quadratic-in-spin terms, we must take the sum in (\ref{Irtildechi}) up to $l=6$, involving parts of $\tilde\chi_{12}$.

This process yields the radial action $I_r$ through the N$^3$LO PN level as an expansion in the inverse canonical orbital angular momentum $L\equiv L_\mr{can}$.  To express the results of that process, it will be advantageous to use the covariant orbital angular momentum $L_\mr{cov}$, which we define for the bound-orbit case by 
\be\label{LcovL}
L_\mr{cov}:=L-\Delta L,
\ee
with $\Delta L(E,a_\mr i)$ still given by the last two lines of (\ref{DeltaL}) or by (\ref{dLxi}), in which we note that everything is still real for bound orbits [unlike in the second line of (\ref{DeltaL}), where we would need to continue to imaginary $b$ to keep $L_\mr{cov}=p_\infty b$ real].  

In fact, the expression of the radial action (mostly) in terms of $L_\mr{cov}$ is simply related to the expression of the scattering angle in terms of $L_\mr{cov}$, as follows.  Taking the form (\ref{masterchi2}) for the scattering angle and eliminating $b$ in favor of $L_\mr{cov}=(\mu/\Gamma)\sqrt{\ve}b$,
\begin{alignat}{3}\label{chiLcov}
\chi&=\Gamma\sum_{k\ge1}\Big(\frac{GM}{b\sqrt{\ve}}\Big)^k\bigg[\ms X_k+\frac{a_\mr i}{b\sqrt{\ve}}\ms X_k{}^{\mr i}+\frac{a_\mr i a_\mr j}{b^2{\ve}}\ms X_k{}^{\mr i\mr j}\bigg]
\\\nnm
&=\Gamma\sum_{k\ge1}\Big(\frac{GM\mu}{\Gamma L_\mr{cov}}\Big)^k\bigg[\ms X_k+\frac{\mu a_\mr i}{\Gamma L_\mr{cov}}\ms X_k{}^{\mr i}+\frac{\mu^2a_\mr i a_\mr j}{\Gamma^2L_\mr{cov}^2}\ms X_k{}^{\mr i\mr j}\bigg],
\end{alignat}
and then using (\ref{WdinvL}), being sure to match up the constant of integration with (\ref{Wto}), we find
\begin{alignat}{3}
\W&=-\frac{ L}{2}-GM\mu\,\ms X_1\frac{\ln L_\mr{cov}}{2\pi }+\frac{1}{2\pi}\sum_{k\ge2}\frac{(GM\mu)^k}{(\Gamma L_\mr{cov})^{k-1}}\frac{\ms X_k}{k-1}
\nnm\\
&\quad+\frac{1}{2\pi}\sum_{k\ge1}\Big(\frac{GM\mu}{\Gamma L_\mr{cov}}\Big)^k\bigg[\mu a_\mr i\frac{\ms X_k{}^{\mr i}}{k}+\frac{\mu^2a_\mr i a_\mr j}{\Gamma L_\mr{cov}}\frac{\ms X_k{}^{\mr i\mr j}}{k+1}\bigg].
\end{alignat}
Then applying (\ref{WtoIr}), as we did between (\ref{Wto}) and (\ref{Irtildechi}), noting $L_\mr{cov}\to-L_\mr{cov}$ under $(L,a_\mr i)\to(-L,-a_\mr i)$, we are left with
\begin{alignat}{3}\label{Ir}
I_r&=-L+GM\mu\frac{1+2\ve}{\sqrt{-\ve}}+\frac{1}{\pi}\sum_{l\ge1}\frac{(GM\mu)^{2l}}{(\Gamma{L_\mr{cov}})^{2l-1}}\bigg[\frac{\ms X_{2l}}{2l-1}
\nnm\\
&\qquad+\frac{\mu a_\mr i}{\Gamma L_\mr{cov}}\frac{\ms X_{2l}{}^\mr i}{2l}+\frac{\mu^2 a_\mr ia_\mr j}{(\Gamma L_\mr{cov})^2}\frac{\ms X_{2l}{}^{\mr i\mr j}}{2l+1}\bigg],
\end{alignat}
where these $\ms X_k{}^{\cdots}(\ve,\nu)$ are precisely the same coefficients from the scattering angle in (\ref{chiLcov}).  These coefficients up through $k=2l=4$ are those we gave or parametrized above in (\ref{X0s}) and (\ref{X1s}), with (\ref{joinXs}).  Recollecting them here, while using the constraints (\ref{lowPNchiSO}) and (\ref{lowPNchiS1S2}) obtained in matching between the scattering angle and the canonical mass shell, we have the $G^2$ coefficients which are independent of $\nu$ and are known exactly,
\bse\label{X_2l}
\begin{alignat}{3}
\ms X_{2}&=\frac{3\pi}{4}(4+5\ve),
\\\nnm
\ms X_{2}{}^{\mr i}a_\mr i&=-\frac{\pi}{2}\gamma(2+5\ve)(4a_\mr b+3a_\mr t),
\\\nnm
\ms X_{2}{}^{\times}&=\frac{3\pi}{2}(2+19\ve+20\ve^2),
\end{alignat}
and the $G^4$ coefficients which are linear in $\nu$,
\begin{alignat}{3}
\ms X_{4}&=\frac{105\pi}{64}(16+48\ve+33\ve^2)
\\\nnm
&\quad+\pi \bigg[{-}\frac{15}{2}+\bigg(\frac{123}{128}\pi^2-\frac{557}{8}\bigg)\ve+\mc O(\ve^2)\bigg]\nu,
\\\nnm
\ms X_{4}{}^{\mr i}a_\mr i&=-\frac{21\pi}{16}\gamma(8+36\ve+33\ve^2)(8a_\mr b+5\mr a_t)
\\\nnm
&\quad+\pi\gamma\bigg[\frac{21}{2}a_\mr b+9a_\mr t+\frac{3}{4}\Big(68a_\mr b+49 a_\mr t+2 \ms X^{1\mr i}_{32}a_\mr i\Big)\ve
\\\nnm
&\qquad\qquad+\ms X^{1\mr i}_{43}a_\mr i\ve^2+\mc O(\ve^3)\bigg]\nu,
\\\nnm
\ms X_{4}{}^{\times}&=\frac{105\pi}{16} (24+212\ve+447\ve^2+264\ve^3)
\\\nnm
&\quad+\pi\bigg[\frac{15}{2}+\frac{45}{32}\Big({-}22+\ms X^{1{\times}}_{32}\Big)\ve+\ms X^{1\times}_{43}\ve^2+\mc O(\ve^3)\bigg]\nu.
\end{alignat}
As mentioned above, for the complete expression of the radial action at the N$^3$LO PN level, we need the low orders in the PN expansions of $\ms X_6{}^{\cdots}$ and (for the spin terms) $\ms X_8{}^{\cdots}$.  We have obtained these from the procedure to compute the radial action described in the paragraph containing (\ref{Irtildechi}) and the following paragraph, in which the inputs are the $f_k^{\cdots}$ up to $k=4$ found in the previous subsection, finally changing variables using (\ref{LcovL}) to bring the result into the form (\ref{chiLcov}).  At $G^6$, we find the nonspinning
\begin{alignat}{3}\label{X6a0}
\frac{\ms X_6}{5\pi}&=\frac{231}{4}+\Big(\frac{123}{128}\pi^2-\frac{125}{2}\Big)\nu+\frac{21}{8}\nu^2+\mc O(\ve),
\end{alignat}
spin-orbit,
\begin{alignat}{3}\label{X6a1}
\frac{\ms X_6{}^\mr ia_\mr i}{{15\pi}}&=\Big({-}99+\frac{127}{4}\nu-\frac{5}{4}\nu^2\Big)a_\mr b
\\\nnm
&\qquad+\Big({-}\frac{231}{4}+\frac{167}{8}\nu-\frac{9}{8}\nu^2\Big)a_\mr t
+\frac{1}{4} \nu\ms X_{32}^{1\mr i}a_\mr i
\\\nnm
&\quad+\bigg[\Big({-}693+\frac{4989}{16}\nu-\frac{123}{32}\pi^2\nu-\frac{225}{16}\nu^2\Big)a_\mr b
\\\nnm
&\qquad+\Big({-}\frac{1617}{4}+\frac{733}{4}\nu-\frac{123}{64}\pi^2\nu-\frac{182}{16}\nu^2\Big)a_\mr t
\\\nnm
&\qquad+\nu\Big(\frac{7-3\nu}{8}\ms X_{32}^{1\mr i}-\frac{5}{4}\ms X_{33}^{1\mr i}+\ms X_{43}^{1\mr i}\Big)a_\mr i\bigg]\ve+\mc O(\ve^2),
\end{alignat}
and bilinear-in-spin, 
\begin{alignat}{3}\label{X6a2}
\frac{\ms X_6{}^\times}{35\pi}&=\frac{495}{4}-\frac{123\nu}{16}-\frac{9}{8}\nu^2+\frac{3}{32}\nu\ms X_{32}^{1{\times}}
\\\nnm
&\quad+\bigg[\frac{10197}{8}-\frac{4835}{32}\nu+\frac{123}{128}\pi^2\nu-\frac{399}{32}\nu^2
\\\nnm
&\qquad-\frac{3}{8}\nu(1+2\nu)\ms X_{32}^{1\mr b}-\frac{3}{4}\nu(1-\nu)\ms X_{32}^{1\mr t}
\\\nnm
&\quad+\frac{9}{64}\nu(2-\nu)\ms X_{32}^{1{\times}}-\frac{15}{32}\nu\ms X_{33}^{1{\times}}+\frac{2}{5}\nu\ms X_{43}^{1{\times}}\bigg]\ve+\mc O(\ve^2).
\end{alignat}
At $G^8$, spin-orbit,
\begin{alignat}{3}
\frac{\ms X_8{}^{\mr i}a_\mr i}{35\pi}&=\Big({-}715+\frac{23947}{48}\nu-\frac{41}{8}\pi^2\nu-\frac{97}{2}\nu^2+\frac{13}{16}\nu^3\Big)a_\mr b
\nnm\\\nnm
&\quad+\Big({-}\frac{6435}{16}+\frac{6883}{24}\nu-\frac{41}{16}\pi^2\nu-\frac{277}{8}\nu^2+\frac{3}{4}\nu^3\Big)a_\mr t
\\
&\quad+\nu\Big(\frac{2-\nu}{2}\ms X_{32}^{1\mr i}- \ms X_{33}^{1\mr i}+\frac{2}{3}\ms X_{43}^{1\mr i}\Big)a_\mr i+\mc O(\ve),
\end{alignat}
and bilinear-in-spin
\begin{alignat}{3}
\frac{\ms X_8{}^\times}{315\pi}&=\frac{5005}{16}-\frac{6599}{96}\nu+\frac{41}{128}\pi^2\nu-\frac{199}{32}\nu^2+\frac{5}{16}\nu^3
\nnm\\\nnm
&\quad-\frac{1}{8}\nu(1+2\nu)\ms X_{32}^{1\mr b}-\frac{1}{4}\nu(1-\nu)\ms X_{32}^{1\mr t}
\\
&\quad+\frac{3}{64}\nu(2-\nu)\ms X_{32}^{1{\times}}-\frac{3}{32}\nu\ms X_{33}^{1{\times}}+\frac{1}{15}\nu\ms X_{43}^{1{\times}}+\mc O(\ve).
\end{alignat}
\ese

Note that the $\ms X_6{}^{\cdots}$ coefficients in (\ref{X6a1}) and (\ref{X6a2}) are exactly quadratic in $\nu$, in spite of the fact that the $f$'s from which they are constructed, in (\ref{f34SO}) and (\ref{f34S1S2}), are cubic in $\nu$.  Less surprisingly, the $\ms X_6{}^{\cdots}$ are cubic in $\nu$, and more surprisingly the $\ms X_4{}^{\cdots}$ are linear in $\nu$ and the $\ms X_2{}^{\cdots}$ are independent of $\nu$.  This is all in fact a simple consequence of (i) the link (\ref{Ir}) between the scattering-angle coefficients $\ms X_k{}^{\cdots}$ and the radial-action coefficients, and (ii) the (straight-forward) extension of the predicted mass-ratio dependence (\ref{joinXs}) to $k$PM order:  $\ms X_k{}^{\cdots}$ is a polynomial of degree $\lfloor{\frac{k-1}{2}}\rfloor$ in $\nu$.  This is the spinning analog of the ``hidden simplicity'' of the mass dependence of (the local-in-time part of) the radial action (which is the complete radial action through the N$^3$LO PN level) emphasized in Ref.~\cite{Bini:2020wpo}; here in the spin terms, this is crucially dependent on expressing $I_r$ in (\ref{Ir}) in terms the covariant $L_\mr{cov}$ rather than the canonical $L$.

Finally, we can make the PN order counting explicit by restoring factors of $1/c$.  Through N$^3$LO, (\ref{Ir}) reads
\begin{alignat}{3}\label{Irexp}
I_r&=\Bigg[{-}L+GM\mu\frac{1+2\ve}{c\sqrt{-\ve}}
+\frac{1}{c^2}\frac{(GM\mu)^2}{\pi\Gamma{L_\mr{cov}}}\ms X_2
\\\nnm
&\quad
+\frac{1}{c^4}\frac{(GM\mu)^4}{3\pi(\Gamma L_\mr{cov})^3}\ms X_4+\frac{1}{c^6}\frac{(GM\mu)^6}{5\pi(\Gamma L_\mr{cov})^5}\ms X_6+\mc O(\frac{1}{c^8})\Bigg]
\\\nnm
&+\frac{\mu}{c}  a_\mr i\Bigg[\frac{(GM\mu)^2}{2\pi(\Gamma L_\mr{cov})^2}\ms X_{2}{}^\mr i
+\frac{1}{c^2}\frac{(GM\mu)^4}{4\pi(\Gamma L_\mr{cov})^4}\ms X_{4}{}^\mr i
\\\nnm
&\quad
+\frac{1}{c^4}\frac{(GM\mu)^6}{6\pi(\Gamma L_\mr{cov})^6}\ms X_{6}{}^\mr i
+\frac{1}{c^6}\frac{(GM\mu)^8}{8\pi(\Gamma L_\mr{cov})^8}\ms X_{8}{}^\mr i+\mc O(\frac{1}{c^8})\Bigg]
\\\nnm
&+\mu^2  {a_\mr ia_\mr j}\Bigg[
 \frac{(GM\mu)^2}{3\pi(\Gamma L_\mr{cov})^3}\ms X_2{}^{\mr i \mr j}
+\frac{1}{c^2}\frac{(GM\mu)^4}{5\pi(\Gamma L_\mr{cov})^5}\ms X_{4}{}^{\mr i \mr j}
\\\nnm
&\quad
+\frac{1}{c^4}\frac{(GM\mu)^6}{7\pi(\Gamma L_\mr{cov})^7}\ms X_{6}{}^{\mr i \mr j}
+\frac{1}{c^6}\frac{(GM\mu)^8}{9\pi(\Gamma L_\mr{cov})^9}\ms X_{8}{}^{\mr i \mr j}+\mc O(\frac{1}{c^8})\Bigg],
\end{alignat}
with all the coefficients, to the orders in $\ve=\gamma^2-1=\mc O(c^{-2})$ contributing here at N$^3$LO, relative $\mc O(c^{-6})$, given explicitly by (\ref{X_2l}). These depend on the remaining unknowns $\ms X_{kn}^{1\mr A}$ from the parametrization of the scattering angle, at $k$PM order and relative $n$PN order. 

In all the above manipulations, it was consistent to keep the nonspinning, spin-orbit, and bilinear-in-spin terms all through the same relative PN orders, here relative 3PN order, N$^3$LO.  However, in matching to self-force results, due to certain changes of variables discussed below, the treatment of the N$^3$LO spin-orbit and bilinear-in-spin terms will require the inclusion of the 4PN nonspinning terms.  We thus need to add to (\ref{Irexp}) the 4PN nonspinning part of the radial action for bound orbits, which includes contributions from the nonlocal-in-time tail integrals.  We present in Appendix \ref{app:4PN} the additional terms at 4PN order, which have been computed from (\ref{defIr}) applied to the 4PN EOB Hamiltonian derived in \cite{Damour:2015isa}, valid in an expansion in eccentricity (about the circular orbit limit) to sixth order.  Replacing the first two lines of (\ref{Irexp}) with (\ref{Ir4PN}) yields the final form of the radial-action function which we will use to compute the gauge-invariant quantities to be compared with self-force calculations.

\section{Third-subleading post-Newtonian spin-orbit and spin$_1$-spin$_2$ couplings}
\label{sec:N3LOSO}

The remaining unknowns in the parametrization of the scattering-angle function~(\ref{X1s}) can be fixed with available self-force results. The key feature here is the existence of a Hamiltonian/radial action allowing us to connect the scattering-angle to the redshift and spin-precession invariants that, in the small-mass-ratio limit, can be matched to expressions independently calculated in GSF literature. A vital step in this calculation is the first law of BBH mechanics, which we extend to aligned-spins and eccentric orbits.

\subsection{The first law of BBH mechanics}\label{firstlaw}

The first law of BBH mechanics~\cite{LeTiec:2011ab} was first derived for nonspinning point particles in circular orbits in Ref.~\cite{LeTiec:2011ab}, then generalized to spinning particles on circular orbits in Ref.~\cite{Blanchet:2012at}, to nonspinning particles in eccentric orbits in Refs.~\cite{Tiec:2015cxa,Blanchet:2017rcn}, and to precessing eccentric orbits of a point-mass in the small mass-ratio approximation~\cite{Fujita:2016igj}.
In the following, we briefly review the arguments leading to these incarnations of the first law for binaries, making explicit how they apply to generic mass-ratio aligned-spin systems on eccentric orbits.

Let us follow Ref.~\cite{Blanchet:2012at} and start out with an action $\mathcal{S}$ for the binary,
\begin{equation}
  \label{fullaction}
  \mathcal{S} = \mathcal{S}_\text{grav} + \mathcal{S}_1 + \mathcal{S}_2 \,,
\end{equation}
where the compact objects are approximated by effective point-particles moving along worldlines $x_{\mr i}^\mu(\tau_{\mr i})$,
\begin{equation}
  \label{ppaction}
  \mathcal{S}_{\mr i} = \int \mr d \tau_{\mr i} \! \left[ - m_{\mr i} + \frac{1}{2} S_{\mr i\mu\nu} \Lambda_{\mr i c}{}^\mu \frac{D \Lambda_{\mr i}^{c\nu}}{d \tau} + \lambda_{\mr i}^\mu S_{\mr i\mu\nu} \dot{x}_{\mr i}^{\nu} + \dots \right] \!,
\end{equation}
and the gravitational action $\mathcal{S}_\text{grav}$ is given by the Einstein-Hilbert one with appropriate gauge-fixing and boundary terms.
Here $\Lambda_{\mr i}^{c\mu}$ are frame transformations between the coordinate frame and a body-fixed frame (labeled by $c=0,1,2,3$) that is Lorentz-orthonormal ($\Lambda_{\mr i c}{}^{\mu} \Lambda_{\mr i d \mu} = \eta_{cd}$).
We take $\tau_{\mr i}$ to be the (full-metric) proper times from now on.
The equations of motion are obtained by varying the action with respect to the dynamical variables $X_A = \{ x_{\mr i}, S_{\mu\nu}, \Lambda_{\mr i}^{c \mu}, \lambda_{\mr i}^\mu, g_{\mu\nu} \}$, leading to Eqs.~\eqref{MPD}--\eqref{Einstein}, see, e.g., Refs.~\cite{Steinhoff:2014kwa,Vines:2016unv}.
The dots in Eq.~\eqref{ppaction} represent nonminimal (curvature) couplings to the worldline that may carry undetermined coefficients.
These terms also include couplings of quadratic and higher orders in spin related to spin-induced multipole moments of the body~\cite{Steinhoff:2014kwa}.

Let us write the action as an integral of a Lagrangian $L$ over coordinate time $t$ as
\begin{equation}
  \label{defL}
  \mathcal{S} = \int \mr dt \, L \,.
\end{equation}
We can vary the Lagrangian $L$ not only with respect to the dynamical variables $X_A$, but also vary certain constants appearing in the action, e.g., the masses $C_B = \{ m_1, m_2 \}$.
Furthermore, taking the dynamical variables $X_A$ on-shell (fulfilling their equations of motion) after variation, we arrive at (using summation convention for $A$, $B$)
\begin{equation}
  \delta L = \frac{\partial L}{\partial C_B} \delta C_B + \underbrace{\frac{\delta L}{\delta X_A}}_{=0 \text{ (on-shell)}} \delta X_A+ \text{(td)} \,,
\end{equation}
with a total time derivative (td).
Now, if one performs a transformation of the dynamical variables $X_A \rightarrow X'_{A'}$, which may depend on the $C_B$, then on-shell it holds
\begin{align}
  \delta L &= \frac{\partial L}{\partial C_B} \delta C_B + \bigg[ \underbrace{\frac{\delta L}{\delta X_A}}_{0} \frac{\delta X_A}{\delta X'_{A'}} \frac{\partial X'_{A'}}{\partial C_B} + \text{(td)} \bigg] \delta C_B \nonumber \nl
  + \underbrace{\frac{\delta L}{\delta X_A}}_{0} \frac{\delta X_A}{\delta X'_{A'}} \delta X'_{A'} + \text{(td)} \,.
\end{align}
Also allowing for changes of the Lagrangian of the form $L = L' + \text{(td)}$, we arrive at
\begin{equation}
  \label{firstlawmaster}
\left \langle \left( \frac{\partial L'}{\partial C_B} \right)_{X'_{A'}} \right \rangle = \left \langle \left( \frac{\partial L}{\partial C_B} \right)_{X_{A}} \right \rangle \quad \text{(on-shell)}\,,
\end{equation}
where the subscripts indicate quantities that are kept fixed during differentiation and with $\langle \dots \rangle$ an appropriate on-shell averaging that removes the total time derivatives.

For generic bound orbits, one can average the conservative motion in Eq.~\eqref{firstlawmaster} over an infinite time in order to remove total time derivatives, which can be traded for a phase-space average in regions where the motion is ergodic; see, e.g., Refs.~\cite{Hinderer:2008dm,Fujita:2016igj}.
For the aligned-spin case where the motion is confined to a plane, all oscillatory behavior can be removed by an average over a single orbit~\cite{Tiec:2015cxa} (defined as an oscillation cycle of the radial distance $r$); this is the averaging used in the present paper.
Further specializing to circular orbits, the radial distance is constant and hence the average becomes trivial~\cite{Blanchet:2012at}.
Finally, note that another benefit of the averaging in Eq.~\eqref{firstlawmaster} is that it helps to make expressions manifestly gauge-invariant~\cite{Fujita:2016igj}, which is important when matching PN Hamiltonians to (eccentric-orbit) self-force results.

It is straightforward to generalize the discussion from Lagrangians $L'$ to Hamiltonians $H'$.
Hamilton's dynamical equations for some pairs of canonical variables $(q^{\ms c}, p_{\ms c})$ are equivalently encoded by Hamilton's action principle,
\begin{equation}
  \label{HamP}
  0=\delta \mathcal{S} = \delta \int \mr dt \bigg[ \underbrace{\sum_{\ms c} p_{\ms c} \frac{d q^{\ms c}}{dt} - H'}_{L'} \bigg] \,.
\end{equation}
Noting that the dynamical variables are now $X'_{A'} = \{q^{\ms c}, p_{\ms c}\} $, and that the kinematic $p \dot{q}$-terms in $L'$ are independent of the $C_B$, we see that either Lagrangian in Eq.~\eqref{firstlawmaster} can be replaced by \emph{minus} a Hamiltonian (i.e., it can be applied also to canonical transformations between two Hamiltonians).
The rather general on-shell relation~\eqref{firstlawmaster} is interesting on its own, aside from facilitating the derivation of the first law of binary dynamics as demonstrated below.

We are now in a position to elaborate on the redshift variables $z_{\mr i}$~\cite{LeTiec:2011ab, Blanchet:2012at, Tiec:2015cxa},
\begin{equation}
  \label{redshiftdef}
  z_{\mr i} \equiv \left \langle \frac{\mr d \tau_{\mr i}}{\mr d t} \right \rangle = - \left \langle \frac{\partial L}{\partial m_{\mr i}} \right \rangle \,,
\end{equation}
where the first equality is the definition of $z_{\mr i}$ adopted by us and the second equality is a consequence of the definition of $L$~\eqref{defL} together with the original point-particle action~\eqref{ppaction}, $\int \mr dt \, L \sim - m_{\mr i} \int \mr dt \, \mr d\tau_{\mr i} / \mr d t$.
We note that this relation holds to all orders in spin if the coefficients in the nonminimal couplings (the dots) in Eq.~\eqref{ppaction} are normalized such that no further explicit dependence on the masses $m_{\mr i}$ arises~\cite{Bini:2020zqy}.
Now, several nontrivial transformations of the original action~\eqref{fullaction} are performed to arrive at a PN Hamiltonian (see, e.g., Refs.~\cite{Levi:2015msa,Blanchet:2012at,Blanchet:2017rcn}):
a transformation to SO(3)-canonical (Newton-Wigner) variables for the spin degrees of freedom, integrating out the orbital/near-zone metric or tetrad field (calculating the ``Fokker action''), reduction of higher-order time derivatives via further variable transformations, a Legendre transform to the Hamiltonian $H$, specialization to the COM system, and eventually reducing nonlocal-in-time tail contributions to local ones.
However, all of these transformations fall into the class of transformations $(X_A, L) \rightarrow (X'_{A'}, L')$ discussed above, so we may apply Eq.~\eqref{firstlawmaster} (with $L' \rightarrow - H$) to Eq.~\eqref{redshiftdef} and conclude that the redshift variables $z_{\mr i}$ can be obtained from a PN Hamiltonian $H$ via
\begin{equation}
  \label{zham}
  z_{\mr i} = \left \langle \frac{\partial H}{\partial m_{\mr i}} \right \rangle \,.
\end{equation}

Beside the redshift, let us introduce the (averaged) spin precession frequency $\Omega_{\mr i}$ as another important observable~\cite{Blanchet:2012at},
\begin{equation}
  \label{OmegaSdef}
\Omega_{\mr i} \equiv \left \langle \left| \vec{\Omega}_{\mr i}^\text{inst} \right| \right \rangle \,.
\end{equation}
The (instantaneous, directed) precession frequency $\vec{\Omega}_{S_{\mr i}}^\text{inst}$ can be read off from the equations of motion for the SO(3)-canonical spin vectors $S_{\mr i}^i$ generated by the Hamiltonian $H$,
\begin{equation}
  \label{OmegaSinsetdef}
  \frac{\mr d \vec{S}_{\mr i}}{\mr d t} = \vec{\Omega}_{\mr i}^\text{inst} \times \vec{S}_{\mr i} \,, \qquad
  \vec{\Omega}_{\mr i}^\text{inst} \equiv \frac{\partial H}{\partial \vec{S}_{\mr i}} \,.
\end{equation}
Indeed, this describes a precession of the spin vector;
it is straightforward to see that the spin length $S_{\mr i} \equiv (\vec{S}_{\mr i} \cdot \vec{S}_{\mr i})^{1/2}$ is constant,
\begin{equation}
\frac{\mr d ( \vec{S}_{\mr i} \cdot \vec{S}_{\mr i} )}{\mr d t} = 2 \vec{S}_{\mr i} \cdot \vec{\Omega}_{\mr i}^\text{inst} \times \vec{S}_{\mr i} = 0 \,.
\end{equation}

From now on, as in previous sections, we simplify the discussion to nonprecessing (aligned or anti-aligned) spins, so that $\vec{\Omega}_{\mr i}^\text{inst} \parallel \vec{S}_{\mr i}$ and $d \vec{S}_{\mr i} / dt = 0$.
That is, the spin degrees of freedom become nondynamical and can be dropped from the set of dynamical variables.\footnote{More precisely, their contribution to the kinematic terms in Hamilton's principle~\eqref{HamP} (have to) vanish or turn into total time derivatives.}
We can now include the spin lengths into our set of constants, $C_B = \{ m_{\mr i}, S_{\mr i} \}$.
Furthermore, the spin-direction component of the defining relation for $\vec{\Omega}_{\mr i}^\text{inst}$~\eqref{OmegaSinsetdef} reads $| \vec{\Omega}_{\mr i}^\text{inst} | = \partial H / \partial S_{\mr i}$.
Hence Eq.~\eqref{OmegaSdef} becomes
\begin{equation}
  \label{precessham}
\Omega_{S_{\mr i}} = \left \langle \frac{\partial H}{\partial S_{\mr i}} \right \rangle \qquad \text{(nonprecessing)} .
\end{equation}
We have now arrived at the important Eqs.~\eqref{zham} and~\eqref{precessham} for the (gauge-invariant) observables $z_{\mr i}$ and $\Omega_{\mr i}$, that could be used to relate a PN Hamiltonian $H$ to self-force results~\cite{Bini:2019lcd,Bini:2019lkm}.
But here, for the purpose of matching to self force, we perform a canonical transformation to different phase-space variables that simplify explicit calculations and connects to the radial action introduced above.

As a first step in that direction, we choose the (nonprecessing) motion to be in the equatorial plane $\theta = \pi /2$, removing the polar angle $\theta$ and its canonical conjugate momentum $p_\theta$ from the phase space;
the Hamiltonian is now of the form discussed in Sec.~\ref{sec:massshell}.
Furthermore, since we consider a system where the Hamilton-Jacobi equation is separable, one can construct a special canonical transformation (for bound orbits) where the \emph{constant} action variables
\begin{align}
  I_r &= \frac{1}{2\pi}\oint \mr dr \, p_r , &
  I_\phi &= \frac{1}{2\pi}\oint \mr d\phi \, p_\phi = L \,,
\end{align}
are the new momenta~\cite{Goldstein:2000}, with the COM orbital angular momentum of the binary $p_\phi \equiv L = \text{const}$ conjugate to the azimuthal angle $\phi$.
The advantage of these variables for our purpose is that the averaging $\langle \dots \rangle$ over one radial period becomes trivial due to the integral over one radial period $\oint$ in their definition.
The canonical conjugates to $I_r$, $I_\phi$ are the so-called angle variables $q_r$, $q_\phi$ and evolve linear in time, i.e., their angular frequencies $\Omega_r = \dot{q}_r$, $\Omega_\phi = \dot{q}_\phi$ are constant~\cite{Goldstein:2000}; overall Hamilton's equations of motion for the new, canonically transformed, Hamiltonian $H'(I_r, I_\phi = L; C_B)$ read
\begin{align}
  \label{hamactionangle}
  \Omega_r &= \frac{\partial H'}{\partial I_r} = \text{const} \,, &
  \Omega_\phi &= \frac{\partial H'}{\partial L} = \text{const} \,, \\
  \dot{I}_r &= -\frac{\partial H'}{\partial q_r} = 0 \,, &
  \dot{L} &= -\frac{\partial H'}{\partial q_\phi} = 0 \,.
\end{align}
Recalling that $C_B = \{ m_{\mr i}, S_{\mr i} \}$, we can apply Eq.~\eqref{firstlawmaster} (with both Lagrangians replaced by Hamiltonians) for the canonical transformation to action-angle variables as well.
Equations~\eqref{zham} and~\eqref{precessham} then turn into
\begin{equation}
  \label{observables}
  z_{\mr i} = \frac{\partial H'}{\partial m_{\mr i}} \,, \qquad
  \Omega_{\mr i} = \frac{\partial H'}{\partial S_{\mr i}} \,,
\end{equation}
where the averaging over one radial period is inconsequential and can be dropped.
Collecting Eqs.~\eqref{hamactionangle} and~\eqref{observables}, we see that the differential of the COM energy $E \equiv H'$ can be written as
\begin{equation}\label{1law}
\mr d E = \Omega_r \mr d I_r + \Omega_\phi \mr d L + \sum_{\mr i} ( z_{\mr i} \mr d m_{\mr i} + \Omega_{\mr i} \mr d S_{\mr i} ) .
\end{equation}
In analogy to the first law of thermodynamics for the differential of the internal energy, this can be called the first law of conservative spinning binary dynamics for nonprecessing bound orbits (covering eccentric orbits and generic mass ratios).
It also resembles the first law of BH thermodynamics, which provides a relation for the differential of the Arnowitt-Deser-Misner (ADM) energy $\mr d m_{\mr i}$ of an isolated BH and can be generalized to other compact objects as well~\cite{Carter:2010}.
Recall that Eq.~\eqref{1law} is valid to all orders in spin, if the coefficients of possible nonminimal coupling terms denoted by dots in Eq.~\eqref{ppaction} are normalized such that no additional dependence on $m_{\mr i}$ arises.
It would be interesting to consider these coefficients as part of the constants $C_B$ in future work.

Since the fundamental function introduced in the last section that generates observables for bound orbits is the radial action $I_r(E, L; m_{\mr i}, S_{\mr i})$, we consider the first law~\eqref{1law} in the form
\begin{equation}
2\pi \, \mr dI_r = T_r \mr dE - \Phi \mr dL - \sum_{\mr i} ( \mathcal{T}_{\mr i} \mr d m_{\mr i} + \Phi_{\mr i} \mr d S_{\mr i} ) \,,
\end{equation}
where we have introduced
\begin{align}
  T_r&= \frac{2\pi}{\Omega_r} =\oint \mr dt\,, &
  \Phi&=\Omega_\phi T_r=\oint \mr d \phi \,, \\
  \mathcal{T}_{\mr i}&=z_{\mr i} T_r=\oint \mr d \tau_{\mr i} \,, &
  \Phi_{\mr i} &= \Omega_{\mr i} T_r \,.
\end{align}
As a consequence of the first law, we hence obtain
\bse
\label{periods}
\begin{align}
&\frac{T_r}{2\pi}=\bigg(\frac{\partial I_r}{\partial E}\bigg)_{L,m_{\mr i},S_{\mr i}},\\
&\frac{\Phi}{2\pi}=-\bigg(\frac{\partial I_r}{\partial L}\bigg)_{E,m_{\mr i},S_{\mr i}},\\
&\frac{\mathcal{T}_{\mr i}}{2\pi}=-\bigg(\frac{\partial I_r}{\partial m_{\mr i}}\bigg)_{E,L,m_{\mr j},S_{\mr i}},\\
&\frac{\Phi_{\mr i}}{2\pi}=-\bigg(\frac{\partial I_r}{\partial S_{\mr i}}\bigg)_{E,L,m_{\mr i},S_{\mr j}}.
\end{align}
\ese
Now the redshift variables can be calculated, from a given radial action $I_r$, as the ratio of proper and coordinate times,
\begin{equation}\label{redgen}
z_{\mr i}= \frac{\mathcal{T}_{\mr i}}{T_r}\,,
\end{equation}
which manifestly agrees with the (inverse of the) Detweiler-Barack-Sago redshift invariant calculated in GSF literature~\cite{Detweiler:2008ft,Barack:2011ed}.
The spin-precession frequency $\Omega_{\mr i}$ is given by $\Omega_{\mr i}= \Phi_{\mr i} / T_r$ from which we obtain the spin-precession invariant~\cite{Dolan:2013roa}
\begin{equation}\label{spininv}
\psi_{\mr i} = \frac{\Omega_{\mr i}}{\Omega_\phi} = \frac{\Phi_{\mr i}}{\Phi} \,.
\end{equation}

\subsection{Comparison with self-force results}
\label{sec:smallq}

Starting from the radial action~\eqref{Ir}, we calculate the redshift $z_1$ and spin-precession invariants $\psi_1$ of the small body  using Eqs.~\eqref{redgen} and~\eqref{spininv}. To compare with results available in the literature, we express them in terms of the gauge-invariant variables
\footnote{
Note that the denominator for $\iota$ in Eq.~\eqref{xiota} is of 1PN order, which effectively scales down the PN ordering in such a way that manifestly nonlocal-in-time (4PN nonspinning) terms appear in the N$^3$LO correction to the spin-precession invariant. For this reason, we have included the 4PN nonspinning tail terms in the radial action as discussed at the end of the previous section.
}
\begin{equation}\label{xiota}
x = (G M \Omega_\phi)^{2/3},   \quad  \iota = \frac{3 x}{\Phi/(2\pi)-1}\,.
\end{equation}
which are linked to $(\ve,L)$ via Eqs.~\eqref{hamactionangle} and~\eqref{periods}. 
The expressions we obtain for $z_1(x,\iota)$ and $\psi (x,\iota)$ agree up to N$^2$LO with those in Eq.~(50) of Ref.~\cite{Bini:2019lcd} and Eq.~(83) of Ref.~\cite{Bini:2019lkm}. The full expressions up to N$^3$LO are lengthy, which is why we provide them as a \texttt{Mathematica} file in the Supplemental Material~\cite{ancmaterial}.

Next, we expand $U_1\equiv z_1^{-1}$ and $\psi_1$ to first order in the mass ratio $q$, first order in the massive body's  spin $ a_2$, and zeroth order in the spin of the smaller companion $a_1$,
\bse
\begin{align}
&U_1=  U^\text{(0)}_{1a^0} +\hat{a}\, U^\text{(0)}_{1a}+q\left(\delta U^\text{GSF}_{1a^0} +\hat{a}\, \delta U^\text{GSF}_{1a}\right)+\mathcal{O}(q^2,\hat{a}^2)\,,\\
&\nonumber\\
&\psi_1=  \psi^\text{(0)}_{1a^0} +\hat{a}\, \psi^\text{(0)}_{1a}+q\left(\delta \psi^\text{GSF}_{1a^0} +\hat{a}\, \delta \psi^\text{GSF}_{1a}\right)+\mathcal{O}(q^2,\hat{a}^2)\,,
\end{align}
\ese
with $\hat a= a_2/m_2$. 
In performing that expansion, we make use of the gauge-independent variables $y$ and $\lambda$, which are related to $x$ and $\iota$ via
\bse
\begin{align}\label{ylambda}
y&= (Gm_2 \Omega_\phi)^{2/3} = \frac{x}{(1 + q)^{2/3}} \,,\\
\lambda &=\frac{3y}{\Phi/(2\pi)-1} = \frac{\iota}{(1 + q)^{2/3}} \,. 
\end{align}
\ese

To compare the 1SF corrections $\delta U^{\text{GSF}}_{1\cdots}$ and $\delta \psi^{\text{GSF}}_{1\cdots}$ with those derived in the literature, we express the redshift and spin-precession invariants in terms of the Kerr-geodesic variables $(u_p,e)$, where $e$ is the eccentricity and $u_p$ is the inverse of the dimensionless semilatus rectum (see Appendix~\ref{Kerrvar} for details.)
The terms needed to solve for the N$^3$LO SO unknowns are $\delta U^\text{GSF}_{1a}$ and  $\delta\psi_{1\, a^0}^\text{GSF}$, for which we obtain
\begin{widetext}
\bse
\begin{align}\label{ua0}
\delta U^\text{GSF}_{1a}&=\left(3-\frac{7 e^2}{2}-\frac{e^4}{8}\right) u_p^{5/2}+\left(18-4
e^2-\frac{117 e^4}{4}\right) u_p^{7/2} +\bigg[\frac{251}{4}+\frac{1}{2}\ms X^{1\mr b}_{32}+
\frac{287 e^2}{2}-e^4 \left(\frac{11099}{32}+\frac{15 }{16}\ms X^{1\mr b}_{32}\right)\bigg]u_p^{9/2}\nonumber\\
&\quad +\bigg[\frac{239}{2}-\frac{5 }{4}\ms X^{1\mr b}_{32}-\frac{5
	}{2}\ms X^{1\mr b}_{33}+\frac{4 }{3}\ms X^{1\mr b}_{43} +e^2 \left(\frac{35441}{24}-\frac{41 \pi ^2}{8}-\frac{11
	}{4}\ms X^{1\mr b}_{32}-\frac{5 }{2}\ms X^{1\mr b}_{33}+2 \ms X^{1\mr b}_{43}\right)\nonumber\\
&\qquad +e^4 \left(-\frac{230497}{96}+\frac{205 \pi ^2}{32}+\frac{195
	}{32}\ms X^{1\mr b}_{32}+\frac{135 }{16}\ms X^{1\mr b}_{33}-5
\ms X^{1\mr b}_{43}\right)\bigg]u_p^{11/2}\,, \\
\label{psia0}
\delta\psi_{1\, a^0}^\text{GSF} &=-u_p+\left(\frac{9}{4}+e^2\right) u_p^2+
\left[\frac{933}{16}-\frac{123 \pi ^2}{64}-\frac{1}{4}\ms X^{1\mr t}_{32}+e^2 \left(\frac{79}{2}-\frac{123
	\pi ^2}{256}-\frac{3 }{8}\ms X^{1\mr t}_{32}\right)\right]u_p^3\nonumber\\
&\quad+\bigg[-\frac{277031}{2880}+\frac{1256 \gamma_E }{15}+\frac{15953 \pi
	^2}{6144}+\frac{11 }{8}\ms X^{1\mr t}_{32}+\frac{5
	}{4}\ms X^{1\mr t}_{33}-\frac{2 }{3}\ms X^{1\mr t}_{43}+\frac{296 }{15}\ln 2+\frac{729}{5}\ln 3+\frac{628}{15}\ln u_p\nonumber\\
&\quad+e^2
\bigg(\frac{20557}{480}+\frac{536 \gamma_E }{5}-\frac{55217 \pi
	^2}{4096}+\frac{55 }{16}\ms X^{1\mr t}_{32}+\frac{25
}{8}\ms X^{1\mr t}_{33}-2 \ms X^{1\mr t}_{43}+\frac{11720 }{3}\ln 2-\frac{10206 }{5}\ln 3+\frac{268 }{5}\ln u_p\bigg)\bigg]u_p^4\,.
\end{align}
\ese
\end{widetext}
These results can be directly compared with the GSF results in Eq.~(4.1) of Ref.~\cite{Kavanagh:2016idg}, Eq.~(23) of Ref.~\cite{Bini:2016dvs} and Eq.~(20) of Ref.~\cite{Bini:2019lcd} for the redshift, and Eq.~(3.33) of Ref.~\cite{Kavanagh:2017wot} for the precession frequency.
At N$^2$LO, as expected, our expressions depend on the scattering-angle coefficients. Upon matching these with the above-mentioned equations in the literature, we get the following four constraints (at each order in eccentricity):
\bse
\begin{align}
&u_p^{9/2} \bigg[\frac{1}{2}\ms X^{1\mr b}_{32}-\frac{97}{4}+e^4\left(\frac{1455}{32}-\frac{15}{16}\ms X^{1\mr b}_{32}\right)\bigg]=0\,, \\
&u_p^3 \bigg[\frac{97}{8}-\frac{1}{4}\ms X^{1\mr t}_{32}
+e^2 \left(\frac{291}{16}-\frac{3}{8}\ms X^{1\mr t}_{32}\right)\bigg]=0\,,
\end{align}
\ese
which can be consistently solved for the two unknowns
\begin{equation}
\label{NNLOsol}
\ms X^{1\mr b}_{32}=\ms X^{1\mr t}_{32}=\frac{97}{2}\,.
\end{equation}
Note that the special constraint~\eqref{speccons}, due to symmetry under interchanging the two bodies' labels $1\leftrightarrow 2$, is thus satisfied. Similarly, at N$^3$LO order, after substituting in the N$^2$LO coefficients, it holds that
\bse
\begin{align}
&u_p^{11/2} \bigg[-\frac{26881}{72}+\frac{241 \pi ^2}{96}-\frac{5
	}{2}X^{1\mr b}_{33}+\frac{4 }{3}X^{1\mr b}_{43}\\
&\qquad +e^2 \bigg(-\frac{1846}{3}+\frac{241 \pi ^2}{64}-\frac{5
	}{2}X^{1\mr b}_{33}+2 X^{1\mr b}_{43}\bigg)\nonumber\\
&\qquad +e^4 \bigg(\frac{276775}{192}-\frac{1205 \pi ^2}{128}+\frac{135
	}{16}X^{1\mr b}_{33}-5 X^{1\mr b}_{43}\bigg)\bigg]=0, \nonumber\\
&u_p^4 \bigg[\frac{8381}{48}-\frac{41 \pi ^2}{16}+\frac{5
	}{4}\ms X^{1\mr t}_{33}-\frac{2}{3}\ms X^{1\mr t}_{43} \\
&\qquad +e^2 \bigg(\frac{17647}{32}-\frac{123 \pi ^2}{16}+\frac{25}{8}\ms X^{1\mr t}_{33}-2 \ms X^{1\mr t}_{43}\bigg)\bigg]=0\nonumber\,.
\end{align}
\ese
These five equations can be consistently solved for the remaining four unknowns in the N$^3$LO SO scattering angle,
\begin{gather}\label{NNNLOsol}
\ms X^{1\mr b}_{33}=\ms X^{1\mr t}_{33}=\frac{177}{4},  \\ \ms X^{1\mr b}_{43}=\frac{17423}{48}-\frac{241 \pi ^2}{128}, \quad \ms X^{1\mr t}_{43}=\frac{2759}{8}-\frac{123 }{32}\pi ^2 .\nonumber
\end{gather}
Again, the special constraint~\eqref{speccons} is satisfied by $\ms X^{1\mr b}_{33}$ and $\ms X^{1\mr t}_{33}$. 
Considering the {\sonestwo} dynamics, the relevant constraints can be obtained from the linear-in-spin correction to the spin-precession invariant, which in terms of the remaining unknown coefficients $\ms X^{1\times}_{ij}$ reads
\begin{widetext}
	\begin{align}\label{psia1}
	\delta\psi_{1\, a^1}^\text{GSF}& = -\frac{u_p^{3/2}}{2}-\left(\frac{41}{8}+\frac{e^2}{8}\right)
	u_p^{5/2}- \left[\frac{63}{32}+\frac{123 \pi ^2}{64}+\frac{3 }{16}\ms X^{1\times}_{32}+e^2 \left(\frac{71}{4}+\frac{123 \pi ^2}{256}+\frac{9 }{32}\ms X^{1\times}_{32}\right)\right]u_p^{7/2}\\
	&+\bigg[\frac{75841 \pi ^2}{6144}-\frac{4496717}{5760}+\frac{1256\gamma_E }{15}+\frac{39 }{32}\ms X^{1\times}_{32}+\frac{15}{16}\ms X^{1\times}_{33}-\frac{8}{15}\ms X^{1\times}_{43}+\frac{296}{15}\ln 2+\frac{729}{5}\ln 3+\frac{628 }{15}\ln u_p\nonumber\\
	&+e^2 \left(\frac{7703 \pi
			^2}{4096} -\frac{1016249}{640}+\frac{536 \gamma_E}{5}+\frac{195 }{64}\ms X^{1\times}_{32}+\frac{75
	}{32}\ms X^{1\times}_{33}-\frac{8 }{5}\ms X^{1\times}_{43}+\frac{11720 }{3}\ln 2-\frac{10206 }{5}\ln 3+\frac{268}{5}\ln u_p\right)\bigg]u_p^{9/2}\,. \nonumber
	\end{align}
\end{widetext}	
At N$^2$LO, this can be matched to Eqs.~(52) and (56) of Ref.~\cite{Bini:2019lkm} to get the two constraints (at each order in $e$)
\begin{equation}
u_p^{7/2} \bigg[\frac{75}{8}+\frac{3 }{16}\ms X^{1\times}_{32}+e^2\left(\frac{225}{16}+\frac{9}{32}\ms X^{1\times}_{32}\right)\bigg]=0\,,
\end{equation}
which can be solved  for
 \begin{equation}\label{NNLOS1S2}
 \ms X^{1\times}_{32} = -50\,.
 \end{equation}
Similarly, at N$^3$LO it holds that
\begin{align}
u_p^{9/2} &\bigg[-\frac{6299}{16}+\frac{123 \pi ^2}{32}+\frac{15
 }{16}\ms X^{1\times}_{33}-\frac{8 }{15}\ms X^{1\times}_{43}+\\
\qquad &e^2\bigg(-\frac{41943}{32}+\frac{369 \pi ^2}{32}+\frac{75
 }{32}\ms X^{1\times}_{33}-\frac{8 }{5}\ms X^{1\times}_{43}\bigg)\bigg]=0\,. \nonumber
\end{align}
Each order in eccentricity is solved for the remaining {\sonestwo} unknown coefficients
\begin{equation}\label{NNNLOS1S2}
\ms X^{1\times}_{33} =-\frac{1383}{5}, \qquad \ms X^{1\times}_{43}=-\frac{9795}{8}+\frac{1845 \pi ^2}{256}\,.
\end{equation}

Combining the solutions obtained in this section with the results of Sec.~\ref{sec:massdep} yields the scattering angle containing the complete local-in-time conservative SO and {\sonestwo} dynamics through the third-subleading PN order
\begin{widetext}
 	\begin{align}
 	\frac{\chi}{\Gamma}=\quad &\left(\frac{GM}{b\sqrt{\ve}}\right)2\frac{1+2\ve}{\sqrt{\ve}}+\left(\frac{GM}{b\sqrt{\ve}}\right)^2\frac{3\pi}{4}(4+5\ve)\label{chifin}\\
 	-&\left(\frac{GM}{b\sqrt{\ve}}\right)^3\frac{1}{\sqrt{\ve}}\left[2\frac{1-12\ve-72\ve^2-64\ve^3}{3\ve}+\nu\left(\frac{8+94\ve+313\ve^2}{12}+\mc O(\ve^3)\right)\right]\nonumber\\
 	+&\left(\frac{GM}{b\sqrt{\ve}}\right)^4\pi\left[\frac{105}{64}(16+48\ve+33\ve^2)+\nu \bigg({-}\frac{15}{2}+\bigg(\frac{123}{128}\pi^2-\frac{557}{8}\bigg)\ve+\mc O(\ve^2)\bigg)\right]\nonumber\\
 	-\left(\frac{a_b}{b\sqrt{\ve}}\right)\bigg\{& \left(\frac{GM}{b\sqrt{\ve}}\right)4\gamma \sqrt{\ve}+\left(\frac{GM}{b\sqrt{\ve}}\right)^22\pi \gamma\, (2+5\ve)\, \nonumber\\
 	+&\left(\frac{GM}{b\sqrt{\ve}}\right)^3\frac{\gamma}{\sqrt{\ve}}\left[12(1+12\ve+16\ve^2)-\nu\left(10\ve+\frac{97}{2}\ve^2+\frac{177}{4}\ve^3+\mc O(\ve^4)\right)\right]\nonumber\\
 	+&\left(\frac{GM}{b\sqrt{\ve}}\right)^4\pi\gamma\left[\frac{21}{2}(8+36\ve+33\ve^2)-\nu\left(\frac{21}{2}+\frac{495}{4}\ve+\left(\frac{17423}{48}-\frac{241\pi^2}{128}\right)\ve^2+\mc O(\ve^3)\right)\right]\bigg\}\nonumber\\
 	-\left(\frac{a_t}{b\sqrt{\ve}}\right)\bigg\{&\left(\frac{GM}{b\sqrt{\ve}}\right)4\gamma \sqrt{\ve}+\left(\frac{GM}{b\sqrt{\ve}}\right)^2\frac{3\pi}{2} \gamma\, (2+5\ve)\, \nonumber\\
 	+&\left(\frac{GM}{b\sqrt{\ve}}\right)^3\frac{\gamma}{\sqrt{\ve}}\left[8(1+12\ve+16\ve^2)-\nu\left(10\ve+\frac{97}{2}\ve^2+\frac{177}{4}\ve^3+\mc O(\ve^4)\right)\right]\nonumber\\
 	+&\left(\frac{GM}{b\sqrt{\ve}}\right)^4\pi\gamma\left[\frac{105}{16}(8+36\ve+33\ve^2)-\nu\left(9+\frac{219}{2}\ve+\left(\frac{2759}{8}-\frac{123}{32}\pi^2\right)\ve^2+\mc O(\ve^3)\right)\right]\bigg\}\nonumber\\
+\left(\frac{a_1a_2}{b^2\ve}\right)\bigg\{&\left(\frac{GM}{b\sqrt{\ve}}\right)\frac{4}{\sqrt{\ve}}(\ve+2\ve^2)+\left(\frac{GM}{b\sqrt{\ve}}\right)^2\frac{3\pi}{2}(2+19\ve+20\ve^2)\nonumber\\
+&\left(\frac{GM}{b\sqrt{\ve}}\right)^3\frac{1}{\sqrt{\ve}}\left[8(1+38\ve+128\ve^2+96\ve^3)+\nu\left(8\ve-50\ve^2-\frac{1383}{5}\ve^3+\mc O(\ve^4)\right)\right]\nonumber\\
+&\left(\frac{GM}{b\sqrt{\ve}}\right)^4\pi\left[
\frac{105}{16} (24+212\ve+447\ve^2+264\ve^3)
+\nu\left(\frac{15}{2}-\frac{405}{4}\ve+\left(-\frac{9795}{8}+\frac{1845\pi^2}{256}\right)\ve^2+\mc O(\ve^3)\right)\right]\bigg\}\nonumber\,.
\end{align}	
\end{widetext}
Importantly, we have checked that all the above results can be reproduced by starting from a Hamiltonian ansatz (rather than a radial action), constraining it via the mass-ratio dependence of the scattering angle (calculated via~{(\ref{chidef})}), and obtaining the redshift and spin-precession invariants through Eqs.~\eqref{zham} and~\eqref{precessham}.

\section{Effective-one-body Hamiltonian and comparison with numerical relativity}
\label{sec:NRcomp}
In this section, we quantify the improvement in accuracy from the new N$^3$LO SO and S$_1$S$_2$ corrections using numerical relativity (NR) simulations as means of comparison. We do this using an EOB Hamiltonian, calculated using the scattering angle obtained above, since the resummation of PN results it grants is expected to improve the agreement with NR in the high-frequency regime.

The EOB Hamiltonian is calculated from an effective Hamiltonian $H^\text{eff}$ via the energy map
\begin{equation}
\label{EOBHam}
H^\text{EOB} = M \sqrt{1 + 2\nu \left(\frac{H^\text{eff}}{\mu} - 1\right)},
\end{equation}
where we use for the effective Hamiltonian an aligned-spin version of the Hamiltonian for a nonspinning test mass in a Kerr background (denoted SEOB$_\text{TM}$ in Ref.~\cite{Khalil:2020mmr}) with SO and S$_1$S$_2$ PN corrections. The effective Hamiltonian is given by
\begin{align}
\label{Heff}
{H}^\text{eff} &= \bigg[
A
\left( \mu^2 + p^2  + B_{p_r} p_r^2  + B_{L} \frac{L^2a^2}{r^2} + \mu^2Q  
\Big)\right]^{1/2}  \nonumber\\
&\quad  
+ \frac{GMr}{\Lambda} L \left(g_S S + g_{S^*} S^*\right),
\end{align}
where $\Lambda = (r^2+a^2)^2 - \Delta a^2$ with $\Delta = r^2 - 2GMr + a^2$. The Kerr spin $a$ is mapped to the binary's spins via $a = a_1 + a_2$, and the potentials are taken to be
\begin{subequations}
\begin{align}
A &= \frac{\Delta r^2}{\Lambda} \left(A^0 +  A^\text{SS}\right), \\
B_{p_r} &= \left(1 - \frac{2GM}{r} + \frac{a^2}{r^2}\right) \left(A^0 D^0 +  B_{p_r}^\text{SS}\right)  - 1, \\
B_L &= - \frac{r^2 + 2GM r}{\Lambda}, \\
Q &= Q^0 + Q^\text{SS},
\end{align}
\end{subequations}
i.e., we factorize the PN corrections to the Kerr potentials. 
The zero-spin corrections $A^0(r),~D^0(r)$ and $Q^0(r)$ are given by Eq.~(28) of Ref.~\cite{Khalil:2020mmr} and are based on the 4PN nonspinning Hamiltonian derived in Ref.~\cite{Damour:2015isa}.
The SO corrections are encoded in the gyro-gravitomagnetic factors $g_S$, and $g_{S^*}$, while the S$_1$S$_2$ corrections are added through $A^\text{SS}, \, B_{p_r}^\text{SS}$, and $Q^\text{SS}$.

For those PN corrections, we choose a gauge such that $g_S$, and $g_{S^*}$ are independent of $L$~\cite{Damour:2008qf,Nagar:2011fx,Barausse:2011ys}; we write an ansatz such that, up to N$^3$LO,
\begin{align}
g_S(r, p_r) &= 2 \sum_{i=0}^{3}\sum_{j=0}^{i} \alpha_{ij}\frac{p_r^{2(i-j)}}{c^{2i} r^j}, \nonumber\\
g_{S^*}(r, p_r) &= \frac{3}{2} \sum_{i=0}^{3}\sum_{j=0}^{i} \alpha_{ij}^*\frac{p_r^{2(i-j)}}{c^{2i} r^j},
\end{align}
for some unknown coefficients $\alpha_{ij}$ and $\alpha_{ij}^*$.
For the S$_1$S$_2$ corrections, $A^\text{SS}$ and $B_{p_r}^\text{SS}$ start at NLO and are independent of $p_r$, while $Q^\text{SS}$ starts at NNLO and depends on $p_r^4$ or higher powers of $p_r$, i.e., we use an ansatz of the form
\begin{align}
A^\text{SS} &= S_1S_2 \left(\frac{\alpha_4^A}{c^6r^4} + \frac{\alpha_5^A}{c^8r^5} + \frac{\alpha_6^A}{c^{10}r^6}\right), \nonumber\\
B^\text{SS} &= S_1S_2 \left(\frac{\alpha_3^B}{c^4r^3} + \frac{\alpha_4^B}{c^6r^4} + \frac{\alpha_5^B}{c^8r^5}\right), \nonumber\\
Q^\text{SS} &= S_1S_2 \left(\alpha_{34}^Q \frac{p_r^4}{c^6r^3} + \alpha_{44}^Q \frac{p_r^4}{c^8r^4} + \alpha_{36}^Q \frac{p_r^6}{c^8r^3}\right).
\end{align}
To determine those unknowns, we calculate the scattering angle from such an ansatz using Eq.~\eqref{chidef} (which entails inverting the EOB Hamiltonian for $p_r$ in a PN expansion, differentiating with respect to $L$, and integrating with respect to $r$). We then match the result of that calculation to the scattering angle calculated in the previous section and solve for the unknown coefficients in the Hamiltonian ansatz. This uniquely determines all the coefficients of the spinning part of the Hamiltonian since our choice for the ansatz fixes the gauge dependence of the Hamiltonian. (See Sec.~\ref{sec:massshell} for a discussion of the gauge freedom in the Hamiltonian.)

We obtain the gyro-gravitomagnetic factors
\begin{widetext}
\bse
\label{gyros}
\begin{align}
g_S &= 2 \bigg\lbrace
1 + \frac{\nu}{c^2}\left[-\frac{27}{16} \frac{p_r^2}{\mu^2} - \frac{5 }{16}\frac{GM}{r}\right]
+\frac{\nu}{c^4}\left[
\left(\frac{5}{16}+\frac{35 \nu}{16}\right) \frac{p_r^4}{\mu^4}
-\left(\frac{21}{4}-\frac{23 \nu}{16}\right)\frac{p_r^2}{\mu^2}\frac{GM}{r}
-\left(\frac{51}{8}+\frac{\nu}{16}\right)\frac{(GM)^2}{r^2}
\right] \nonumber\\
&\quad +\frac{\nu}{c^6} \bigg[
\left(-\frac{80399}{2304}+\frac{241 \pi ^2}{384}+\frac{379 \nu }{64}-\frac{7 \nu ^2}{256}\right)\frac{(GM)^3}{r^3} 
+ \left(-\frac{5283}{128}+\frac{1557 \nu }{32}+\frac{69 \nu ^2}{128}\right) \frac{p_r^2}{\mu^2} \frac{(GM)^2}{r^2} \nonumber\\
&\quad\qquad + \left(\frac{781}{256}+\frac{831 \nu }{64}-\frac{771 \nu ^2}{256}\right) \frac{p_r^4}{\mu^4} \frac{GM}{r} 
+  \left(\frac{7}{256}-\frac{63 \nu }{64}-\frac{665 \nu ^2}{256}\right) \frac{p_r^6}{\mu^6}
\bigg]
\bigg\rbrace, \\
g_{S^*} &= \frac{3}{2} \bigg\lbrace
1 + \frac{1}{c^2} \left[-\left(\frac{3 \nu }{2}+\frac{5}{4}\right) \frac{p_r^2}{\mu^2}
-\left(\frac{3}{4}+\frac{\nu }{2}\right)\frac{GM}{r}\right]
 \nonumber\\
&\quad
+\frac{1}{c^4} \left[
\left(\frac{15 \nu ^2}{8}+\frac{5 \nu }{3}+\frac{35}{24}\right) \frac{p_r^4}{\mu^4}
+ \left(\frac{19 \nu ^2}{8}-\frac{3 \nu }{2}+\frac{23}{8}\right)\frac{p_r^2}{\mu^2}\frac{GM}{r}
+ \left(-\frac{\nu ^2}{8}-\frac{13 \nu }{2}-\frac{9}{8}\right) \frac{(GM)^2}{r^2}
\right] \nonumber\\
&\quad 
+\frac{1}{c^6}
 \bigg[
\left(-\frac{135}{64}+\frac{41 \pi ^2 \nu }{48}-\frac{7627 \nu }{288} +\frac{237 \nu ^2}{32} -\frac{\nu ^3}{16}\right) \frac{(GM)^3}{r^3} 
+ \left(-\frac{15}{32}-\frac{279 \nu }{16}+\frac{787 \nu ^2}{16} + \frac{9 \nu ^3}{8}\right) \frac{p_r^2}{\mu^2} \frac{(GM)^2}{r^2} \nonumber\\
&\quad\qquad 
+ \left(-\frac{1105}{192}-\frac{53 \nu }{96}+\frac{117 \nu ^2}{32}-\frac{81 \nu ^3}{16}\right) \frac{p_r^4}{\mu^4} \frac{GM}{r} 
+ \left(-\frac{105}{64}-\frac{175 \nu }{96}-\frac{77 \nu ^2}{32}-\frac{35 \nu ^3}{16}\right) \frac{p_r^6}{\mu^6}
\bigg]
\bigg\rbrace,
\end{align}
\ese
and the S$_1$S$_2$ corrections
\bse
\label{SSpots}
\begin{align}
A^\text{SS} &= \frac{S_1S_2}{G^2M^2\mu^2} \bigg\lbrace
\frac{(GM)^4}{c^6r^4}\left(2\nu - \nu^2\right)
+ \frac{(GM)^5}{c^8r^5}\left(\frac{17 \nu }{2}+\frac{113 \nu ^2}{8}+\frac{3 \nu^3}{4}\right) \nonumber\\
&\quad\qquad\qquad
+ \frac{(GM)^6}{c^{10}r^6}\left(\frac{61 \nu}{2}-\frac{41 \pi^2 \nu^2}{16}+\frac{3791 \nu^2}{48}+\frac{25 \nu^3}{4}+\frac{21 \nu^4}{32}\right) 
\bigg\rbrace, \\
B_{p_r}^\text{SS} &=\frac{S_1S_2}{G^2M^2\mu^2} \bigg\lbrace
\frac{(GM)^3}{c^4r^3}\left(6\nu + \frac{9}{2} \nu^2\right)
+ \frac{(GM)^4}{c^6r^4}\left(20 \nu+26 \nu ^2+10 \nu ^3\right)
+ \frac{(GM)^5}{c^8r^5}\left(67 \nu+328 \nu ^2-\frac{375 \nu ^3}{4} -\frac{37 \nu ^4}{16}\right)\bigg\rbrace, \\
Q^\text{SS} &= \frac{S_1S_2}{G^2M^2\mu^2} \bigg\lbrace
\frac{1}{c^6}\frac{p_r^4}{\mu^4}\frac{(GM)^3}{r^3} \left(+\frac{5 \nu }{2}+\frac{45 \nu ^2}{8}-\frac{25 \nu^3}{4}\right) 
+ \frac{1}{c^8} \bigg[
\frac{p_r^6}{\mu^6}\frac{(GM)^3}{r^3} \left(-\frac{7 \nu }{4}-\frac{63 \nu^2}{16}-\frac{35 \nu^3}{8}+\frac{245 \nu^4}{32}\right) \nonumber\\
&\quad\qquad\qquad
+ \frac{p_r^4}{\mu^4}\frac{(GM)^4}{r^4} \left(-13 \nu+\frac{183 \nu^2}{4} -63 \nu^3 -\frac{437 \nu^4}{16}\right)
\bigg]\bigg\rbrace.
\end{align}
\ese
\end{widetext}
Importantly, the factors $g_S$ and $g_{S^*}$, obtained here for the aligned-spin case, also fix the generic-spin case by simply writing the odd-in-spin part of the effective Hamiltonian as
\begin{equation}
H_\text{eff}^\text{odd} = \frac{GMr}{\Lambda} \bm{L} \cdot \left(g_S \bm{S} + g_{S^*} \bm{S}^*\right),
\end{equation}
with $g_S$ and $g_{S^*}$ unmodified since they are independent of the spins (see Ref.~\cite{Antonelli:2020aeb} for more details.)
However, the spin$_1$-spin$_2$ corrections in Eq.~\eqref{SSpots} are only for aligned spins since the generic-spins case has additional contributions proportional to $(\bm{n}\cdot \bm{S}_1)(\bm{n}\cdot \bm{S}_2)$, where $\bm{n}=\bm{r}/r$.
Such terms vanish for aligned spins and cannot be fixed from aligned-spin self-force results or be removed by canonical transformations.

\begin{figure*}
\centering
\includegraphics[width=0.49\linewidth]{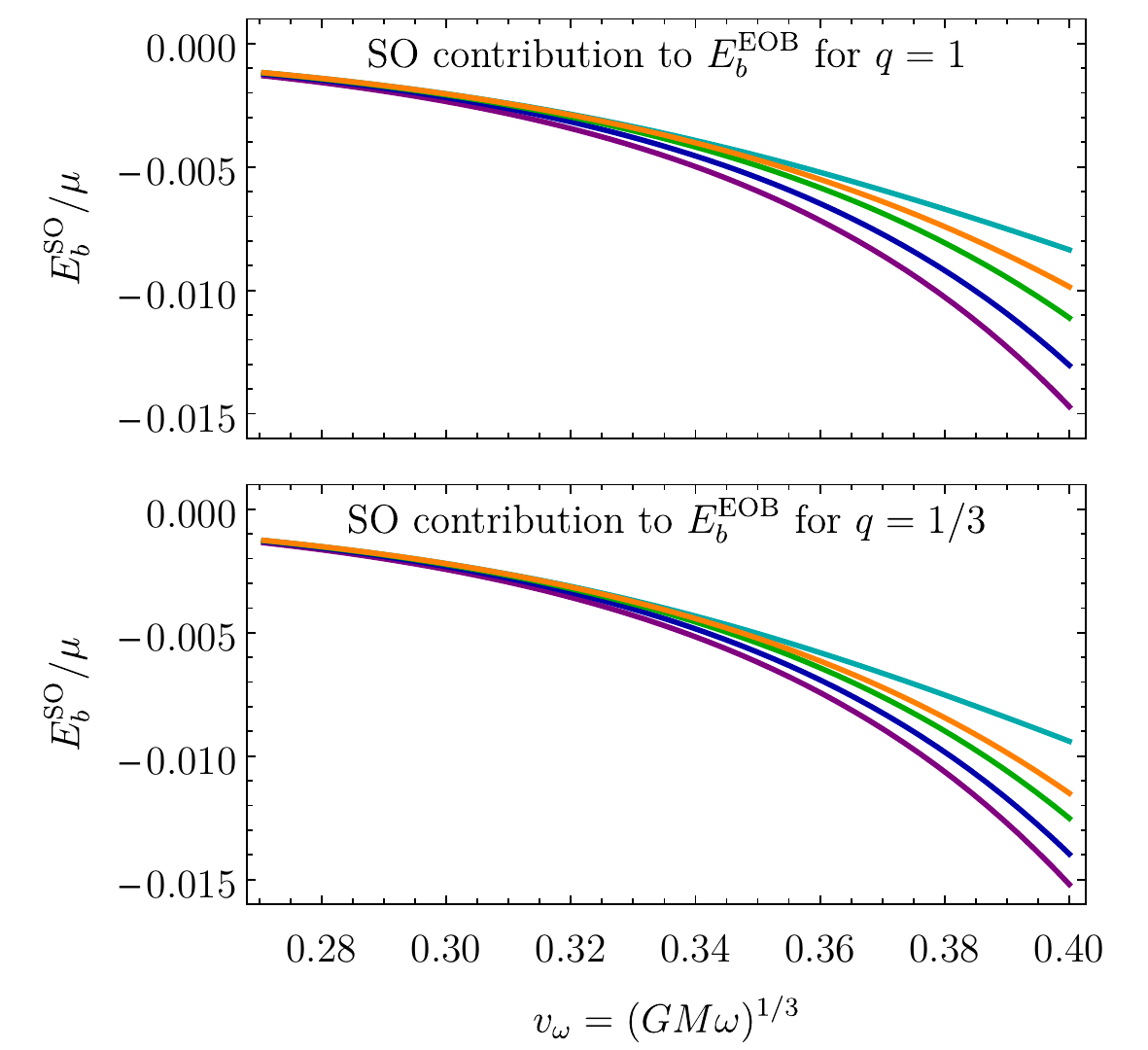}
\includegraphics[width=0.49\linewidth]{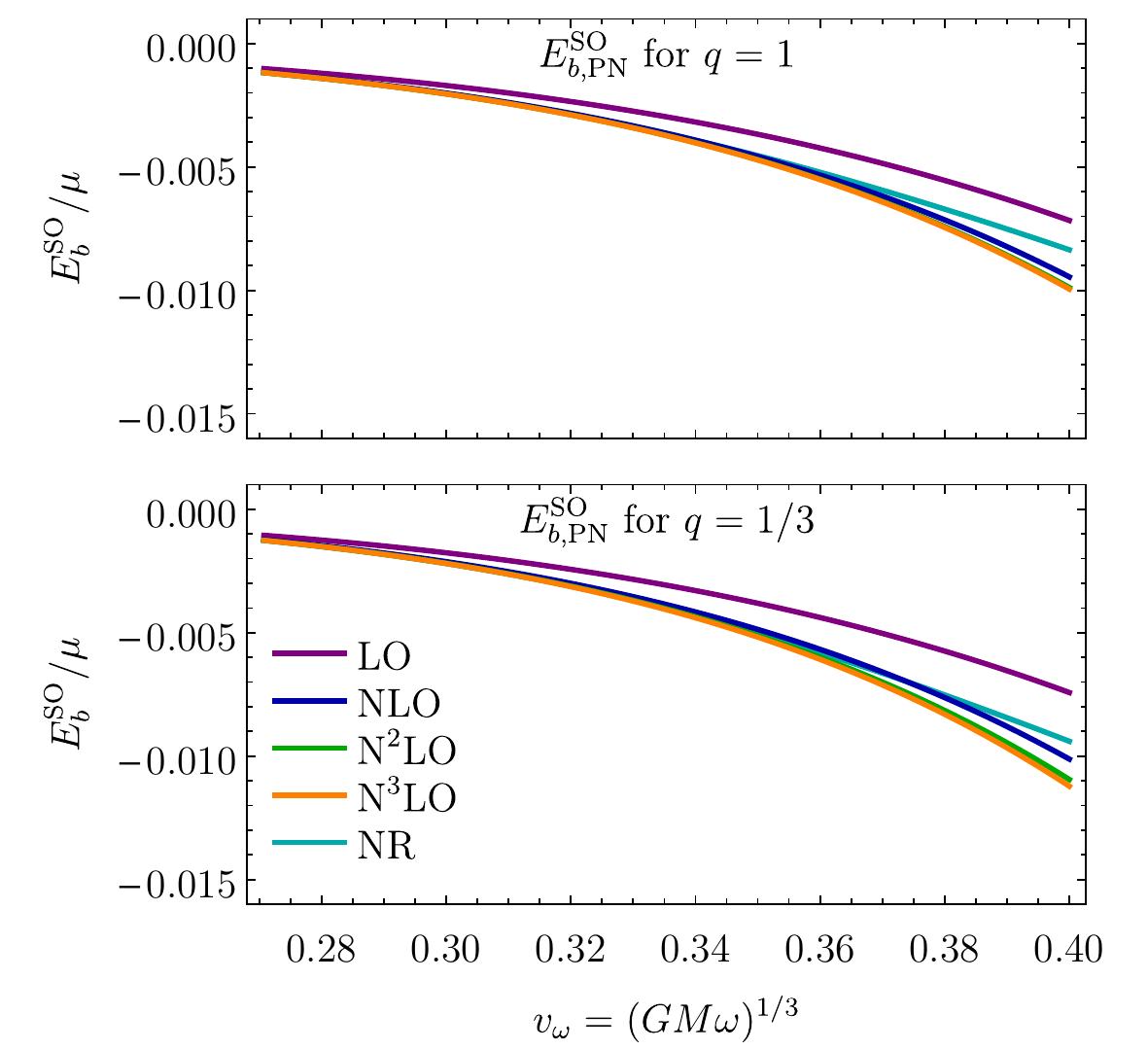}
\caption{Binding energy versus the velocity parameter $v_\omega$ for the SO contribution to the EOB (left panels) and PN-expanded (right panels) binding energies for mass ratios $q=1$ (top panels) and $q=1/3$ (bottom panels).}
\label{fig:bindEnSO} 
\end{figure*}

\begin{figure*}
\centering
\includegraphics[width=0.49\linewidth]{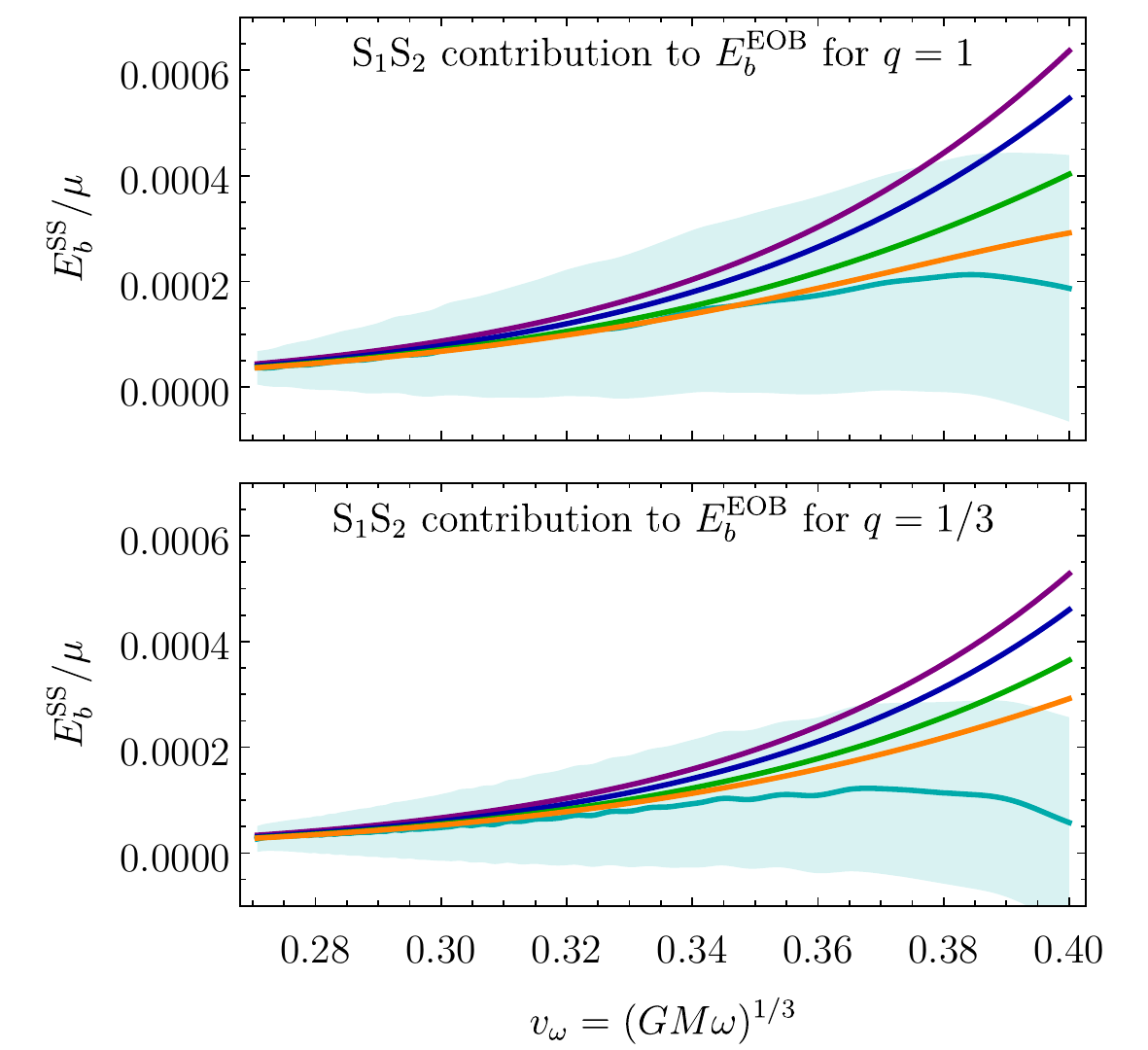}
\includegraphics[width=0.49\linewidth]{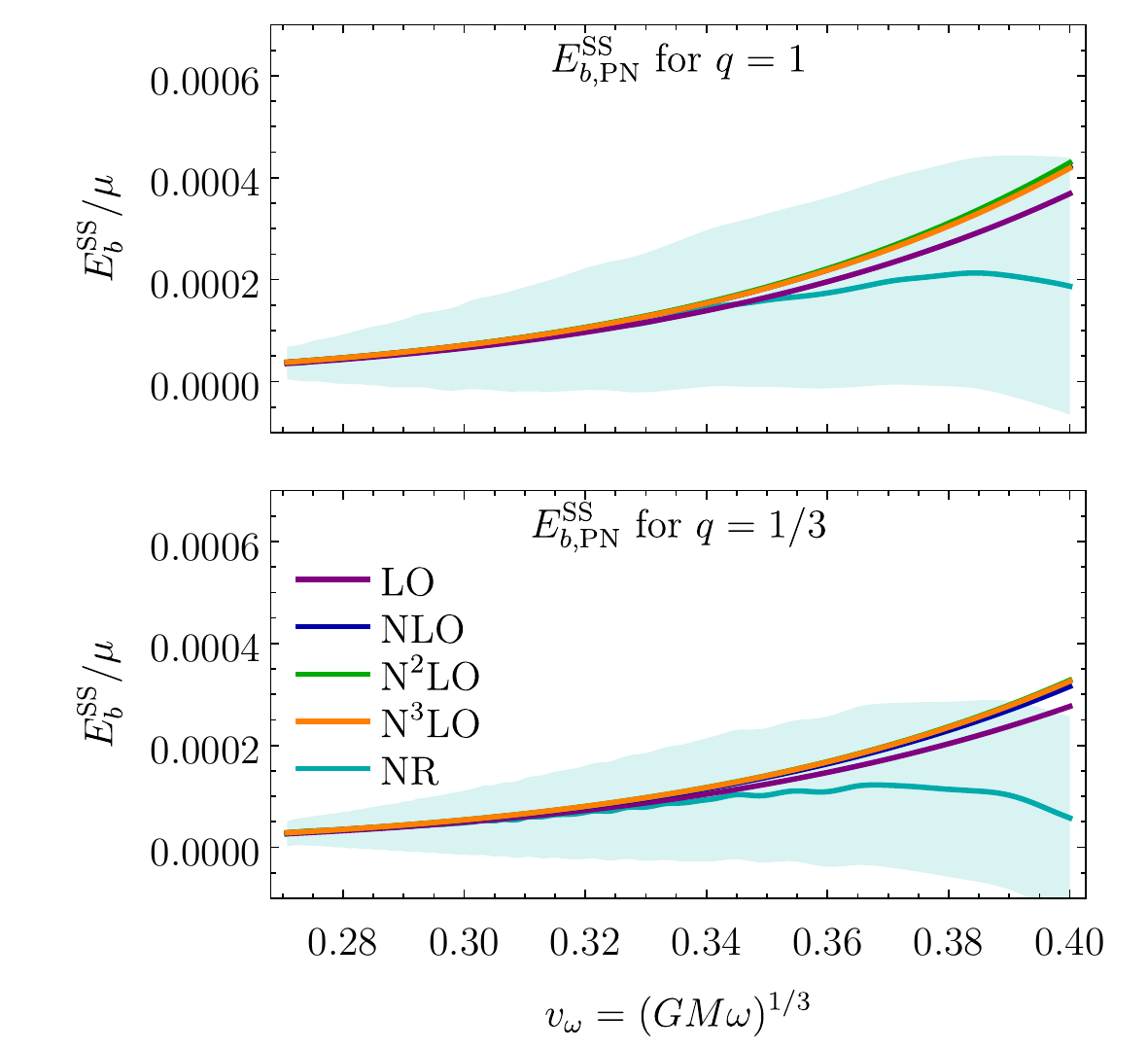}
\caption{Binding energy versus the velocity parameter $v_\omega$ for the S$_1$S$_2$  contribution to the EOB (left panels) and PN-expanded (right panels) binding energies for mass ratios $q=1$ (top panels) and $q=1/3$ (bottom panels). The NR error is indicated by the shaded regions.}
\label{fig:bindEnSS} 
\end{figure*}

For comparison with NR, a particularly good quantity to consider is the binding energy, since it encapsulates the conservative dynamics of analytical models, and can be obtained from accurate NR simulations~\cite{Damour:2011fu,Nagar:2015xqa}. The NR data for binding energy that we use were extracted in Ref.~\cite{Ossokine:2017dge} from the Simulating eXtreme Spacetimes (SXS) catalog \cite{SXS}.
The binding energy calculated from NR simulations is defined by
\begin{equation}
E_b^\text{NR} = E_\text{ADM} - E_\text{rad} - Mc^2,
\end{equation}
where $E_\text{rad}$ is the radiated energy, and $E_\text{ADM}$ is the ADM energy at the beginning of the simulation. 
We then calculate the binding energy from the EOB conservative Hamiltonian using $E_b = H^\text{EOB} - Mc^2$ for exact circular orbits at different orbital separations, i.e., we neglect the radiation-reaction due to the emitted GWs. As a result of this assumption, the circular-orbit binding energy we calculate is not expected to agree  with NR in the last few orbits.

To obtain the binding energy from a Hamiltonian in an analytical PN expansion, we set $p_r=0$ for circular orbits and perturbatively solve $\dot{p}_r=0=-\partial H/\partial r$ for the angular momentum $L$. The orbital frequency $\omega$ is given by $\omega = \partial H / \partial L$ from which we define the velocity parameter
\begin{equation}
v_\omega = (GM\omega)^{1/3}.
\end{equation}
Expressing the PN-expanded Hamiltonian in terms of $v_\omega$ yields, for the SO part,
\begin{widetext}
\begin{align}
\label{EbPNSO}
E_{b,\text{PN}}^\text{SO} &= \frac{\nu}{G M} \bigg\lbrace
v_\omega^5 \left[-\frac{4}{3} S - S^* \right]
+ v_\omega^7 \left[ S \left(\frac{31 \nu }{18}-4\right) + S^* \left(\frac{5 \nu }{3}-\frac{3}{2} \right)\right] \nonumber\\
&\quad
+ v_\omega^9 \left[\frac{S}{24}  \left(- 324+ 633 \nu  - 14 \nu ^2\right) + \frac{S^*}{8} \left( - 27 + 156 \nu -5 \nu ^2\right) \right] \nonumber\\
&\quad + v_\omega^{11} \bigg[ S \left(- 45 + \frac{19679+174\pi^2}{144} \nu - \frac{1979}{36} \nu^2 - \frac{265}{3888} \nu^3 \right)
- \frac{S^*}{8} \left(\frac{135}{2} - 565 \nu + \frac{1109}{3} \nu^2 + \frac{50}{81} \nu^3 \right) \bigg] \bigg\rbrace,
\end{align}
while for the S$_1$S$_2$ part,
\begin{align}
\label{EbindS1S2}
E_{b,\text{PN}}^\text{SS} &= \frac{S_1 S_2}{G^2 M^3} \bigg[v_\omega^6 
+ v_\omega^8 \left(\frac{5}{6} + \frac{5}{18}\nu \! \right)
+ v_\omega^{10} \left(\frac{35}{8}-\frac{1001}{72}\nu -\frac{371 }{216}\nu ^2 \! \right)  + v_\omega^{12} \left( \frac{243}{16} + \frac{123 \pi^2-4214}{32}\nu + \frac{147}{8}\nu ^2 + \frac{13}{16}\nu^3 \! \right) \! \bigg].
\end{align}
\end{widetext}
The same steps can be performed numerically to obtain the EOB binding energy without a PN expansion.

To examine the effect of the new N$^3$LO terms on the binding energy, we isolate the SO and the S$_1$S$_2$ contributions to the binding energy by combining configurations with different spin orientations (parallel or anti-parallel to the orbital angular momentum), as explained in Refs.~\cite{Dietrich:2016lyp,Ossokine:2017dge}. For the SO contribution, we use
\begin{equation}
\label{EbSO}
E_b^\text{SO}(\nu,\hat{a},\hat{a}) = \frac{1}{2} \left[ E_b(\nu,\hat{a},\hat{a}) - E_b(\nu,-\hat{a},-\hat{a})\right] + \Order(\hat{a}^3),
\end{equation}
while for the S$_1$S$_2$ contribution, we use
\begin{align}
\label{EbSS}
E_b^\text{SS}(\nu,\hat{a},\hat{a}) &= E_b(\nu,\hat{a},0) + E_b(\nu,0,-\hat{a}) - E_b(\nu,\hat{a},-\hat{a}) \nonumber\\
&\quad - E_b(\nu,0,0)  + \Order(\hat{a}^3).
\end{align}

In Fig.~\ref{fig:bindEnSO}, we plot the SO contribution to the EOB and PN-expanded binding energies versus the velocity parameter $v_\omega$ for spin magnitudes $\hat{a} = 0.6$. We also plot the NR results by combining the binding energies of configurations with different spins using results from Refs.~\cite{SXS,Ossokine:2017dge}.
From the figure, we see that, adding each PN order improves agreement of the EOB binding energy with NR, especially in the high-frequency regime, with better improvement for equal masses than for unequal masses.  In contrast, the PN binding energy, plotted using Eq.~\eqref{EbPNSO}, seems \emph{not} to converge towards NR in the high-frequency regime, with little difference between the N$^2$LO and N$^3$LO SO orders. 
Figure~\ref{fig:bindEnSS} shows the S$_1$S$_2$ contribution to the EOB and PN binding energies. As in the SO case, adding the new N$^3$LO significantly improves agreement of the EOB binding energy to NR, especially for equal masses, but there is little difference between PN orders for the PN binding energy.

Note that Figs.~\ref{fig:bindEnSO} and \ref{fig:bindEnSS} should not be interpreted as a direct comparison between PN and EOB dynamics since our results were obtained for simplicity using exact circular-orbits, which leads to a very different behavior than for an \emph{inspiraling} binary;  Refs.~\cite{Ossokine:2017dge,Antonelli:2019fmq,Nagar:2015xqa}, for example, show that EOB results are significantly better than PN when taking into account the binary evolution.
Let us also stress that while the EOB and PN curves are based on the same PN information, the EOB Hamiltonian represents a particular resummation of the PN results.
We leave the exploration of other resummations and a calibration to NR for future work.

\section{Conclusions}
\label{sec:conc}

GW astronomy allows a multitude of applications in fundamental and astrophysics \cite{Abbott:2019yzh,LIGOScientific:2019fpa,LIGOScientific:2018jsj,LIGOScientific:2018mvr} that rely on accurate waveform models for inferring the source parameters.  In this paper, we improved the PN description of spinning compact binaries using information from relativistic scattering and self-force theory, which is an extension of the approach introduced and used in Refs.~\cite{Bini:2019nra,Damour:2019lcq,Bini:2020wpo} for the nonspinning case.
We started by extending the arguments from Ref.~\cite{Damour:2019lcq} to show that the scattering angle for an aligned-spin binary has a simple dependence on the masses. This allowed us to determine the SO and aligned {\sonestwo} couplings through N$^3$LO in a PN expansion using GSF results for the redshift and precession frequency of a small body on an eccentric orbit in a Kerr background.
This result is neatly encapsulated in the gauge-invariant aligned-spin scattering-angle function, given explicitly in Eq.~\eqref{chifin}.  The derivation presented here provides the full details for the recently reported result at SO level in Ref.~\cite{Antonelli:2020aeb}, while extending the analysis to aligned {\sonestwo} couplings. 

Using these new PN results, we calculated the circular-orbit binding energy, the EOB gyro-gravitomagnetic factors, and implemented these results in an EOB Hamiltonian. 
To illustrate the effect of the new N$^3$LO terms, we compared the binding energy with NR simulations (see Figs.~\ref{fig:bindEnSO} and \ref{fig:bindEnSS},) showing an improvement over the N$^2$LO.
These results could be implemented in state-of-the-art \texttt{SEOBNR}~\cite{Bohe:2016gbl,Babak:2016tgq,Cotesta:2018fcv,Ossokine:2020kjp} and \texttt{TEOBResumS}~\cite{Nagar:2018plt,Nagar:2018zoe} waveform models used in LIGO-Virgo searches and inference analyses~\cite{LIGOScientific:2018mvr}.

While it is arguable whether PM results already provide a useful resummation of the PN ones~\cite{Antonelli:2019ytb}, the present work shows that, with the crucial contribution of GSF theory, advances in PM theory already allow one to advance the PN knowledge in the spin sector. We thus beseech further research to explore synergies between GSF, PM, and PN theory, along the lines of Refs.~\cite{Bini:2019nra,Siemonsen:2019dsu,Antonelli:2020aeb,Bini:2020wpo,Bini:2020nsb,Bini:2020uiq} and the present paper. 
One could, for instance, extend the results in this paper to N$^3$LO S$^2$ couplings, i.e.\ at quadratic order in each spin.  This is an important step to complete the aligned-spin 5PN dynamics for BBHs. However, we leave such a calculation for future work, since it would require currently unavailable GSF results.

One can envision further important work at the interface between the PM and GSF approximations. With knowledge of first-order GSF theory, one can in principle determine the full 3PM and 4PM scattering angle in a completely independent way from techniques employed, e.g., in Ref.~\cite{Bern:2019nnu}. To this end, one could calculate the PM expansion of GSF gauge-invariant quantities for bound orbits directly (e.g., expansions in $u_p$ valid at all orders in the eccentricity $e$). This enterprise would have to take great care in the inclusion of tail terms in the dynamics, as well as in the analytical continuation of such results to scattering systems. Should these quantities be calculated, one could exploit the method herein presented to fix the 3PM and 4PM scattering angles without further PN re-expansions.
Even better would be a direct GSF treatment of scattering orbits and the scattering angle. This is likely to first come in the form of numerical calculations at first-order in the mass ratio. It will however be worth exploring whether ``experimental mathematics'' techniques can be used to obtain analytic expressions for the 4PM scattering angle by pushing such numerical calculations to extreme precision (see, e.g., Ref.~\cite{Johnson-McDaniel:2015vva} for an example along these lines in the GSF literature).

Finally, we stress that it is paramount to check our results with more established PN calculations (e.g., with the EFT approach, as was done partially at N$^3$LO in Refs.~\cite{Levi:2020kvb,Levi:2020uwu}), as they have been obtained with a so-far completely unexplored method in the spinning sector that is begging to be further scrutinized.

\section*{Acknowledgments}
We are grateful to Alessandra Buonanno and Maarten van de Meent for helpful discussions.
We also thank Sergei Ossokine and Tim Dietrich for providing NR data for the binding energy and for related useful suggestions.

\appendix

\section{The nonspinning 4PN terms in the bound radial action through sixth order in eccentricity}\label{app:4PN}

Here we present the additional 4PN-order terms in the radial action for bound orbits, computed via (\ref{defIr}) applied to the 4PN EOB Hamiltonian given in \cite{Damour:2015isa}, valid to sixth order in the orbital eccentricity $e$.  Note that the expansion in eccentricity has occurred only in the 4PN terms, at $\mc O(c^{-8})$, where it is sufficient to use the Newtonian relation $e=\sqrt{1+\ve(L/GM\mu)^2}+\mc O(c^{-2})$.  The complete radial action we employ above, through 4PN order for the nonspinning terms and through NNNLO for the spin terms, is obtained by replacing the first two lines of (\ref{Irexp}) with 
\begin{alignat}{3}\label{Ir4PN}
I_r&={-}L+GM\mu\frac{1+2\ve}{c\sqrt{-\ve}}
+\frac{1}{c^2}\frac{(GM\mu)^2}{\pi\Gamma{L_\mr{cov}}}\ms X_2
\\\nnm
&\quad
+\frac{1}{\pi}\sum_{l=2}^4\frac{1}{c^{2l}}\frac{(GM\mu)^{2l}}{(\Gamma L_\mr{cov})^{2l-1}}\frac{\bar{\ms X}_{2l}}{2l-1}+\frac{1}{c^8}\mc O(e^8)+\mc O(\frac{1}{c^{10}}),
\end{alignat}
where
\begin{alignat}{3}
\frac{\bar{\ms X}_4}{3\pi}&=\frac{5}{4}(7-2\nu)+\bigg[\frac{105}{4}+\Big(\frac{41}{128}\pi^2-\frac{557}{24}\Big)\nu\bigg]\ve
\\\nnm
&\quad+\bigg[\frac{1155}{64}+\Big(\frac{65383}{1440}+\frac{33601}{24576}\pi^2-\frac{74}{15}\gamma_\mr E
\\\nnm
&\qquad-\frac{6122}{3}\ln 2+\frac{24057}{20}\ln 3+\frac{74}{15}\ln\frac{cL_\mr{cov}}{GM\mu}\Big)\nu
\\\nnm
&\qquad-\frac{81}{32}\nu^2+\frac{45}{16}\nu^3\bigg]\ve^2+\mc O(\ve^3),
\end{alignat}
\begin{alignat}{3}
\frac{\bar{\ms X}_6}{5\pi}&=\frac{231}{4}+\Big(\frac{123}{128}\pi^2-\frac{125}{2}\Big)\nu+\frac{21}{8}\nu^2
\\\nnm
&\quad+\bigg[\frac{9009}{32}+\Big({-}\frac{64739}{240}+\frac{51439}{4096}\pi^2-\frac{244}{5}\gamma_\mr E
\\\nnm
&\qquad-\frac{60172}{15}\ln 2+\frac{22599}{10}\ln 3+\frac{244}{5}\ln\frac{cL_\mr{cov}}{GM\mu}\Big)\nu
\\\nnm
&\qquad+\Big(\frac{483}{8}-\frac{369}{256}\pi^2\Big)\nu^2+\frac{45}{16}\nu^3\bigg]\ve+\mc O(\ve^2),
\end{alignat}
and
\begin{alignat}{3}
\frac{\bar{\ms X}_8}{7\pi}&=\frac{32175}{64}
+\Big({-}\frac{534089}{720}+\frac{425105}{24576}\pi^2-\frac{170}{3}\gamma_\mr E
\\\nnm
&\qquad-\frac{9982}{5}\ln 2+\frac{21141}{20}\ln 3+\frac{170}{3}\ln\frac{cL_\mr{cov}}{GM\mu}\Big)\nu
\\\nnm
&\qquad+\Big(\frac{4711}{24}-\frac{1025}{256}\pi^2\Big)\nu^2-\frac{15}{8}\nu^3+\mc O(\ve).
\end{alignat}

\section{Kerr-geodesic variables}
\label{Kerrvar}

We provide here the relevant details to compute the change of variables from $(y,\lambda)$ to $(u_p,e)$ needed for comparison with the 1SF calculations of the perturbed redshift and spin precession invariants. Since we are working with perturbed quantities we need only compute this change of variables at the geodesic level. 

The geodesic equations in Kerr spacetime when specialized to the equator $\theta=\tfrac{\pi}{2}$ are
\begin{align}
	\dot{t}&=\frac{1}{\Sigma}\left[E\left(\frac{(r^2+a^2)^2}{\Delta}-a^2\right)+a L\left(1-\frac{r^2+a^2}{\Delta}\right)\right],\\
	\dot{r}&=\frac{1}{\Sigma}\sqrt{\left(E(r^2+a^2)-a L\right)^2-\Delta(r^2+(L-aE)^2)}, \\
	\dot{\phi}&=\frac{1}{\Sigma}\left[(L-aE)+\frac{a}{\Delta}\left(r^2 E-a (L-aE)\right)\right],
\end{align}
where $\dot{{}}\equiv\frac{d}{d\tau}$. The radial motion is commonly parametrized using the Darwin relativistic anomaly $\chi$ as
\begin{equation}
r=\frac{m_2 p}{(1+e\cos \chi)}\,,
\end{equation} 
where $e$  is the eccentricity and  $p$ the (dimensionless) semilatus rectum. This defines the turning points of the orbit to be at $\chi={0,\pi}$. Note that here we use $p$ instead of $u_p\equiv 1/p$ from the text since it makes the equations below simpler.
To determine the constants of motion $E,L$ as functions of $(p,e)$ we set $\dot{r}=0$ at the turning points. While these simultaneous equations can be solved fully, we give their expansion in $a$, which will be sufficient for this work,
\begin{align}
	E&=\sqrt{\frac{(p-2)^2-4e^2}{p(p-3-e^2)}}-\frac{(e^2-1)^2}{p(p-3-e^2)^{3/2}}a+\mathcal{O}(a^2), \\
	L&=\frac{p}{\sqrt{p-3-e^2}}+(3+e^2)\sqrt{\frac{(p-2)^2-4e^2}{p(p-3-e^2)^3}}a+\mathcal{O}(a^2).
\end{align}

Next, we calculate the radial and azimuthal periods $T_{r0}$ and $\Phi_0$ in the Kerr background geometry
\begin{align}
T_{r0}=&\oint dt=\int_{0}^{2\pi}\frac{dt}{d\chi} d\chi\,,\\
\Phi_0=&\oint d\phi=\int_{0}^{2\pi}\frac{d\phi}{d\chi} d\chi\,,
\end{align} 
where 
\begin{align}
	\frac{dt}{d\chi}=\frac{\dot{t}}{\dot{r}}\frac{dr}{d\chi}, \qquad 
	\frac{d\phi}{d\chi}=\frac{\dot{\phi}}{\dot{r}}\frac{dr}{d\chi}.
\end{align} 
Further expanding the integrands in eccentricity, and integrating order by order in $a$ and $e$ gives for the periods a result of the form
\begin{align}
T_{r0}(p,e)=&T_{r0}^{0}(p,e)+\, T_{r0}^{1}(p,e)a+\mathcal{O}(a^2)\,,\\
\Phi_0(p,e)=&\Phi_0^{0}(p,e)+ \, \Phi_0^{1}(p,e)a+\mathcal{O}(a^2)\,,
\end{align}
with
\begin{widetext}
\begin{align}
	T_{r0}^{0}=&\frac{2 \pi p^2}{\sqrt{p-6}}\bigg(1+\frac{3 \left(2 p^3-32 p^2+165 p-266\right)}{4 (p-6)^2 (p-2)}e^2  \nonumber\\
	&+\frac{3  \left(40 p^7-1296 p^6+17556 p^5-128448 p^4+546523 p^3-1350786 p^2+1803396
   p-1016920\right)}{64 (p-6)^4 (p-2)^3}e^4+\mathcal{O}(e^6)\bigg), \\
  	T_{r0}^{1}=&-\frac{6 \pi  \sqrt{p} (p+2)}{(p-6)^{3/2}}\bigg(1+\frac{ \left(2 p^3-32 p^2+139 p+6\right)}{4 (p-6)^2 (p+2)}e^2+\frac{\left(24 p^5-656 p^4+6844 p^3-32576 p^2+60889 p+210\right)}{64 (p-6)^4
   (p+2)}e^4\nonumber \\
   &+\mathcal{O}(e^6) \bigg),
\end{align}
and
\begin{align}
	\Phi_0^{0}=&2 \pi  \sqrt{\frac{p}{p-6}}\bigg(1+\frac{3 }{4 (p-6)^2}e^2+\frac{105 }{64 (p-6)^4}e^4+\mathcal{O}(e^6) \bigg),\\
	\Phi_0^{1}=&-\frac{8 \pi }{(p-6)^{3/2}}\bigg(1+\frac{3 (9 p-34)}{4 (p-6)^2 (p-2)}e^2+\frac{3 \left(739 p^3-6962 p^2+23332 p-28824\right)}{64 (p-6)^4 (p-2)^3}e^4+\mathcal{O}(e^6) \bigg).
\end{align}

With these we can use Eq.~\eqref{ylambda} to obtain $(y,\lambda)$ to the desired $4.5$PN accuracy by expanding about small $u_p=1/p$ as
\begin{align}
y(u_p,e)=&y^{0}(u_p,e)+ a\, y^{a}(u_p,e)+\mathcal{O}(a^2,u_p^{6})\,,\\
\lambda(u_p,e)=&\lambda^{0}(u_p,e)+ a\, \lambda^{a}(u_p,e)+\mathcal{O}(a^2,u_p^{5})\,,
\end{align}
with
	\begin{align}
	y^0(u_p,e)=&\left(1-e^2\right) u_p-2 e^2 \left(-1+e^2\right)
	u_p^2+\left(6 e^2-\frac{23 e^4}{8}\right) u_p^3+\left(24
	e^2-\frac{13 e^4}{4}\right) u_p^4-\frac{1}{4} e^2
	\left(-480+e^2\right) u_p^5\,,\\
	y^a(u_p,e)=&\frac{2}{3} \left(-1-2 e^2+3 e^4\right)
	u_p^{5/2}+\frac{1}{3} e^2 \left(-52+37 e^2\right)
	u_p^{7/2}+\left(-102 e^2+\frac{353 e^4}{12}\right)
	u_p^{9/2}+\left(-704 e^2+\frac{763 e^4}{6}\right)
	u_p^{11/2}\,,\nonumber\\
	\lambda^0(u_p,e)=&1-e^2+\frac{1}{4} \left(-18+25 e^2-7 e^4\right) u_p+\frac{1}{16}
	\left(-36-36 e^2+115 e^4\right) u_p^2+\frac{3}{64}
	\left(-144-220 e^2+421 e^4\right) u_p^3\nonumber\\
	&+\frac{1}{16}
	\left(-405-807 e^2+1007 e^4\right) u_p^4+\frac{3}{256}
	\left(-9072-24772 e^2+22501 e^4\right) u_p^5\,,\\
	\lambda^a(u_p,e)=&\frac{4}{3} \left(1-e^2\right)
	\sqrt{u_p}-\frac{2}{3} \left(1-e^2\right)
	u_p^{3/2}+\left(6-\frac{25 e^2}{6}-\frac{13 e^4}{4}\right)
	u_p^{5/2}+\left(\frac{39}{2}+32 e^2-\frac{202 e^4}{3}\right)
	u_p^{7/2}\nonumber\\
	&+\left(\frac{423}{4}+\frac{1761 e^2}{8}-\frac{26243
		e^4}{96}\right) u_p^{9/2}
	\,.
	\end{align}
\end{widetext}

\bibliography{paper.bbl}

\begin{thebibliography}{140}%
\makeatletter
\providecommand \@ifxundefined [1]{%
 \@ifx{#1\undefined}
}%
\providecommand \@ifnum [1]{%
 \ifnum #1\expandafter \@firstoftwo
 \else \expandafter \@secondoftwo
 \fi
}%
\providecommand \@ifx [1]{%
 \ifx #1\expandafter \@firstoftwo
 \else \expandafter \@secondoftwo
 \fi
}%
\providecommand \natexlab [1]{#1}%
\providecommand \enquote  [1]{``#1''}%
\providecommand \bibnamefont  [1]{#1}%
\providecommand \bibfnamefont [1]{#1}%
\providecommand \citenamefont [1]{#1}%
\providecommand \href@noop [0]{\@secondoftwo}%
\providecommand \href [0]{\begingroup \@sanitize@url \@href}%
\providecommand \@href[1]{\@@startlink{#1}\@@href}%
\providecommand \@@href[1]{\endgroup#1\@@endlink}%
\providecommand \@sanitize@url [0]{\catcode `\\12\catcode `\$12\catcode
  `\&12\catcode `\#12\catcode `\^12\catcode `\_12\catcode `\%12\relax}%
\providecommand \@@startlink[1]{}%
\providecommand \@@endlink[0]{}%
\providecommand \url  [0]{\begingroup\@sanitize@url \@url }%
\providecommand \@url [1]{\endgroup\@href {#1}{\urlprefix }}%
\providecommand \urlprefix  [0]{URL }%
\providecommand \Eprint [0]{\href }%
\providecommand \doibase [0]{http://dx.doi.org/}%
\providecommand \selectlanguage [0]{\@gobble}%
\providecommand \bibinfo  [0]{\@secondoftwo}%
\providecommand \bibfield  [0]{\@secondoftwo}%
\providecommand \translation [1]{[#1]}%
\providecommand \BibitemOpen [0]{}%
\providecommand \bibitemStop [0]{}%
\providecommand \bibitemNoStop [0]{.\EOS\space}%
\providecommand \EOS [0]{\spacefactor3000\relax}%
\providecommand \BibitemShut  [1]{\csname bibitem#1\endcsname}%
\let\auto@bib@innerbib\@empty
\bibitem [{\citenamefont {Abbott}\ \emph
  {et~al.}(2019{\natexlab{a}})\citenamefont {Abbott} \emph
  {et~al.}}]{Abbott:2019yzh}%
  \BibitemOpen
  \bibfield  {author} {\bibinfo {author} {\bibfnamefont {B.~P.}\ \bibnamefont
  {Abbott}} \emph {et~al.} (\bibinfo {collaboration} {LIGO Scientific,
  Virgo}),\ }\href@noop {} {\  (\bibinfo {year} {2019}{\natexlab{a}})},\
  \Eprint {http://arxiv.org/abs/1908.06060} {arXiv:1908.06060 [astro-ph.CO]}
  \BibitemShut {NoStop}%
\bibitem [{\citenamefont {Abbott}\ \emph
  {et~al.}(2019{\natexlab{b}})\citenamefont {Abbott} \emph
  {et~al.}}]{LIGOScientific:2019fpa}%
  \BibitemOpen
  \bibfield  {author} {\bibinfo {author} {\bibfnamefont {B.~P.}\ \bibnamefont
  {Abbott}} \emph {et~al.} (\bibinfo {collaboration} {LIGO Scientific,
  Virgo}),\ }\href {\doibase 10.1103/PhysRevD.100.104036} {\bibfield  {journal}
  {\bibinfo  {journal} {Phys. Rev.}\ }\textbf {\bibinfo {volume} {D100}},\
  \bibinfo {pages} {104036} (\bibinfo {year} {2019}{\natexlab{b}})},\ \Eprint
  {http://arxiv.org/abs/1903.04467} {arXiv:1903.04467 [gr-qc]} \BibitemShut
  {NoStop}%
\bibitem [{\citenamefont {Abbott}\ \emph
  {et~al.}(2019{\natexlab{c}})\citenamefont {Abbott} \emph
  {et~al.}}]{LIGOScientific:2018jsj}%
  \BibitemOpen
  \bibfield  {author} {\bibinfo {author} {\bibfnamefont {B.~P.}\ \bibnamefont
  {Abbott}} \emph {et~al.} (\bibinfo {collaboration} {LIGO Scientific,
  Virgo}),\ }\href {\doibase 10.3847/2041-8213/ab3800} {\bibfield  {journal}
  {\bibinfo  {journal} {Astrophys. J.}\ }\textbf {\bibinfo {volume} {882}},\
  \bibinfo {pages} {L24} (\bibinfo {year} {2019}{\natexlab{c}})},\ \Eprint
  {http://arxiv.org/abs/1811.12940} {arXiv:1811.12940 [astro-ph.HE]}
  \BibitemShut {NoStop}%
\bibitem [{\citenamefont {Abbott}\ \emph
  {et~al.}(2019{\natexlab{d}})\citenamefont {Abbott} \emph
  {et~al.}}]{LIGOScientific:2018mvr}%
  \BibitemOpen
  \bibfield  {author} {\bibinfo {author} {\bibfnamefont {B.~P.}\ \bibnamefont
  {Abbott}} \emph {et~al.} (\bibinfo {collaboration} {LIGO Scientific,
  Virgo}),\ }\href {\doibase 10.1103/PhysRevX.9.031040} {\bibfield  {journal}
  {\bibinfo  {journal} {Phys. Rev.}\ }\textbf {\bibinfo {volume} {X9}},\
  \bibinfo {pages} {031040} (\bibinfo {year} {2019}{\natexlab{d}})},\ \Eprint
  {http://arxiv.org/abs/1811.12907} {arXiv:1811.12907 [astro-ph.HE]}
  \BibitemShut {NoStop}%
\bibitem [{\citenamefont {Blanchet}(2014)}]{Blanchet:2013haa}%
  \BibitemOpen
  \bibfield  {author} {\bibinfo {author} {\bibfnamefont {L.}~\bibnamefont
  {Blanchet}},\ }\href {\doibase 10.12942/lrr-2014-2} {\bibfield  {journal}
  {\bibinfo  {journal} {Living Rev. Rel.}\ }\textbf {\bibinfo {volume} {17}},\
  \bibinfo {pages} {2} (\bibinfo {year} {2014})},\ \Eprint
  {http://arxiv.org/abs/1310.1528} {arXiv:1310.1528 [gr-qc]} \BibitemShut
  {NoStop}%
\bibitem [{\citenamefont {Schäfer}\ and\ \citenamefont
  {Jaranowski}(2018)}]{Schafer:2018kuf}%
  \BibitemOpen
  \bibfield  {author} {\bibinfo {author} {\bibfnamefont {G.}~\bibnamefont
  {Schäfer}}\ and\ \bibinfo {author} {\bibfnamefont {P.}~\bibnamefont
  {Jaranowski}},\ }\href {\doibase 10.1007/s41114-018-0016-5} {\bibfield
  {journal} {\bibinfo  {journal} {Living Rev. Rel.}\ }\textbf {\bibinfo
  {volume} {21}},\ \bibinfo {pages} {7} (\bibinfo {year} {2018})},\ \Eprint
  {http://arxiv.org/abs/1805.07240} {arXiv:1805.07240 [gr-qc]} \BibitemShut
  {NoStop}%
\bibitem [{\citenamefont {Rothstein}(2014)}]{Rothstein:2014sra}%
  \BibitemOpen
  \bibfield  {author} {\bibinfo {author} {\bibfnamefont {I.~Z.}\ \bibnamefont
  {Rothstein}},\ }\href {\doibase 10.1007/s10714-014-1726-y} {\bibfield
  {journal} {\bibinfo  {journal} {Gen. Rel. Grav.}\ }\textbf {\bibinfo {volume}
  {46}},\ \bibinfo {pages} {1726} (\bibinfo {year} {2014})}\BibitemShut
  {NoStop}%
\bibitem [{\citenamefont {Goldberger}(2007)}]{Goldberger:2007hy}%
  \BibitemOpen
  \bibfield  {author} {\bibinfo {author} {\bibfnamefont {W.~D.}\ \bibnamefont
  {Goldberger}},\ }in\ \href
  {http://www.sciencedirect.com/science/bookseries/09248099/86} {\emph
  {\bibinfo {booktitle} {{Les Houches Summer School - Session 86: Particle
  Physics and Cosmology: The Fabric of Spacetime}}}}\ (\bibinfo {year} {2007})\
  pp.\ \bibinfo {pages} {351--353, 355--396},\ \Eprint
  {http://arxiv.org/abs/hep-ph/0701129} {arXiv:hep-ph/0701129 [hep-ph]}
  \BibitemShut {NoStop}%
\bibitem [{\citenamefont {Futamase}\ and\ \citenamefont
  {Itoh}(2007)}]{Futamase:2007zz}%
  \BibitemOpen
  \bibfield  {author} {\bibinfo {author} {\bibfnamefont {T.}~\bibnamefont
  {Futamase}}\ and\ \bibinfo {author} {\bibfnamefont {Y.}~\bibnamefont
  {Itoh}},\ }\href {\doibase 10.12942/lrr-2007-2} {\bibfield  {journal}
  {\bibinfo  {journal} {Living Rev. Rel.}\ }\textbf {\bibinfo {volume} {10}},\
  \bibinfo {pages} {2} (\bibinfo {year} {2007})}\BibitemShut {NoStop}%
\bibitem [{\citenamefont {Pati}\ and\ \citenamefont
  {Will}(2000)}]{Pati:2000vt}%
  \BibitemOpen
  \bibfield  {author} {\bibinfo {author} {\bibfnamefont {M.~E.}\ \bibnamefont
  {Pati}}\ and\ \bibinfo {author} {\bibfnamefont {C.~M.}\ \bibnamefont
  {Will}},\ }\href {\doibase 10.1103/PhysRevD.62.124015} {\bibfield  {journal}
  {\bibinfo  {journal} {Phys. Rev. D}\ }\textbf {\bibinfo {volume} {62}},\
  \bibinfo {pages} {124015} (\bibinfo {year} {2000})},\ \Eprint
  {http://arxiv.org/abs/gr-qc/0007087} {arXiv:gr-qc/0007087} \BibitemShut
  {NoStop}%
\bibitem [{\citenamefont {Porto}(2016)}]{Porto:2016pyg}%
  \BibitemOpen
  \bibfield  {author} {\bibinfo {author} {\bibfnamefont {R.~A.}\ \bibnamefont
  {Porto}},\ }\href {\doibase 10.1016/j.physrep.2016.04.003} {\bibfield
  {journal} {\bibinfo  {journal} {Phys. Rept.}\ }\textbf {\bibinfo {volume}
  {633}},\ \bibinfo {pages} {1} (\bibinfo {year} {2016})},\ \Eprint
  {http://arxiv.org/abs/1601.04914} {arXiv:1601.04914 [hep-th]} \BibitemShut
  {NoStop}%
\bibitem [{\citenamefont {Levi}(2020)}]{Levi:2018nxp}%
  \BibitemOpen
  \bibfield  {author} {\bibinfo {author} {\bibfnamefont {M.}~\bibnamefont
  {Levi}},\ }\href {\doibase 10.1088/1361-6633/ab12bc} {\bibfield  {journal}
  {\bibinfo  {journal} {Rept. Prog. Phys.}\ }\textbf {\bibinfo {volume} {83}},\
  \bibinfo {pages} {075901} (\bibinfo {year} {2020})},\ \Eprint
  {http://arxiv.org/abs/1807.01699} {arXiv:1807.01699 [hep-th]} \BibitemShut
  {NoStop}%
\bibitem [{\citenamefont {Abbott}\ \emph {et~al.}(2016)\citenamefont {Abbott}
  \emph {et~al.}}]{Abbott:2016blz}%
  \BibitemOpen
  \bibfield  {author} {\bibinfo {author} {\bibfnamefont {B.~P.}\ \bibnamefont
  {Abbott}} \emph {et~al.} (\bibinfo {collaboration} {LIGO Scientific,
  Virgo}),\ }\href {\doibase 10.1103/PhysRevLett.116.061102} {\bibfield
  {journal} {\bibinfo  {journal} {Phys. Rev. Lett.}\ }\textbf {\bibinfo
  {volume} {116}},\ \bibinfo {pages} {061102} (\bibinfo {year} {2016})},\
  \Eprint {http://arxiv.org/abs/1602.03837} {arXiv:1602.03837 [gr-qc]}
  \BibitemShut {NoStop}%
\bibitem [{\citenamefont {Damour}\ \emph {et~al.}(2014)\citenamefont {Damour},
  \citenamefont {Jaranowski},\ and\ \citenamefont {Schäfer}}]{Damour:2014jta}%
  \BibitemOpen
  \bibfield  {author} {\bibinfo {author} {\bibfnamefont {T.}~\bibnamefont
  {Damour}}, \bibinfo {author} {\bibfnamefont {P.}~\bibnamefont {Jaranowski}},
  \ and\ \bibinfo {author} {\bibfnamefont {G.}~\bibnamefont {Schäfer}},\
  }\href {\doibase 10.1103/PhysRevD.89.064058} {\bibfield  {journal} {\bibinfo
  {journal} {Phys. Rev.}\ }\textbf {\bibinfo {volume} {D89}},\ \bibinfo {pages}
  {064058} (\bibinfo {year} {2014})},\ \Eprint {http://arxiv.org/abs/1401.4548}
  {arXiv:1401.4548 [gr-qc]} \BibitemShut {NoStop}%
\bibitem [{\citenamefont {Damour}\ \emph {et~al.}(2015)\citenamefont {Damour},
  \citenamefont {Jaranowski},\ and\ \citenamefont {Schäfer}}]{Damour:2015isa}%
  \BibitemOpen
  \bibfield  {author} {\bibinfo {author} {\bibfnamefont {T.}~\bibnamefont
  {Damour}}, \bibinfo {author} {\bibfnamefont {P.}~\bibnamefont {Jaranowski}},
  \ and\ \bibinfo {author} {\bibfnamefont {G.}~\bibnamefont {Schäfer}},\
  }\href {\doibase 10.1103/PhysRevD.91.084024} {\bibfield  {journal} {\bibinfo
  {journal} {Phys. Rev.}\ }\textbf {\bibinfo {volume} {D91}},\ \bibinfo {pages}
  {084024} (\bibinfo {year} {2015})},\ \Eprint
  {http://arxiv.org/abs/1502.07245} {arXiv:1502.07245 [gr-qc]} \BibitemShut
  {NoStop}%
\bibitem [{\citenamefont {Bernard}\ \emph {et~al.}(2017)\citenamefont
  {Bernard}, \citenamefont {Blanchet}, \citenamefont {Bohé}, \citenamefont
  {Faye},\ and\ \citenamefont {Marsat}}]{Bernard:2016wrg}%
  \BibitemOpen
  \bibfield  {author} {\bibinfo {author} {\bibfnamefont {L.}~\bibnamefont
  {Bernard}}, \bibinfo {author} {\bibfnamefont {L.}~\bibnamefont {Blanchet}},
  \bibinfo {author} {\bibfnamefont {A.}~\bibnamefont {Bohé}}, \bibinfo
  {author} {\bibfnamefont {G.}~\bibnamefont {Faye}}, \ and\ \bibinfo {author}
  {\bibfnamefont {S.}~\bibnamefont {Marsat}},\ }\href {\doibase
  10.1103/PhysRevD.95.044026} {\bibfield  {journal} {\bibinfo  {journal} {Phys.
  Rev.}\ }\textbf {\bibinfo {volume} {D95}},\ \bibinfo {pages} {044026}
  (\bibinfo {year} {2017})},\ \Eprint {http://arxiv.org/abs/1610.07934}
  {arXiv:1610.07934 [gr-qc]} \BibitemShut {NoStop}%
\bibitem [{\citenamefont {Bernard}\ \emph {et~al.}(2018)\citenamefont
  {Bernard}, \citenamefont {Blanchet}, \citenamefont {Faye},\ and\
  \citenamefont {Marchand}}]{Bernard:2017ktp}%
  \BibitemOpen
  \bibfield  {author} {\bibinfo {author} {\bibfnamefont {L.}~\bibnamefont
  {Bernard}}, \bibinfo {author} {\bibfnamefont {L.}~\bibnamefont {Blanchet}},
  \bibinfo {author} {\bibfnamefont {G.}~\bibnamefont {Faye}}, \ and\ \bibinfo
  {author} {\bibfnamefont {T.}~\bibnamefont {Marchand}},\ }\href {\doibase
  10.1103/PhysRevD.97.044037} {\bibfield  {journal} {\bibinfo  {journal}
  {Phys.\ Rev.\ D}\ }\textbf {\bibinfo {volume} {97}},\ \bibinfo {pages}
  {044037} (\bibinfo {year} {2018})},\ \Eprint
  {http://arxiv.org/abs/1711.00283} {arXiv:1711.00283 [gr-qc]} \BibitemShut
  {NoStop}%
\bibitem [{\citenamefont {Foffa}\ \emph
  {et~al.}(2019{\natexlab{a}})\citenamefont {Foffa}, \citenamefont {Mastrolia},
  \citenamefont {Sturani}, \citenamefont {Sturm},\ and\ \citenamefont
  {Torres~Bobadilla}}]{Foffa:2019hrb}%
  \BibitemOpen
  \bibfield  {author} {\bibinfo {author} {\bibfnamefont {S.}~\bibnamefont
  {Foffa}}, \bibinfo {author} {\bibfnamefont {P.}~\bibnamefont {Mastrolia}},
  \bibinfo {author} {\bibfnamefont {R.}~\bibnamefont {Sturani}}, \bibinfo
  {author} {\bibfnamefont {C.}~\bibnamefont {Sturm}}, \ and\ \bibinfo {author}
  {\bibfnamefont {W.~J.}\ \bibnamefont {Torres~Bobadilla}},\ }\href {\doibase
  10.1103/PhysRevLett.122.241605} {\bibfield  {journal} {\bibinfo  {journal}
  {Phys. Rev. Lett.}\ }\textbf {\bibinfo {volume} {122}},\ \bibinfo {pages}
  {241605} (\bibinfo {year} {2019}{\natexlab{a}})},\ \Eprint
  {http://arxiv.org/abs/1902.10571} {arXiv:1902.10571 [gr-qc]} \BibitemShut
  {NoStop}%
\bibitem [{\citenamefont {Foffa}\ and\ \citenamefont
  {Sturani}(2019)}]{Foffa:2019rdf}%
  \BibitemOpen
  \bibfield  {author} {\bibinfo {author} {\bibfnamefont {S.}~\bibnamefont
  {Foffa}}\ and\ \bibinfo {author} {\bibfnamefont {R.}~\bibnamefont
  {Sturani}},\ }\href {\doibase 10.1103/PhysRevD.100.024047} {\bibfield
  {journal} {\bibinfo  {journal} {Phys. Rev.}\ }\textbf {\bibinfo {volume}
  {D100}},\ \bibinfo {pages} {024047} (\bibinfo {year} {2019})},\ \Eprint
  {http://arxiv.org/abs/1903.05113} {arXiv:1903.05113 [gr-qc]} \BibitemShut
  {NoStop}%
\bibitem [{\citenamefont {Foffa}\ \emph
  {et~al.}(2019{\natexlab{b}})\citenamefont {Foffa}, \citenamefont {Porto},
  \citenamefont {Rothstein},\ and\ \citenamefont {Sturani}}]{Foffa:2019yfl}%
  \BibitemOpen
  \bibfield  {author} {\bibinfo {author} {\bibfnamefont {S.}~\bibnamefont
  {Foffa}}, \bibinfo {author} {\bibfnamefont {R.~A.}\ \bibnamefont {Porto}},
  \bibinfo {author} {\bibfnamefont {I.}~\bibnamefont {Rothstein}}, \ and\
  \bibinfo {author} {\bibfnamefont {R.}~\bibnamefont {Sturani}},\ }\href
  {\doibase 10.1103/PhysRevD.100.024048} {\bibfield  {journal} {\bibinfo
  {journal} {Phys.\ Rev.\ D}\ }\textbf {\bibinfo {volume} {100}},\ \bibinfo
  {pages} {024048} (\bibinfo {year} {2019}{\natexlab{b}})},\ \Eprint
  {http://arxiv.org/abs/1903.05118} {arXiv:1903.05118 [gr-qc]} \BibitemShut
  {NoStop}%
\bibitem [{\citenamefont {Blümlein}\ \emph
  {et~al.}(2020{\natexlab{a}})\citenamefont {Blümlein}, \citenamefont
  {Maier},\ and\ \citenamefont {Marquard}}]{Blumlein:2019zku}%
  \BibitemOpen
  \bibfield  {author} {\bibinfo {author} {\bibfnamefont {J.}~\bibnamefont
  {Blümlein}}, \bibinfo {author} {\bibfnamefont {A.}~\bibnamefont {Maier}}, \
  and\ \bibinfo {author} {\bibfnamefont {P.}~\bibnamefont {Marquard}},\ }\href
  {\doibase 10.1016/j.physletb.2019.135100} {\bibfield  {journal} {\bibinfo
  {journal} {Phys. Lett.}\ }\textbf {\bibinfo {volume} {B800}},\ \bibinfo
  {pages} {135100} (\bibinfo {year} {2020}{\natexlab{a}})},\ \Eprint
  {http://arxiv.org/abs/1902.11180} {arXiv:1902.11180 [gr-qc]} \BibitemShut
  {NoStop}%
\bibitem [{\citenamefont {Blümlein}\ \emph
  {et~al.}(2020{\natexlab{b}})\citenamefont {Blümlein}, \citenamefont {Maier},
  \citenamefont {Marquard},\ and\ \citenamefont {Schäfer}}]{Blumlein:2020pog}%
  \BibitemOpen
  \bibfield  {author} {\bibinfo {author} {\bibfnamefont {J.}~\bibnamefont
  {Blümlein}}, \bibinfo {author} {\bibfnamefont {A.}~\bibnamefont {Maier}},
  \bibinfo {author} {\bibfnamefont {P.}~\bibnamefont {Marquard}}, \ and\
  \bibinfo {author} {\bibfnamefont {G.}~\bibnamefont {Schäfer}},\ }\href
  {\doibase 10.1016/j.nuclphysb.2020.115041} {\bibfield  {journal} {\bibinfo
  {journal} {Nucl. Phys. B}\ }\textbf {\bibinfo {volume} {955}},\ \bibinfo
  {pages} {115041} (\bibinfo {year} {2020}{\natexlab{b}})},\ \Eprint
  {http://arxiv.org/abs/2003.01692} {arXiv:2003.01692 [gr-qc]} \BibitemShut
  {NoStop}%
\bibitem [{\citenamefont {Bini}\ \emph {et~al.}(2019)\citenamefont {Bini},
  \citenamefont {Damour},\ and\ \citenamefont {Geralico}}]{Bini:2019nra}%
  \BibitemOpen
  \bibfield  {author} {\bibinfo {author} {\bibfnamefont {D.}~\bibnamefont
  {Bini}}, \bibinfo {author} {\bibfnamefont {T.}~\bibnamefont {Damour}}, \ and\
  \bibinfo {author} {\bibfnamefont {A.}~\bibnamefont {Geralico}},\ }\href
  {\doibase 10.1103/PhysRevLett.123.231104} {\bibfield  {journal} {\bibinfo
  {journal} {Phys. Rev. Lett.}\ }\textbf {\bibinfo {volume} {123}},\ \bibinfo
  {pages} {231104} (\bibinfo {year} {2019})},\ \Eprint
  {http://arxiv.org/abs/1909.02375} {arXiv:1909.02375 [gr-qc]} \BibitemShut
  {NoStop}%
\bibitem [{\citenamefont {Bini}\ \emph
  {et~al.}(2020{\natexlab{a}})\citenamefont {Bini}, \citenamefont {Damour},\
  and\ \citenamefont {Geralico}}]{Bini:2020wpo}%
  \BibitemOpen
  \bibfield  {author} {\bibinfo {author} {\bibfnamefont {D.}~\bibnamefont
  {Bini}}, \bibinfo {author} {\bibfnamefont {T.}~\bibnamefont {Damour}}, \ and\
  \bibinfo {author} {\bibfnamefont {A.}~\bibnamefont {Geralico}},\ }\href
  {\doibase 10.1103/PhysRevD.102.024062} {\bibfield  {journal} {\bibinfo
  {journal} {Phys. Rev. D}\ }\textbf {\bibinfo {volume} {102}},\ \bibinfo
  {pages} {024062} (\bibinfo {year} {2020}{\natexlab{a}})},\ \Eprint
  {http://arxiv.org/abs/2003.11891} {arXiv:2003.11891 [gr-qc]} \BibitemShut
  {NoStop}%
\bibitem [{\citenamefont {Bini}\ \emph
  {et~al.}(2020{\natexlab{b}})\citenamefont {Bini}, \citenamefont {Damour},\
  and\ \citenamefont {Geralico}}]{Bini:2020hmy}%
  \BibitemOpen
  \bibfield  {author} {\bibinfo {author} {\bibfnamefont {D.}~\bibnamefont
  {Bini}}, \bibinfo {author} {\bibfnamefont {T.}~\bibnamefont {Damour}}, \ and\
  \bibinfo {author} {\bibfnamefont {A.}~\bibnamefont {Geralico}},\ }\href@noop
  {} {\  (\bibinfo {year} {2020}{\natexlab{b}})},\ \Eprint
  {http://arxiv.org/abs/2007.11239} {arXiv:2007.11239 [gr-qc]} \BibitemShut
  {NoStop}%
\bibitem [{\citenamefont {Bini}\ \emph
  {et~al.}(2020{\natexlab{c}})\citenamefont {Bini}, \citenamefont {Damour},\
  and\ \citenamefont {Geralico}}]{Bini:2020nsb}%
  \BibitemOpen
  \bibfield  {author} {\bibinfo {author} {\bibfnamefont {D.}~\bibnamefont
  {Bini}}, \bibinfo {author} {\bibfnamefont {T.}~\bibnamefont {Damour}}, \ and\
  \bibinfo {author} {\bibfnamefont {A.}~\bibnamefont {Geralico}},\ }\href
  {\doibase 10.1103/PhysRevD.102.024061} {\bibfield  {journal} {\bibinfo
  {journal} {Phys. Rev. D}\ }\textbf {\bibinfo {volume} {102}},\ \bibinfo
  {pages} {024061} (\bibinfo {year} {2020}{\natexlab{c}})},\ \Eprint
  {http://arxiv.org/abs/2004.05407} {arXiv:2004.05407 [gr-qc]} \BibitemShut
  {NoStop}%
\bibitem [{\citenamefont {Bini}\ \emph
  {et~al.}(2020{\natexlab{d}})\citenamefont {Bini}, \citenamefont {Damour},
  \citenamefont {Geralico}, \citenamefont {Laporta},\ and\ \citenamefont
  {Mastrolia}}]{Bini:2020uiq}%
  \BibitemOpen
  \bibfield  {author} {\bibinfo {author} {\bibfnamefont {D.}~\bibnamefont
  {Bini}}, \bibinfo {author} {\bibfnamefont {T.}~\bibnamefont {Damour}},
  \bibinfo {author} {\bibfnamefont {A.}~\bibnamefont {Geralico}}, \bibinfo
  {author} {\bibfnamefont {S.}~\bibnamefont {Laporta}}, \ and\ \bibinfo
  {author} {\bibfnamefont {P.}~\bibnamefont {Mastrolia}},\ }\href@noop {} {\
  (\bibinfo {year} {2020}{\natexlab{d}})},\ \Eprint
  {http://arxiv.org/abs/2008.09389} {arXiv:2008.09389 [gr-qc]} \BibitemShut
  {NoStop}%
\bibitem [{\citenamefont {Hartung}\ and\ \citenamefont
  {Steinhoff}(2011{\natexlab{a}})}]{Hartung:2011te}%
  \BibitemOpen
  \bibfield  {author} {\bibinfo {author} {\bibfnamefont {J.}~\bibnamefont
  {Hartung}}\ and\ \bibinfo {author} {\bibfnamefont {J.}~\bibnamefont
  {Steinhoff}},\ }\href {\doibase 10.1002/andp.201100094} {\bibfield  {journal}
  {\bibinfo  {journal} {Ann. Phys. (Berlin)}\ }\textbf {\bibinfo {volume}
  {523}},\ \bibinfo {pages} {783} (\bibinfo {year} {2011}{\natexlab{a}})},\
  \Eprint {http://arxiv.org/abs/1104.3079} {arXiv:1104.3079 [gr-qc]}
  \BibitemShut {NoStop}%
\bibitem [{\citenamefont {Hartung}\ \emph {et~al.}(2013)\citenamefont
  {Hartung}, \citenamefont {Steinhoff},\ and\ \citenamefont
  {Schafer}}]{Hartung:2013dza}%
  \BibitemOpen
  \bibfield  {author} {\bibinfo {author} {\bibfnamefont {J.}~\bibnamefont
  {Hartung}}, \bibinfo {author} {\bibfnamefont {J.}~\bibnamefont {Steinhoff}},
  \ and\ \bibinfo {author} {\bibfnamefont {G.}~\bibnamefont {Schafer}},\ }\href
  {\doibase 10.1002/andp.201200271} {\bibfield  {journal} {\bibinfo  {journal}
  {Annalen Phys.}\ }\textbf {\bibinfo {volume} {525}},\ \bibinfo {pages} {359}
  (\bibinfo {year} {2013})},\ \Eprint {http://arxiv.org/abs/1302.6723}
  {arXiv:1302.6723 [gr-qc]} \BibitemShut {NoStop}%
\bibitem [{\citenamefont {Marsat}\ \emph {et~al.}(2013)\citenamefont {Marsat},
  \citenamefont {Bohe}, \citenamefont {Faye},\ and\ \citenamefont
  {Blanchet}}]{Marsat:2012fn}%
  \BibitemOpen
  \bibfield  {author} {\bibinfo {author} {\bibfnamefont {S.}~\bibnamefont
  {Marsat}}, \bibinfo {author} {\bibfnamefont {A.}~\bibnamefont {Bohe}},
  \bibinfo {author} {\bibfnamefont {G.}~\bibnamefont {Faye}}, \ and\ \bibinfo
  {author} {\bibfnamefont {L.}~\bibnamefont {Blanchet}},\ }\href {\doibase
  10.1088/0264-9381/30/5/055007} {\bibfield  {journal} {\bibinfo  {journal}
  {Class.\ Quant.\ Grav.}\ }\textbf {\bibinfo {volume} {30}},\ \bibinfo {pages}
  {055007} (\bibinfo {year} {2013})},\ \Eprint {http://arxiv.org/abs/1210.4143}
  {arXiv:1210.4143 [gr-qc]} \BibitemShut {NoStop}%
\bibitem [{\citenamefont {Bohe}\ \emph {et~al.}(2013)\citenamefont {Bohe},
  \citenamefont {Marsat}, \citenamefont {Faye},\ and\ \citenamefont
  {Blanchet}}]{Bohe:2012mr}%
  \BibitemOpen
  \bibfield  {author} {\bibinfo {author} {\bibfnamefont {A.}~\bibnamefont
  {Bohe}}, \bibinfo {author} {\bibfnamefont {S.}~\bibnamefont {Marsat}},
  \bibinfo {author} {\bibfnamefont {G.}~\bibnamefont {Faye}}, \ and\ \bibinfo
  {author} {\bibfnamefont {L.}~\bibnamefont {Blanchet}},\ }\href {\doibase
  10.1088/0264-9381/30/7/075017} {\bibfield  {journal} {\bibinfo  {journal}
  {Class. Quant. Grav.}\ }\textbf {\bibinfo {volume} {30}},\ \bibinfo {pages}
  {075017} (\bibinfo {year} {2013})},\ \Eprint {http://arxiv.org/abs/1212.5520}
  {arXiv:1212.5520 [gr-qc]} \BibitemShut {NoStop}%
\bibitem [{\citenamefont {Levi}\ and\ \citenamefont
  {Steinhoff}(2016{\natexlab{a}})}]{Levi:2015uxa}%
  \BibitemOpen
  \bibfield  {author} {\bibinfo {author} {\bibfnamefont {M.}~\bibnamefont
  {Levi}}\ and\ \bibinfo {author} {\bibfnamefont {J.}~\bibnamefont
  {Steinhoff}},\ }\href {\doibase 10.1088/1475-7516/2016/01/011} {\bibfield
  {journal} {\bibinfo  {journal} {JCAP}\ }\textbf {\bibinfo {volume} {1601}},\
  \bibinfo {pages} {011} (\bibinfo {year} {2016}{\natexlab{a}})},\ \Eprint
  {http://arxiv.org/abs/1506.05056} {arXiv:1506.05056 [gr-qc]} \BibitemShut
  {NoStop}%
\bibitem [{\citenamefont {Levi}\ \emph
  {et~al.}(2020{\natexlab{a}})\citenamefont {Levi}, \citenamefont {McLeod},\
  and\ \citenamefont {von Hippel}}]{Levi:2020kvb}%
  \BibitemOpen
  \bibfield  {author} {\bibinfo {author} {\bibfnamefont {M.}~\bibnamefont
  {Levi}}, \bibinfo {author} {\bibfnamefont {A.~J.}\ \bibnamefont {McLeod}}, \
  and\ \bibinfo {author} {\bibfnamefont {M.}~\bibnamefont {von Hippel}},\
  }\href@noop {} {\  (\bibinfo {year} {2020}{\natexlab{a}})},\ \Eprint
  {http://arxiv.org/abs/2003.02827} {arXiv:2003.02827 [hep-th]} \BibitemShut
  {NoStop}%
\bibitem [{\citenamefont {Hartung}\ and\ \citenamefont
  {Steinhoff}(2011{\natexlab{b}})}]{Hartung:2011ea}%
  \BibitemOpen
  \bibfield  {author} {\bibinfo {author} {\bibfnamefont {J.}~\bibnamefont
  {Hartung}}\ and\ \bibinfo {author} {\bibfnamefont {J.}~\bibnamefont
  {Steinhoff}},\ }\href {\doibase 10.1002/andp.201100163} {\bibfield  {journal}
  {\bibinfo  {journal} {Annalen Phys.}\ }\textbf {\bibinfo {volume} {523}},\
  \bibinfo {pages} {919} (\bibinfo {year} {2011}{\natexlab{b}})},\ \Eprint
  {http://arxiv.org/abs/1107.4294} {arXiv:1107.4294 [gr-qc]} \BibitemShut
  {NoStop}%
\bibitem [{\citenamefont {Levi}(2012)}]{Levi:2011eq}%
  \BibitemOpen
  \bibfield  {author} {\bibinfo {author} {\bibfnamefont {M.}~\bibnamefont
  {Levi}},\ }\href {\doibase 10.1103/PhysRevD.85.064043} {\bibfield  {journal}
  {\bibinfo  {journal} {Phys. Rev.}\ }\textbf {\bibinfo {volume} {D85}},\
  \bibinfo {pages} {064043} (\bibinfo {year} {2012})},\ \Eprint
  {http://arxiv.org/abs/1107.4322} {arXiv:1107.4322 [gr-qc]} \BibitemShut
  {NoStop}%
\bibitem [{\citenamefont {Levi}\ and\ \citenamefont
  {Steinhoff}(2014)}]{Levi:2014sba}%
  \BibitemOpen
  \bibfield  {author} {\bibinfo {author} {\bibfnamefont {M.}~\bibnamefont
  {Levi}}\ and\ \bibinfo {author} {\bibfnamefont {J.}~\bibnamefont
  {Steinhoff}},\ }\href {\doibase 10.1088/1475-7516/2014/12/003} {\bibfield
  {journal} {\bibinfo  {journal} {JCAP}\ }\textbf {\bibinfo {volume} {12}},\
  \bibinfo {pages} {003} (\bibinfo {year} {2014})},\ \Eprint
  {http://arxiv.org/abs/1408.5762} {arXiv:1408.5762 [gr-qc]} \BibitemShut
  {NoStop}%
\bibitem [{\citenamefont {Levi}\ \emph
  {et~al.}(2020{\natexlab{b}})\citenamefont {Levi}, \citenamefont {Mcleod},\
  and\ \citenamefont {Von~Hippel}}]{Levi:2020uwu}%
  \BibitemOpen
  \bibfield  {author} {\bibinfo {author} {\bibfnamefont {M.}~\bibnamefont
  {Levi}}, \bibinfo {author} {\bibfnamefont {A.~J.}\ \bibnamefont {Mcleod}}, \
  and\ \bibinfo {author} {\bibfnamefont {M.}~\bibnamefont {Von~Hippel}},\
  }\href@noop {} {\  (\bibinfo {year} {2020}{\natexlab{b}})},\ \Eprint
  {http://arxiv.org/abs/2003.07890} {arXiv:2003.07890 [hep-th]} \BibitemShut
  {NoStop}%
\bibitem [{\citenamefont {Levi}\ and\ \citenamefont
  {Steinhoff}(2016{\natexlab{b}})}]{Levi:2015ixa}%
  \BibitemOpen
  \bibfield  {author} {\bibinfo {author} {\bibfnamefont {M.}~\bibnamefont
  {Levi}}\ and\ \bibinfo {author} {\bibfnamefont {J.}~\bibnamefont
  {Steinhoff}},\ }\href {\doibase 10.1088/1475-7516/2016/01/008} {\bibfield
  {journal} {\bibinfo  {journal} {JCAP}\ }\textbf {\bibinfo {volume} {1601}},\
  \bibinfo {pages} {008} (\bibinfo {year} {2016}{\natexlab{b}})},\ \Eprint
  {http://arxiv.org/abs/1506.05794} {arXiv:1506.05794 [gr-qc]} \BibitemShut
  {NoStop}%
\bibitem [{\citenamefont {Levi}\ and\ \citenamefont
  {Steinhoff}(2015{\natexlab{a}})}]{Levi:2015msa}%
  \BibitemOpen
  \bibfield  {author} {\bibinfo {author} {\bibfnamefont {M.}~\bibnamefont
  {Levi}}\ and\ \bibinfo {author} {\bibfnamefont {J.}~\bibnamefont
  {Steinhoff}},\ }\href {\doibase 10.1007/JHEP09(2015)219} {\bibfield
  {journal} {\bibinfo  {journal} {JHEP}\ }\textbf {\bibinfo {volume} {09}},\
  \bibinfo {pages} {219} (\bibinfo {year} {2015}{\natexlab{a}})},\ \Eprint
  {http://arxiv.org/abs/1501.04956} {arXiv:1501.04956 [gr-qc]} \BibitemShut
  {NoStop}%
\bibitem [{\citenamefont {Levi}\ and\ \citenamefont
  {Steinhoff}(2016{\natexlab{c}})}]{Levi:2016ofk}%
  \BibitemOpen
  \bibfield  {author} {\bibinfo {author} {\bibfnamefont {M.}~\bibnamefont
  {Levi}}\ and\ \bibinfo {author} {\bibfnamefont {J.}~\bibnamefont
  {Steinhoff}},\ }\href@noop {} {\  (\bibinfo {year} {2016}{\natexlab{c}})},\
  \Eprint {http://arxiv.org/abs/1607.04252} {arXiv:1607.04252 [gr-qc]}
  \BibitemShut {NoStop}%
\bibitem [{\citenamefont {Levi}\ \emph {et~al.}(2019)\citenamefont {Levi},
  \citenamefont {Mougiakakos},\ and\ \citenamefont {Vieira}}]{Levi:2019kgk}%
  \BibitemOpen
  \bibfield  {author} {\bibinfo {author} {\bibfnamefont {M.}~\bibnamefont
  {Levi}}, \bibinfo {author} {\bibfnamefont {S.}~\bibnamefont {Mougiakakos}}, \
  and\ \bibinfo {author} {\bibfnamefont {M.}~\bibnamefont {Vieira}},\
  }\href@noop {} {\  (\bibinfo {year} {2019})},\ \Eprint
  {http://arxiv.org/abs/1912.06276} {arXiv:1912.06276 [hep-th]} \BibitemShut
  {NoStop}%
\bibitem [{\citenamefont {Levi}\ and\ \citenamefont
  {Steinhoff}(2015{\natexlab{b}})}]{Levi:2014gsa}%
  \BibitemOpen
  \bibfield  {author} {\bibinfo {author} {\bibfnamefont {M.}~\bibnamefont
  {Levi}}\ and\ \bibinfo {author} {\bibfnamefont {J.}~\bibnamefont
  {Steinhoff}},\ }\href {\doibase 10.1007/JHEP06(2015)059} {\bibfield
  {journal} {\bibinfo  {journal} {JHEP}\ }\textbf {\bibinfo {volume} {06}},\
  \bibinfo {pages} {059} (\bibinfo {year} {2015}{\natexlab{b}})},\ \Eprint
  {http://arxiv.org/abs/1410.2601} {arXiv:1410.2601 [gr-qc]} \BibitemShut
  {NoStop}%
\bibitem [{\citenamefont {Levi}\ and\ \citenamefont
  {Teng}(2020)}]{Levi:2020lfn}%
  \BibitemOpen
  \bibfield  {author} {\bibinfo {author} {\bibfnamefont {M.}~\bibnamefont
  {Levi}}\ and\ \bibinfo {author} {\bibfnamefont {F.}~\bibnamefont {Teng}},\
  }\href@noop {} {\  (\bibinfo {year} {2020})},\ \Eprint
  {http://arxiv.org/abs/2008.12280} {arXiv:2008.12280 [hep-th]} \BibitemShut
  {NoStop}%
\bibitem [{\citenamefont {Vines}\ and\ \citenamefont
  {Steinhoff}(2018)}]{Vines:2016qwa}%
  \BibitemOpen
  \bibfield  {author} {\bibinfo {author} {\bibfnamefont {J.}~\bibnamefont
  {Vines}}\ and\ \bibinfo {author} {\bibfnamefont {J.}~\bibnamefont
  {Steinhoff}},\ }\href {\doibase 10.1103/PhysRevD.97.064010} {\bibfield
  {journal} {\bibinfo  {journal} {Phys.\ Rev.\ D}\ }\textbf {\bibinfo {volume}
  {97}},\ \bibinfo {pages} {064010} (\bibinfo {year} {2018})},\ \Eprint
  {http://arxiv.org/abs/1606.08832} {arXiv:1606.08832 [gr-qc]} \BibitemShut
  {NoStop}%
\bibitem [{\citenamefont {Siemonsen}\ and\ \citenamefont
  {Vines}(2019)}]{Siemonsen:2019dsu}%
  \BibitemOpen
  \bibfield  {author} {\bibinfo {author} {\bibfnamefont {N.}~\bibnamefont
  {Siemonsen}}\ and\ \bibinfo {author} {\bibfnamefont {J.}~\bibnamefont
  {Vines}},\ }\href@noop {} {\  (\bibinfo {year} {2019})},\ \Eprint
  {http://arxiv.org/abs/1909.07361} {arXiv:1909.07361 [gr-qc]} \BibitemShut
  {NoStop}%
\bibitem [{\citenamefont {Nagar}(2011)}]{Nagar:2011fx}%
  \BibitemOpen
  \bibfield  {author} {\bibinfo {author} {\bibfnamefont {A.}~\bibnamefont
  {Nagar}},\ }\href {\doibase 10.1103/PhysRevD.84.084028,
  10.1103/PhysRevD.88.089901} {\bibfield  {journal} {\bibinfo  {journal} {Phys.
  Rev.}\ }\textbf {\bibinfo {volume} {D84}},\ \bibinfo {pages} {084028}
  (\bibinfo {year} {2011})},\ \bibinfo {note} {[Erratum: Phys.
  Rev.D88,no.8,089901(2013)]},\ \Eprint {http://arxiv.org/abs/1106.4349}
  {arXiv:1106.4349 [gr-qc]} \BibitemShut {NoStop}%
\bibitem [{\citenamefont {Barausse}\ and\ \citenamefont
  {Buonanno}(2011)}]{Barausse:2011ys}%
  \BibitemOpen
  \bibfield  {author} {\bibinfo {author} {\bibfnamefont {E.}~\bibnamefont
  {Barausse}}\ and\ \bibinfo {author} {\bibfnamefont {A.}~\bibnamefont
  {Buonanno}},\ }\href {\doibase 10.1103/PhysRevD.84.104027} {\bibfield
  {journal} {\bibinfo  {journal} {Phys. Rev.}\ }\textbf {\bibinfo {volume}
  {D84}},\ \bibinfo {pages} {104027} (\bibinfo {year} {2011})},\ \Eprint
  {http://arxiv.org/abs/1107.2904} {arXiv:1107.2904 [gr-qc]} \BibitemShut
  {NoStop}%
\bibitem [{\citenamefont {Bohé}\ \emph {et~al.}(2017)\citenamefont {Bohé}
  \emph {et~al.}}]{Bohe:2016gbl}%
  \BibitemOpen
  \bibfield  {author} {\bibinfo {author} {\bibfnamefont {A.}~\bibnamefont
  {Bohé}} \emph {et~al.},\ }\href {\doibase 10.1103/PhysRevD.95.044028}
  {\bibfield  {journal} {\bibinfo  {journal} {Phys. Rev.}\ }\textbf {\bibinfo
  {volume} {D95}},\ \bibinfo {pages} {044028} (\bibinfo {year} {2017})},\
  \Eprint {http://arxiv.org/abs/1611.03703} {arXiv:1611.03703 [gr-qc]}
  \BibitemShut {NoStop}%
\bibitem [{\citenamefont {Babak}\ \emph {et~al.}(2017)\citenamefont {Babak},
  \citenamefont {Taracchini},\ and\ \citenamefont {Buonanno}}]{Babak:2016tgq}%
  \BibitemOpen
  \bibfield  {author} {\bibinfo {author} {\bibfnamefont {S.}~\bibnamefont
  {Babak}}, \bibinfo {author} {\bibfnamefont {A.}~\bibnamefont {Taracchini}}, \
  and\ \bibinfo {author} {\bibfnamefont {A.}~\bibnamefont {Buonanno}},\ }\href
  {\doibase 10.1103/PhysRevD.95.024010} {\bibfield  {journal} {\bibinfo
  {journal} {Phys.\ Rev.\ D}\ }\textbf {\bibinfo {volume} {95}},\ \bibinfo
  {pages} {024010} (\bibinfo {year} {2017})},\ \Eprint
  {http://arxiv.org/abs/1607.05661} {arXiv:1607.05661 [gr-qc]} \BibitemShut
  {NoStop}%
\bibitem [{\citenamefont {Cotesta}\ \emph {et~al.}(2018)\citenamefont
  {Cotesta}, \citenamefont {Buonanno}, \citenamefont {Bohé}, \citenamefont
  {Taracchini}, \citenamefont {Hinder},\ and\ \citenamefont
  {Ossokine}}]{Cotesta:2018fcv}%
  \BibitemOpen
  \bibfield  {author} {\bibinfo {author} {\bibfnamefont {R.}~\bibnamefont
  {Cotesta}}, \bibinfo {author} {\bibfnamefont {A.}~\bibnamefont {Buonanno}},
  \bibinfo {author} {\bibfnamefont {A.}~\bibnamefont {Bohé}}, \bibinfo
  {author} {\bibfnamefont {A.}~\bibnamefont {Taracchini}}, \bibinfo {author}
  {\bibfnamefont {I.}~\bibnamefont {Hinder}}, \ and\ \bibinfo {author}
  {\bibfnamefont {S.}~\bibnamefont {Ossokine}},\ }\href {\doibase
  10.1103/PhysRevD.98.084028} {\bibfield  {journal} {\bibinfo  {journal} {Phys.
  Rev.}\ }\textbf {\bibinfo {volume} {D98}},\ \bibinfo {pages} {084028}
  (\bibinfo {year} {2018})},\ \Eprint {http://arxiv.org/abs/1803.10701}
  {arXiv:1803.10701 [gr-qc]} \BibitemShut {NoStop}%
\bibitem [{\citenamefont {Ossokine}\ \emph {et~al.}(2020)\citenamefont
  {Ossokine} \emph {et~al.}}]{Ossokine:2020kjp}%
  \BibitemOpen
  \bibfield  {author} {\bibinfo {author} {\bibfnamefont {S.}~\bibnamefont
  {Ossokine}} \emph {et~al.},\ }\href@noop {} {\  (\bibinfo {year} {2020})},\
  \Eprint {http://arxiv.org/abs/2004.09442} {arXiv:2004.09442 [gr-qc]}
  \BibitemShut {NoStop}%
\bibitem [{\citenamefont {Nagar}\ \emph {et~al.}(2019)\citenamefont {Nagar},
  \citenamefont {Messina}, \citenamefont {Rettegno}, \citenamefont {Bini},
  \citenamefont {Damour}, \citenamefont {Geralico}, \citenamefont {Akcay},\
  and\ \citenamefont {Bernuzzi}}]{Nagar:2018plt}%
  \BibitemOpen
  \bibfield  {author} {\bibinfo {author} {\bibfnamefont {A.}~\bibnamefont
  {Nagar}}, \bibinfo {author} {\bibfnamefont {F.}~\bibnamefont {Messina}},
  \bibinfo {author} {\bibfnamefont {P.}~\bibnamefont {Rettegno}}, \bibinfo
  {author} {\bibfnamefont {D.}~\bibnamefont {Bini}}, \bibinfo {author}
  {\bibfnamefont {T.}~\bibnamefont {Damour}}, \bibinfo {author} {\bibfnamefont
  {A.}~\bibnamefont {Geralico}}, \bibinfo {author} {\bibfnamefont
  {S.}~\bibnamefont {Akcay}}, \ and\ \bibinfo {author} {\bibfnamefont
  {S.}~\bibnamefont {Bernuzzi}},\ }\href {\doibase 10.1103/PhysRevD.99.044007}
  {\bibfield  {journal} {\bibinfo  {journal} {Phys. Rev.}\ }\textbf {\bibinfo
  {volume} {D99}},\ \bibinfo {pages} {044007} (\bibinfo {year} {2019})},\
  \Eprint {http://arxiv.org/abs/1812.07923} {arXiv:1812.07923 [gr-qc]}
  \BibitemShut {NoStop}%
\bibitem [{\citenamefont {Nagar}\ \emph {et~al.}(2018)\citenamefont {Nagar}
  \emph {et~al.}}]{Nagar:2018zoe}%
  \BibitemOpen
  \bibfield  {author} {\bibinfo {author} {\bibfnamefont {A.}~\bibnamefont
  {Nagar}} \emph {et~al.},\ }\href {\doibase 10.1103/PhysRevD.98.104052}
  {\bibfield  {journal} {\bibinfo  {journal} {Phys. Rev.}\ }\textbf {\bibinfo
  {volume} {D98}},\ \bibinfo {pages} {104052} (\bibinfo {year} {2018})},\
  \Eprint {http://arxiv.org/abs/1806.01772} {arXiv:1806.01772 [gr-qc]}
  \BibitemShut {NoStop}%
\bibitem [{\citenamefont {Khalil}\ \emph {et~al.}(2020)\citenamefont {Khalil},
  \citenamefont {Steinhoff}, \citenamefont {Vines},\ and\ \citenamefont
  {Buonanno}}]{Khalil:2020mmr}%
  \BibitemOpen
  \bibfield  {author} {\bibinfo {author} {\bibfnamefont {M.}~\bibnamefont
  {Khalil}}, \bibinfo {author} {\bibfnamefont {J.}~\bibnamefont {Steinhoff}},
  \bibinfo {author} {\bibfnamefont {J.}~\bibnamefont {Vines}}, \ and\ \bibinfo
  {author} {\bibfnamefont {A.}~\bibnamefont {Buonanno}},\ }\href@noop {} {\
  (\bibinfo {year} {2020})},\ \Eprint {http://arxiv.org/abs/2003.04469}
  {arXiv:2003.04469 [gr-qc]} \BibitemShut {NoStop}%
\bibitem [{\citenamefont {Barack}\ and\ \citenamefont
  {Pound}(2019)}]{Barack:2018yvs}%
  \BibitemOpen
  \bibfield  {author} {\bibinfo {author} {\bibfnamefont {L.}~\bibnamefont
  {Barack}}\ and\ \bibinfo {author} {\bibfnamefont {A.}~\bibnamefont {Pound}},\
  }\href {\doibase 10.1088/1361-6633/aae552} {\bibfield  {journal} {\bibinfo
  {journal} {Rept. Prog. Phys.}\ }\textbf {\bibinfo {volume} {82}},\ \bibinfo
  {pages} {016904} (\bibinfo {year} {2019})},\ \Eprint
  {http://arxiv.org/abs/1805.10385} {arXiv:1805.10385 [gr-qc]} \BibitemShut
  {NoStop}%
\bibitem [{\citenamefont {van~de Meent}(2018)}]{vandeMeent:2017bcc}%
  \BibitemOpen
  \bibfield  {author} {\bibinfo {author} {\bibfnamefont {M.}~\bibnamefont
  {van~de Meent}},\ }\href {\doibase 10.1103/PhysRevD.97.104033} {\bibfield
  {journal} {\bibinfo  {journal} {Phys. Rev.}\ }\textbf {\bibinfo {volume}
  {D97}},\ \bibinfo {pages} {104033} (\bibinfo {year} {2018})},\ \Eprint
  {http://arxiv.org/abs/1711.09607} {arXiv:1711.09607 [gr-qc]} \BibitemShut
  {NoStop}%
\bibitem [{\citenamefont {Pound}\ \emph {et~al.}(2020)\citenamefont {Pound},
  \citenamefont {Wardell}, \citenamefont {Warburton},\ and\ \citenamefont
  {Miller}}]{Pound:2019lzj}%
  \BibitemOpen
  \bibfield  {author} {\bibinfo {author} {\bibfnamefont {A.}~\bibnamefont
  {Pound}}, \bibinfo {author} {\bibfnamefont {B.}~\bibnamefont {Wardell}},
  \bibinfo {author} {\bibfnamefont {N.}~\bibnamefont {Warburton}}, \ and\
  \bibinfo {author} {\bibfnamefont {J.}~\bibnamefont {Miller}},\ }\href
  {\doibase 10.1103/PhysRevLett.124.021101} {\bibfield  {journal} {\bibinfo
  {journal} {Phys. Rev. Lett.}\ }\textbf {\bibinfo {volume} {124}},\ \bibinfo
  {pages} {021101} (\bibinfo {year} {2020})},\ \Eprint
  {http://arxiv.org/abs/1908.07419} {arXiv:1908.07419 [gr-qc]} \BibitemShut
  {NoStop}%
\bibitem [{\citenamefont {Le~Tiec}\ \emph {et~al.}(2012)\citenamefont
  {Le~Tiec}, \citenamefont {Blanchet},\ and\ \citenamefont
  {Whiting}}]{LeTiec:2011ab}%
  \BibitemOpen
  \bibfield  {author} {\bibinfo {author} {\bibfnamefont {A.}~\bibnamefont
  {Le~Tiec}}, \bibinfo {author} {\bibfnamefont {L.}~\bibnamefont {Blanchet}}, \
  and\ \bibinfo {author} {\bibfnamefont {B.~F.}\ \bibnamefont {Whiting}},\
  }\href {\doibase 10.1103/PhysRevD.85.064039} {\bibfield  {journal} {\bibinfo
  {journal} {Phys. Rev.}\ }\textbf {\bibinfo {volume} {D85}},\ \bibinfo {pages}
  {064039} (\bibinfo {year} {2012})},\ \Eprint {http://arxiv.org/abs/1111.5378}
  {arXiv:1111.5378 [gr-qc]} \BibitemShut {NoStop}%
\bibitem [{\citenamefont {Detweiler}(2008)}]{Detweiler:2008ft}%
  \BibitemOpen
  \bibfield  {author} {\bibinfo {author} {\bibfnamefont {S.~L.}\ \bibnamefont
  {Detweiler}},\ }\href {\doibase 10.1103/PhysRevD.77.124026} {\bibfield
  {journal} {\bibinfo  {journal} {Phys. Rev.}\ }\textbf {\bibinfo {volume}
  {D77}},\ \bibinfo {pages} {124026} (\bibinfo {year} {2008})},\ \Eprint
  {http://arxiv.org/abs/0804.3529} {arXiv:0804.3529 [gr-qc]} \BibitemShut
  {NoStop}%
\bibitem [{\citenamefont {Bini}\ and\ \citenamefont
  {Damour}(2014{\natexlab{a}})}]{Bini:2013rfa}%
  \BibitemOpen
  \bibfield  {author} {\bibinfo {author} {\bibfnamefont {D.}~\bibnamefont
  {Bini}}\ and\ \bibinfo {author} {\bibfnamefont {T.}~\bibnamefont {Damour}},\
  }\href {\doibase 10.1103/PhysRevD.89.064063} {\bibfield  {journal} {\bibinfo
  {journal} {Phys. Rev. D}\ }\textbf {\bibinfo {volume} {89}},\ \bibinfo
  {pages} {064063} (\bibinfo {year} {2014}{\natexlab{a}})},\ \Eprint
  {http://arxiv.org/abs/1312.2503} {arXiv:1312.2503 [gr-qc]} \BibitemShut
  {NoStop}%
\bibitem [{\citenamefont {Kavanagh}\ \emph {et~al.}(2015)\citenamefont
  {Kavanagh}, \citenamefont {Ottewill},\ and\ \citenamefont
  {Wardell}}]{Kavanagh:2015lva}%
  \BibitemOpen
  \bibfield  {author} {\bibinfo {author} {\bibfnamefont {C.}~\bibnamefont
  {Kavanagh}}, \bibinfo {author} {\bibfnamefont {A.~C.}\ \bibnamefont
  {Ottewill}}, \ and\ \bibinfo {author} {\bibfnamefont {B.}~\bibnamefont
  {Wardell}},\ }\href {\doibase 10.1103/PhysRevD.92.084025} {\bibfield
  {journal} {\bibinfo  {journal} {Phys. Rev.}\ }\textbf {\bibinfo {volume}
  {D92}},\ \bibinfo {pages} {084025} (\bibinfo {year} {2015})},\ \Eprint
  {http://arxiv.org/abs/1503.02334} {arXiv:1503.02334 [gr-qc]} \BibitemShut
  {NoStop}%
\bibitem [{\citenamefont {Johnson-McDaniel}\ \emph {et~al.}(2015)\citenamefont
  {Johnson-McDaniel}, \citenamefont {Shah},\ and\ \citenamefont
  {Whiting}}]{Johnson-McDaniel:2015vva}%
  \BibitemOpen
  \bibfield  {author} {\bibinfo {author} {\bibfnamefont {N.~K.}\ \bibnamefont
  {Johnson-McDaniel}}, \bibinfo {author} {\bibfnamefont {A.~G.}\ \bibnamefont
  {Shah}}, \ and\ \bibinfo {author} {\bibfnamefont {B.~F.}\ \bibnamefont
  {Whiting}},\ }\href {\doibase 10.1103/PhysRevD.92.044007} {\bibfield
  {journal} {\bibinfo  {journal} {Phys. Rev. D}\ }\textbf {\bibinfo {volume}
  {92}},\ \bibinfo {pages} {044007} (\bibinfo {year} {2015})},\ \Eprint
  {http://arxiv.org/abs/1503.02638} {arXiv:1503.02638 [gr-qc]} \BibitemShut
  {NoStop}%
\bibitem [{\citenamefont {Hopper}\ \emph {et~al.}(2016)\citenamefont {Hopper},
  \citenamefont {Kavanagh},\ and\ \citenamefont {Ottewill}}]{Hopper:2015icj}%
  \BibitemOpen
  \bibfield  {author} {\bibinfo {author} {\bibfnamefont {S.}~\bibnamefont
  {Hopper}}, \bibinfo {author} {\bibfnamefont {C.}~\bibnamefont {Kavanagh}}, \
  and\ \bibinfo {author} {\bibfnamefont {A.~C.}\ \bibnamefont {Ottewill}},\
  }\href {\doibase 10.1103/PhysRevD.93.044010} {\bibfield  {journal} {\bibinfo
  {journal} {Phys. Rev.}\ }\textbf {\bibinfo {volume} {D93}},\ \bibinfo {pages}
  {044010} (\bibinfo {year} {2016})},\ \Eprint
  {http://arxiv.org/abs/1512.01556} {arXiv:1512.01556 [gr-qc]} \BibitemShut
  {NoStop}%
\bibitem [{\citenamefont {Bini}\ \emph
  {et~al.}(2016{\natexlab{a}})\citenamefont {Bini}, \citenamefont {Damour},\
  and\ \citenamefont {Geralico}}]{Bini:2015bfb}%
  \BibitemOpen
  \bibfield  {author} {\bibinfo {author} {\bibfnamefont {D.}~\bibnamefont
  {Bini}}, \bibinfo {author} {\bibfnamefont {T.}~\bibnamefont {Damour}}, \ and\
  \bibinfo {author} {\bibfnamefont {A.}~\bibnamefont {Geralico}},\ }\href
  {\doibase 10.1103/PhysRevD.93.064023} {\bibfield  {journal} {\bibinfo
  {journal} {Phys. Rev. D}\ }\textbf {\bibinfo {volume} {93}},\ \bibinfo
  {pages} {064023} (\bibinfo {year} {2016}{\natexlab{a}})},\ \Eprint
  {http://arxiv.org/abs/1511.04533} {arXiv:1511.04533 [gr-qc]} \BibitemShut
  {NoStop}%
\bibitem [{\citenamefont {Kavanagh}\ \emph {et~al.}(2016)\citenamefont
  {Kavanagh}, \citenamefont {Ottewill},\ and\ \citenamefont
  {Wardell}}]{Kavanagh:2016idg}%
  \BibitemOpen
  \bibfield  {author} {\bibinfo {author} {\bibfnamefont {C.}~\bibnamefont
  {Kavanagh}}, \bibinfo {author} {\bibfnamefont {A.~C.}\ \bibnamefont
  {Ottewill}}, \ and\ \bibinfo {author} {\bibfnamefont {B.}~\bibnamefont
  {Wardell}},\ }\href {\doibase 10.1103/PhysRevD.93.124038} {\bibfield
  {journal} {\bibinfo  {journal} {Phys. Rev.}\ }\textbf {\bibinfo {volume}
  {D93}},\ \bibinfo {pages} {124038} (\bibinfo {year} {2016})},\ \Eprint
  {http://arxiv.org/abs/1601.03394} {arXiv:1601.03394 [gr-qc]} \BibitemShut
  {NoStop}%
\bibitem [{\citenamefont {Bini}\ \emph
  {et~al.}(2018{\natexlab{a}})\citenamefont {Bini}, \citenamefont {Damour},
  \citenamefont {Geralico},\ and\ \citenamefont {Kavanagh}}]{Bini:2018zde}%
  \BibitemOpen
  \bibfield  {author} {\bibinfo {author} {\bibfnamefont {D.}~\bibnamefont
  {Bini}}, \bibinfo {author} {\bibfnamefont {T.}~\bibnamefont {Damour}},
  \bibinfo {author} {\bibfnamefont {A.}~\bibnamefont {Geralico}}, \ and\
  \bibinfo {author} {\bibfnamefont {C.}~\bibnamefont {Kavanagh}},\ }\href
  {\doibase 10.1103/PhysRevD.97.104022} {\bibfield  {journal} {\bibinfo
  {journal} {Phys. Rev. D}\ }\textbf {\bibinfo {volume} {97}},\ \bibinfo
  {pages} {104022} (\bibinfo {year} {2018}{\natexlab{a}})},\ \Eprint
  {http://arxiv.org/abs/1801.09616} {arXiv:1801.09616 [gr-qc]} \BibitemShut
  {NoStop}%
\bibitem [{\citenamefont {Bini}\ and\ \citenamefont
  {Geralico}(2019{\natexlab{a}})}]{Bini:2019lcd}%
  \BibitemOpen
  \bibfield  {author} {\bibinfo {author} {\bibfnamefont {D.}~\bibnamefont
  {Bini}}\ and\ \bibinfo {author} {\bibfnamefont {A.}~\bibnamefont
  {Geralico}},\ }\href {\doibase 10.1103/PhysRevD.100.104002} {\bibfield
  {journal} {\bibinfo  {journal} {Phys. Rev.}\ }\textbf {\bibinfo {volume}
  {D100}},\ \bibinfo {pages} {104002} (\bibinfo {year} {2019}{\natexlab{a}})},\
  \Eprint {http://arxiv.org/abs/1907.11080} {arXiv:1907.11080 [gr-qc]}
  \BibitemShut {NoStop}%
\bibitem [{\citenamefont {Bini}\ \emph
  {et~al.}(2020{\natexlab{e}})\citenamefont {Bini}, \citenamefont {Geralico},\
  and\ \citenamefont {Steinhoff}}]{Bini:2020zqy}%
  \BibitemOpen
  \bibfield  {author} {\bibinfo {author} {\bibfnamefont {D.}~\bibnamefont
  {Bini}}, \bibinfo {author} {\bibfnamefont {A.}~\bibnamefont {Geralico}}, \
  and\ \bibinfo {author} {\bibfnamefont {J.}~\bibnamefont {Steinhoff}},\
  }\href@noop {} {\  (\bibinfo {year} {2020}{\natexlab{e}})},\ \Eprint
  {http://arxiv.org/abs/2003.12887} {arXiv:2003.12887 [gr-qc]} \BibitemShut
  {NoStop}%
\bibitem [{\citenamefont {Dolan}\ \emph {et~al.}(2014)\citenamefont {Dolan},
  \citenamefont {Warburton}, \citenamefont {Harte}, \citenamefont {Le~Tiec},
  \citenamefont {Wardell},\ and\ \citenamefont {Barack}}]{Dolan:2013roa}%
  \BibitemOpen
  \bibfield  {author} {\bibinfo {author} {\bibfnamefont {S.~R.}\ \bibnamefont
  {Dolan}}, \bibinfo {author} {\bibfnamefont {N.}~\bibnamefont {Warburton}},
  \bibinfo {author} {\bibfnamefont {A.~I.}\ \bibnamefont {Harte}}, \bibinfo
  {author} {\bibfnamefont {A.}~\bibnamefont {Le~Tiec}}, \bibinfo {author}
  {\bibfnamefont {B.}~\bibnamefont {Wardell}}, \ and\ \bibinfo {author}
  {\bibfnamefont {L.}~\bibnamefont {Barack}},\ }\href {\doibase
  10.1103/PhysRevD.89.064011} {\bibfield  {journal} {\bibinfo  {journal} {Phys.
  Rev.}\ }\textbf {\bibinfo {volume} {D89}},\ \bibinfo {pages} {064011}
  (\bibinfo {year} {2014})},\ \Eprint {http://arxiv.org/abs/1312.0775}
  {arXiv:1312.0775 [gr-qc]} \BibitemShut {NoStop}%
\bibitem [{\citenamefont {Bini}\ and\ \citenamefont
  {Damour}(2014{\natexlab{b}})}]{Bini:2014ica}%
  \BibitemOpen
  \bibfield  {author} {\bibinfo {author} {\bibfnamefont {D.}~\bibnamefont
  {Bini}}\ and\ \bibinfo {author} {\bibfnamefont {T.}~\bibnamefont {Damour}},\
  }\href {\doibase 10.1103/PhysRevD.90.024039} {\bibfield  {journal} {\bibinfo
  {journal} {Phys. Rev.}\ }\textbf {\bibinfo {volume} {D90}},\ \bibinfo {pages}
  {024039} (\bibinfo {year} {2014}{\natexlab{b}})},\ \Eprint
  {http://arxiv.org/abs/1404.2747} {arXiv:1404.2747 [gr-qc]} \BibitemShut
  {NoStop}%
\bibitem [{\citenamefont {Akcay}\ \emph {et~al.}(2017)\citenamefont {Akcay},
  \citenamefont {Dempsey},\ and\ \citenamefont {Dolan}}]{Akcay:2016dku}%
  \BibitemOpen
  \bibfield  {author} {\bibinfo {author} {\bibfnamefont {S.}~\bibnamefont
  {Akcay}}, \bibinfo {author} {\bibfnamefont {D.}~\bibnamefont {Dempsey}}, \
  and\ \bibinfo {author} {\bibfnamefont {S.~R.}\ \bibnamefont {Dolan}},\ }\href
  {\doibase 10.1088/1361-6382/aa61d6} {\bibfield  {journal} {\bibinfo
  {journal} {Class. Quant. Grav.}\ }\textbf {\bibinfo {volume} {34}},\ \bibinfo
  {pages} {084001} (\bibinfo {year} {2017})},\ \Eprint
  {http://arxiv.org/abs/1608.04811} {arXiv:1608.04811 [gr-qc]} \BibitemShut
  {NoStop}%
\bibitem [{\citenamefont {Kavanagh}\ \emph {et~al.}(2017)\citenamefont
  {Kavanagh}, \citenamefont {Bini}, \citenamefont {Damour}, \citenamefont
  {Hopper}, \citenamefont {Ottewill},\ and\ \citenamefont
  {Wardell}}]{Kavanagh:2017wot}%
  \BibitemOpen
  \bibfield  {author} {\bibinfo {author} {\bibfnamefont {C.}~\bibnamefont
  {Kavanagh}}, \bibinfo {author} {\bibfnamefont {D.}~\bibnamefont {Bini}},
  \bibinfo {author} {\bibfnamefont {T.}~\bibnamefont {Damour}}, \bibinfo
  {author} {\bibfnamefont {S.}~\bibnamefont {Hopper}}, \bibinfo {author}
  {\bibfnamefont {A.~C.}\ \bibnamefont {Ottewill}}, \ and\ \bibinfo {author}
  {\bibfnamefont {B.}~\bibnamefont {Wardell}},\ }\href {\doibase
  10.1103/PhysRevD.96.064012} {\bibfield  {journal} {\bibinfo  {journal} {Phys.
  Rev.}\ }\textbf {\bibinfo {volume} {D96}},\ \bibinfo {pages} {064012}
  (\bibinfo {year} {2017})},\ \Eprint {http://arxiv.org/abs/1706.00459}
  {arXiv:1706.00459 [gr-qc]} \BibitemShut {NoStop}%
\bibitem [{\citenamefont {Akcay}(2017)}]{Akcay:2017azq}%
  \BibitemOpen
  \bibfield  {author} {\bibinfo {author} {\bibfnamefont {S.}~\bibnamefont
  {Akcay}},\ }\href {\doibase 10.1103/PhysRevD.96.044024} {\bibfield  {journal}
  {\bibinfo  {journal} {Phys. Rev.}\ }\textbf {\bibinfo {volume} {D96}},\
  \bibinfo {pages} {044024} (\bibinfo {year} {2017})},\ \Eprint
  {http://arxiv.org/abs/1705.03282} {arXiv:1705.03282 [gr-qc]} \BibitemShut
  {NoStop}%
\bibitem [{\citenamefont {Bini}\ \emph
  {et~al.}(2018{\natexlab{b}})\citenamefont {Bini}, \citenamefont {Damour},
  \citenamefont {Geralico}, \citenamefont {Kavanagh},\ and\ \citenamefont
  {van~de Meent}}]{Bini:2018ylh}%
  \BibitemOpen
  \bibfield  {author} {\bibinfo {author} {\bibfnamefont {D.}~\bibnamefont
  {Bini}}, \bibinfo {author} {\bibfnamefont {T.}~\bibnamefont {Damour}},
  \bibinfo {author} {\bibfnamefont {A.}~\bibnamefont {Geralico}}, \bibinfo
  {author} {\bibfnamefont {C.}~\bibnamefont {Kavanagh}}, \ and\ \bibinfo
  {author} {\bibfnamefont {M.}~\bibnamefont {van~de Meent}},\ }\href {\doibase
  10.1103/PhysRevD.98.104062} {\bibfield  {journal} {\bibinfo  {journal} {Phys.
  Rev.}\ }\textbf {\bibinfo {volume} {D98}},\ \bibinfo {pages} {104062}
  (\bibinfo {year} {2018}{\natexlab{b}})},\ \Eprint
  {http://arxiv.org/abs/1809.02516} {arXiv:1809.02516 [gr-qc]} \BibitemShut
  {NoStop}%
\bibitem [{\citenamefont {Blanchet}\ \emph {et~al.}(2011)\citenamefont
  {Blanchet}, \citenamefont {Detweiler}, \citenamefont {Le~Tiec},\ and\
  \citenamefont {Whiting}}]{Blanchet:2011aha}%
  \BibitemOpen
  \bibfield  {author} {\bibinfo {author} {\bibfnamefont {L.}~\bibnamefont
  {Blanchet}}, \bibinfo {author} {\bibfnamefont {S.}~\bibnamefont {Detweiler}},
  \bibinfo {author} {\bibfnamefont {A.}~\bibnamefont {Le~Tiec}}, \ and\
  \bibinfo {author} {\bibfnamefont {B.~F.}\ \bibnamefont {Whiting}},\
  }\bibfield  {booktitle} {\emph {\bibinfo {booktitle} {{Mass and motion in
  general relativity. Proceedings, School on Mass, Orleans, France, June 23-25,
  2008}}},\ }\href {\doibase 10.1007/978-90-481-3015-3_15} {\bibfield
  {journal} {\bibinfo  {journal} {Fundam. Theor. Phys.}\ }\textbf {\bibinfo
  {volume} {162}},\ \bibinfo {pages} {415} (\bibinfo {year} {2011})},\ \bibinfo
  {note} {[,415(2010)]},\ \Eprint {http://arxiv.org/abs/1007.2614}
  {arXiv:1007.2614 [gr-qc]} \BibitemShut {NoStop}%
\bibitem [{\citenamefont {Akcay}\ \emph {et~al.}(2015)\citenamefont {Akcay},
  \citenamefont {Le~Tiec}, \citenamefont {Barack}, \citenamefont {Sago},\ and\
  \citenamefont {Warburton}}]{Akcay:2015pza}%
  \BibitemOpen
  \bibfield  {author} {\bibinfo {author} {\bibfnamefont {S.}~\bibnamefont
  {Akcay}}, \bibinfo {author} {\bibfnamefont {A.}~\bibnamefont {Le~Tiec}},
  \bibinfo {author} {\bibfnamefont {L.}~\bibnamefont {Barack}}, \bibinfo
  {author} {\bibfnamefont {N.}~\bibnamefont {Sago}}, \ and\ \bibinfo {author}
  {\bibfnamefont {N.}~\bibnamefont {Warburton}},\ }\href {\doibase
  10.1103/PhysRevD.91.124014} {\bibfield  {journal} {\bibinfo  {journal} {Phys.
  Rev. D}\ }\textbf {\bibinfo {volume} {91}},\ \bibinfo {pages} {124014}
  (\bibinfo {year} {2015})},\ \Eprint {http://arxiv.org/abs/1503.01374}
  {arXiv:1503.01374 [gr-qc]} \BibitemShut {NoStop}%
\bibitem [{\citenamefont {Barausse}\ \emph {et~al.}(2012)\citenamefont
  {Barausse}, \citenamefont {Buonanno},\ and\ \citenamefont
  {Le~Tiec}}]{Barausse:2011dq}%
  \BibitemOpen
  \bibfield  {author} {\bibinfo {author} {\bibfnamefont {E.}~\bibnamefont
  {Barausse}}, \bibinfo {author} {\bibfnamefont {A.}~\bibnamefont {Buonanno}},
  \ and\ \bibinfo {author} {\bibfnamefont {A.}~\bibnamefont {Le~Tiec}},\ }\href
  {\doibase 10.1103/PhysRevD.85.064010} {\bibfield  {journal} {\bibinfo
  {journal} {Phys. Rev.}\ }\textbf {\bibinfo {volume} {D85}},\ \bibinfo {pages}
  {064010} (\bibinfo {year} {2012})},\ \Eprint {http://arxiv.org/abs/1111.5610}
  {arXiv:1111.5610 [gr-qc]} \BibitemShut {NoStop}%
\bibitem [{\citenamefont {Akcay}\ \emph {et~al.}(2012)\citenamefont {Akcay},
  \citenamefont {Barack}, \citenamefont {Damour},\ and\ \citenamefont
  {Sago}}]{Akcay:2012ea}%
  \BibitemOpen
  \bibfield  {author} {\bibinfo {author} {\bibfnamefont {S.}~\bibnamefont
  {Akcay}}, \bibinfo {author} {\bibfnamefont {L.}~\bibnamefont {Barack}},
  \bibinfo {author} {\bibfnamefont {T.}~\bibnamefont {Damour}}, \ and\ \bibinfo
  {author} {\bibfnamefont {N.}~\bibnamefont {Sago}},\ }\href {\doibase
  10.1103/PhysRevD.86.104041} {\bibfield  {journal} {\bibinfo  {journal} {Phys.
  Rev.}\ }\textbf {\bibinfo {volume} {D86}},\ \bibinfo {pages} {104041}
  (\bibinfo {year} {2012})},\ \Eprint {http://arxiv.org/abs/1209.0964}
  {arXiv:1209.0964 [gr-qc]} \BibitemShut {NoStop}%
\bibitem [{\citenamefont {Akcay}\ and\ \citenamefont {van~de
  Meent}(2016)}]{Akcay:2015pjz}%
  \BibitemOpen
  \bibfield  {author} {\bibinfo {author} {\bibfnamefont {S.}~\bibnamefont
  {Akcay}}\ and\ \bibinfo {author} {\bibfnamefont {M.}~\bibnamefont {van~de
  Meent}},\ }\href {\doibase 10.1103/PhysRevD.93.064063} {\bibfield  {journal}
  {\bibinfo  {journal} {Phys. Rev.}\ }\textbf {\bibinfo {volume} {D93}},\
  \bibinfo {pages} {064063} (\bibinfo {year} {2016})},\ \Eprint
  {http://arxiv.org/abs/1512.03392} {arXiv:1512.03392 [gr-qc]} \BibitemShut
  {NoStop}%
\bibitem [{\citenamefont {Antonelli}\ \emph
  {et~al.}(2020{\natexlab{a}})\citenamefont {Antonelli}, \citenamefont {van~de
  Meent}, \citenamefont {Buonanno}, \citenamefont {Steinhoff},\ and\
  \citenamefont {Vines}}]{Antonelli:2019fmq}%
  \BibitemOpen
  \bibfield  {author} {\bibinfo {author} {\bibfnamefont {A.}~\bibnamefont
  {Antonelli}}, \bibinfo {author} {\bibfnamefont {M.}~\bibnamefont {van~de
  Meent}}, \bibinfo {author} {\bibfnamefont {A.}~\bibnamefont {Buonanno}},
  \bibinfo {author} {\bibfnamefont {J.}~\bibnamefont {Steinhoff}}, \ and\
  \bibinfo {author} {\bibfnamefont {J.}~\bibnamefont {Vines}},\ }\href
  {\doibase 10.1103/PhysRevD.101.024024} {\bibfield  {journal} {\bibinfo
  {journal} {Phys. Rev.}\ }\textbf {\bibinfo {volume} {D101}},\ \bibinfo
  {pages} {024024} (\bibinfo {year} {2020}{\natexlab{a}})},\ \Eprint
  {http://arxiv.org/abs/1907.11597} {arXiv:1907.11597 [gr-qc]} \BibitemShut
  {NoStop}%
\bibitem [{\citenamefont {Damour}(2016)}]{Damour:2016gwp}%
  \BibitemOpen
  \bibfield  {author} {\bibinfo {author} {\bibfnamefont {T.}~\bibnamefont
  {Damour}},\ }\href {\doibase 10.1103/PhysRevD.94.104015} {\bibfield
  {journal} {\bibinfo  {journal} {Phys. Rev.}\ }\textbf {\bibinfo {volume}
  {D94}},\ \bibinfo {pages} {104015} (\bibinfo {year} {2016})},\ \Eprint
  {http://arxiv.org/abs/1609.00354} {arXiv:1609.00354 [gr-qc]} \BibitemShut
  {NoStop}%
\bibitem [{\citenamefont {Damour}(2018)}]{Damour:2017zjx}%
  \BibitemOpen
  \bibfield  {author} {\bibinfo {author} {\bibfnamefont {T.}~\bibnamefont
  {Damour}},\ }\href {\doibase 10.1103/PhysRevD.97.044038} {\bibfield
  {journal} {\bibinfo  {journal} {Phys. Rev.}\ }\textbf {\bibinfo {volume}
  {D97}},\ \bibinfo {pages} {044038} (\bibinfo {year} {2018})},\ \Eprint
  {http://arxiv.org/abs/1710.10599} {arXiv:1710.10599 [gr-qc]} \BibitemShut
  {NoStop}%
\bibitem [{\citenamefont {Bjerrum-Bohr}\ \emph {et~al.}(2018)\citenamefont
  {Bjerrum-Bohr}, \citenamefont {Damgaard}, \citenamefont {Festuccia},
  \citenamefont {Plant\'e},\ and\ \citenamefont
  {Vanhove}}]{Bjerrum-Bohr:2018xdl}%
  \BibitemOpen
  \bibfield  {author} {\bibinfo {author} {\bibfnamefont {N.~J.}\ \bibnamefont
  {Bjerrum-Bohr}}, \bibinfo {author} {\bibfnamefont {P.~H.}\ \bibnamefont
  {Damgaard}}, \bibinfo {author} {\bibfnamefont {G.}~\bibnamefont {Festuccia}},
  \bibinfo {author} {\bibfnamefont {L.}~\bibnamefont {Plant\'e}}, \ and\
  \bibinfo {author} {\bibfnamefont {P.}~\bibnamefont {Vanhove}},\ }\href
  {\doibase 10.1103/PhysRevLett.121.171601} {\bibfield  {journal} {\bibinfo
  {journal} {Phys. Rev. Lett.}\ }\textbf {\bibinfo {volume} {121}},\ \bibinfo
  {pages} {171601} (\bibinfo {year} {2018})},\ \Eprint
  {http://arxiv.org/abs/1806.04920} {arXiv:1806.04920 [hep-th]} \BibitemShut
  {NoStop}%
\bibitem [{\citenamefont {Cheung}\ \emph {et~al.}(2018)\citenamefont {Cheung},
  \citenamefont {Rothstein},\ and\ \citenamefont {Solon}}]{Cheung:2018wkq}%
  \BibitemOpen
  \bibfield  {author} {\bibinfo {author} {\bibfnamefont {C.}~\bibnamefont
  {Cheung}}, \bibinfo {author} {\bibfnamefont {I.~Z.}\ \bibnamefont
  {Rothstein}}, \ and\ \bibinfo {author} {\bibfnamefont {M.~P.}\ \bibnamefont
  {Solon}},\ }\href {\doibase 10.1103/PhysRevLett.121.251101} {\bibfield
  {journal} {\bibinfo  {journal} {Phys. Rev. Lett.}\ }\textbf {\bibinfo
  {volume} {121}},\ \bibinfo {pages} {251101} (\bibinfo {year} {2018})},\
  \Eprint {http://arxiv.org/abs/1808.02489} {arXiv:1808.02489 [hep-th]}
  \BibitemShut {NoStop}%
\bibitem [{\citenamefont {Bern}\ \emph
  {et~al.}(2019{\natexlab{a}})\citenamefont {Bern}, \citenamefont {Cheung},
  \citenamefont {Roiban}, \citenamefont {Shen}, \citenamefont {Solon},\ and\
  \citenamefont {Zeng}}]{Bern:2019nnu}%
  \BibitemOpen
  \bibfield  {author} {\bibinfo {author} {\bibfnamefont {Z.}~\bibnamefont
  {Bern}}, \bibinfo {author} {\bibfnamefont {C.}~\bibnamefont {Cheung}},
  \bibinfo {author} {\bibfnamefont {R.}~\bibnamefont {Roiban}}, \bibinfo
  {author} {\bibfnamefont {C.-H.}\ \bibnamefont {Shen}}, \bibinfo {author}
  {\bibfnamefont {M.~P.}\ \bibnamefont {Solon}}, \ and\ \bibinfo {author}
  {\bibfnamefont {M.}~\bibnamefont {Zeng}},\ }\href {\doibase
  10.1103/PhysRevLett.122.201603} {\bibfield  {journal} {\bibinfo  {journal}
  {Phys. Rev. Lett.}\ }\textbf {\bibinfo {volume} {122}},\ \bibinfo {pages}
  {201603} (\bibinfo {year} {2019}{\natexlab{a}})},\ \Eprint
  {http://arxiv.org/abs/1901.04424} {arXiv:1901.04424 [hep-th]} \BibitemShut
  {NoStop}%
\bibitem [{\citenamefont {Bern}\ \emph
  {et~al.}(2019{\natexlab{b}})\citenamefont {Bern}, \citenamefont {Cheung},
  \citenamefont {Roiban}, \citenamefont {Shen}, \citenamefont {Solon},\ and\
  \citenamefont {Zeng}}]{Bern:2019crd}%
  \BibitemOpen
  \bibfield  {author} {\bibinfo {author} {\bibfnamefont {Z.}~\bibnamefont
  {Bern}}, \bibinfo {author} {\bibfnamefont {C.}~\bibnamefont {Cheung}},
  \bibinfo {author} {\bibfnamefont {R.}~\bibnamefont {Roiban}}, \bibinfo
  {author} {\bibfnamefont {C.-H.}\ \bibnamefont {Shen}}, \bibinfo {author}
  {\bibfnamefont {M.~P.}\ \bibnamefont {Solon}}, \ and\ \bibinfo {author}
  {\bibfnamefont {M.}~\bibnamefont {Zeng}},\ }\href {\doibase
  10.1007/JHEP10(2019)206} {\bibfield  {journal} {\bibinfo  {journal} {JHEP}\
  }\textbf {\bibinfo {volume} {10}},\ \bibinfo {pages} {206} (\bibinfo {year}
  {2019}{\natexlab{b}})},\ \Eprint {http://arxiv.org/abs/1908.01493}
  {arXiv:1908.01493 [hep-th]} \BibitemShut {NoStop}%
\bibitem [{\citenamefont {Westpfahl}\ and\ \citenamefont
  {Hoyler}(1980)}]{Westpfahl:1980mk}%
  \BibitemOpen
  \bibfield  {author} {\bibinfo {author} {\bibfnamefont {K.}~\bibnamefont
  {Westpfahl}}\ and\ \bibinfo {author} {\bibfnamefont {H.}~\bibnamefont
  {Hoyler}},\ }\href {\doibase 10.1007/BF02750304} {\bibfield  {journal}
  {\bibinfo  {journal} {Lett. Nuovo Cim.}\ }\textbf {\bibinfo {volume} {27}},\
  \bibinfo {pages} {581} (\bibinfo {year} {1980})}\BibitemShut {NoStop}%
\bibitem [{\citenamefont {Westpfahl}\ and\ \citenamefont
  {Goller}(1979)}]{Westpfahl:1979gu}%
  \BibitemOpen
  \bibfield  {author} {\bibinfo {author} {\bibfnamefont {K.}~\bibnamefont
  {Westpfahl}}\ and\ \bibinfo {author} {\bibfnamefont {M.}~\bibnamefont
  {Goller}},\ }\href {\doibase 10.1007/BF02817047} {\bibfield  {journal}
  {\bibinfo  {journal} {Lett. Nuovo Cim.}\ }\textbf {\bibinfo {volume} {26}},\
  \bibinfo {pages} {573} (\bibinfo {year} {1979})}\BibitemShut {NoStop}%
\bibitem [{\citenamefont {Antonelli}\ \emph {et~al.}(2019)\citenamefont
  {Antonelli}, \citenamefont {Buonanno}, \citenamefont {Steinhoff},
  \citenamefont {van~de Meent},\ and\ \citenamefont
  {Vines}}]{Antonelli:2019ytb}%
  \BibitemOpen
  \bibfield  {author} {\bibinfo {author} {\bibfnamefont {A.}~\bibnamefont
  {Antonelli}}, \bibinfo {author} {\bibfnamefont {A.}~\bibnamefont {Buonanno}},
  \bibinfo {author} {\bibfnamefont {J.}~\bibnamefont {Steinhoff}}, \bibinfo
  {author} {\bibfnamefont {M.}~\bibnamefont {van~de Meent}}, \ and\ \bibinfo
  {author} {\bibfnamefont {J.}~\bibnamefont {Vines}},\ }\href {\doibase
  10.1103/PhysRevD.99.104004} {\bibfield  {journal} {\bibinfo  {journal} {Phys.
  Rev.}\ }\textbf {\bibinfo {volume} {D99}},\ \bibinfo {pages} {104004}
  (\bibinfo {year} {2019})},\ \Eprint {http://arxiv.org/abs/1901.07102}
  {arXiv:1901.07102 [gr-qc]} \BibitemShut {NoStop}%
\bibitem [{\citenamefont {Kälin}\ and\ \citenamefont
  {Porto}(2020{\natexlab{a}})}]{Kalin:2019rwq}%
  \BibitemOpen
  \bibfield  {author} {\bibinfo {author} {\bibfnamefont {G.}~\bibnamefont
  {Kälin}}\ and\ \bibinfo {author} {\bibfnamefont {R.~A.}\ \bibnamefont
  {Porto}},\ }\href {\doibase 10.1007/JHEP01(2020)072} {\bibfield  {journal}
  {\bibinfo  {journal} {JHEP}\ }\textbf {\bibinfo {volume} {01}},\ \bibinfo
  {pages} {072} (\bibinfo {year} {2020}{\natexlab{a}})},\ \Eprint
  {http://arxiv.org/abs/1910.03008} {arXiv:1910.03008 [hep-th]} \BibitemShut
  {NoStop}%
\bibitem [{\citenamefont {Kälin}\ and\ \citenamefont
  {Porto}(2020{\natexlab{b}})}]{Kalin:2019inp}%
  \BibitemOpen
  \bibfield  {author} {\bibinfo {author} {\bibfnamefont {G.}~\bibnamefont
  {Kälin}}\ and\ \bibinfo {author} {\bibfnamefont {R.~A.}\ \bibnamefont
  {Porto}},\ }\href {\doibase 10.1007/JHEP02(2020)120} {\bibfield  {journal}
  {\bibinfo  {journal} {JHEP}\ }\textbf {\bibinfo {volume} {02}},\ \bibinfo
  {pages} {120} (\bibinfo {year} {2020}{\natexlab{b}})},\ \Eprint
  {http://arxiv.org/abs/1911.09130} {arXiv:1911.09130 [hep-th]} \BibitemShut
  {NoStop}%
\bibitem [{\citenamefont {Cheung}\ and\ \citenamefont
  {Solon}(2020)}]{Cheung:2020gyp}%
  \BibitemOpen
  \bibfield  {author} {\bibinfo {author} {\bibfnamefont {C.}~\bibnamefont
  {Cheung}}\ and\ \bibinfo {author} {\bibfnamefont {M.~P.}\ \bibnamefont
  {Solon}},\ }\href@noop {} {\  (\bibinfo {year} {2020})},\ \Eprint
  {http://arxiv.org/abs/2003.08351} {arXiv:2003.08351 [hep-th]} \BibitemShut
  {NoStop}%
\bibitem [{\citenamefont {Blümlein}\ \emph
  {et~al.}(2020{\natexlab{c}})\citenamefont {Blümlein}, \citenamefont {Maier},
  \citenamefont {Marquard},\ and\ \citenamefont {Schäfer}}]{Blumlein:2020znm}%
  \BibitemOpen
  \bibfield  {author} {\bibinfo {author} {\bibfnamefont {J.}~\bibnamefont
  {Blümlein}}, \bibinfo {author} {\bibfnamefont {A.}~\bibnamefont {Maier}},
  \bibinfo {author} {\bibfnamefont {P.}~\bibnamefont {Marquard}}, \ and\
  \bibinfo {author} {\bibfnamefont {G.}~\bibnamefont {Schäfer}},\ }\href@noop
  {} {\  (\bibinfo {year} {2020}{\natexlab{c}})},\ \Eprint
  {http://arxiv.org/abs/2003.07145} {arXiv:2003.07145 [gr-qc]} \BibitemShut
  {NoStop}%
\bibitem [{\citenamefont {K\"alin}\ \emph {et~al.}(2020)\citenamefont
  {K\"alin}, \citenamefont {Liu},\ and\ \citenamefont {Porto}}]{Kalin:2020fhe}%
  \BibitemOpen
  \bibfield  {author} {\bibinfo {author} {\bibfnamefont {G.}~\bibnamefont
  {K\"alin}}, \bibinfo {author} {\bibfnamefont {Z.}~\bibnamefont {Liu}}, \ and\
  \bibinfo {author} {\bibfnamefont {R.~A.}\ \bibnamefont {Porto}},\ }\href@noop
  {} {\  (\bibinfo {year} {2020})},\ \Eprint {http://arxiv.org/abs/2007.04977}
  {arXiv:2007.04977 [hep-th]} \BibitemShut {NoStop}%
\bibitem [{\citenamefont {Bini}\ and\ \citenamefont
  {Damour}(2017{\natexlab{a}})}]{Bini:2017xzy}%
  \BibitemOpen
  \bibfield  {author} {\bibinfo {author} {\bibfnamefont {D.}~\bibnamefont
  {Bini}}\ and\ \bibinfo {author} {\bibfnamefont {T.}~\bibnamefont {Damour}},\
  }\href {\doibase 10.1103/PhysRevD.96.104038} {\bibfield  {journal} {\bibinfo
  {journal} {Phys. Rev.}\ }\textbf {\bibinfo {volume} {D96}},\ \bibinfo {pages}
  {104038} (\bibinfo {year} {2017}{\natexlab{a}})},\ \Eprint
  {http://arxiv.org/abs/1709.00590} {arXiv:1709.00590 [gr-qc]} \BibitemShut
  {NoStop}%
\bibitem [{\citenamefont {Bini}\ and\ \citenamefont
  {Damour}(2018)}]{Bini:2018ywr}%
  \BibitemOpen
  \bibfield  {author} {\bibinfo {author} {\bibfnamefont {D.}~\bibnamefont
  {Bini}}\ and\ \bibinfo {author} {\bibfnamefont {T.}~\bibnamefont {Damour}},\
  }\href {\doibase 10.1103/PhysRevD.98.044036} {\bibfield  {journal} {\bibinfo
  {journal} {Phys. Rev.}\ }\textbf {\bibinfo {volume} {D98}},\ \bibinfo {pages}
  {044036} (\bibinfo {year} {2018})},\ \Eprint
  {http://arxiv.org/abs/1805.10809} {arXiv:1805.10809 [gr-qc]} \BibitemShut
  {NoStop}%
\bibitem [{\citenamefont {Bern}\ \emph {et~al.}(2020)\citenamefont {Bern},
  \citenamefont {Luna}, \citenamefont {Roiban}, \citenamefont {Shen},\ and\
  \citenamefont {Zeng}}]{Bern:2020buy}%
  \BibitemOpen
  \bibfield  {author} {\bibinfo {author} {\bibfnamefont {Z.}~\bibnamefont
  {Bern}}, \bibinfo {author} {\bibfnamefont {A.}~\bibnamefont {Luna}}, \bibinfo
  {author} {\bibfnamefont {R.}~\bibnamefont {Roiban}}, \bibinfo {author}
  {\bibfnamefont {C.-H.}\ \bibnamefont {Shen}}, \ and\ \bibinfo {author}
  {\bibfnamefont {M.}~\bibnamefont {Zeng}},\ }\href@noop {} {\  (\bibinfo
  {year} {2020})},\ \Eprint {http://arxiv.org/abs/2005.03071} {arXiv:2005.03071
  [hep-th]} \BibitemShut {NoStop}%
\bibitem [{\citenamefont {Aoude}\ \emph {et~al.}(2020)\citenamefont {Aoude},
  \citenamefont {Haddad},\ and\ \citenamefont {Helset}}]{Aoude:2020onz}%
  \BibitemOpen
  \bibfield  {author} {\bibinfo {author} {\bibfnamefont {R.}~\bibnamefont
  {Aoude}}, \bibinfo {author} {\bibfnamefont {K.}~\bibnamefont {Haddad}}, \
  and\ \bibinfo {author} {\bibfnamefont {A.}~\bibnamefont {Helset}},\ }\href
  {\doibase 10.1007/JHEP05(2020)051} {\bibfield  {journal} {\bibinfo  {journal}
  {JHEP}\ }\textbf {\bibinfo {volume} {05}},\ \bibinfo {pages} {051} (\bibinfo
  {year} {2020})},\ \Eprint {http://arxiv.org/abs/2001.09164} {arXiv:2001.09164
  [hep-th]} \BibitemShut {NoStop}%
\bibitem [{\citenamefont {Chung}\ \emph {et~al.}(2020)\citenamefont {Chung},
  \citenamefont {Huang}, \citenamefont {Kim},\ and\ \citenamefont
  {Lee}}]{Chung:2020rrz}%
  \BibitemOpen
  \bibfield  {author} {\bibinfo {author} {\bibfnamefont {M.-Z.}\ \bibnamefont
  {Chung}}, \bibinfo {author} {\bibfnamefont {Y.-t.}\ \bibnamefont {Huang}},
  \bibinfo {author} {\bibfnamefont {J.-W.}\ \bibnamefont {Kim}}, \ and\
  \bibinfo {author} {\bibfnamefont {S.}~\bibnamefont {Lee}},\ }\href {\doibase
  10.1007/JHEP05(2020)105} {\bibfield  {journal} {\bibinfo  {journal} {JHEP}\
  }\textbf {\bibinfo {volume} {05}},\ \bibinfo {pages} {105} (\bibinfo {year}
  {2020})},\ \Eprint {http://arxiv.org/abs/2003.06600} {arXiv:2003.06600
  [hep-th]} \BibitemShut {NoStop}%
\bibitem [{\citenamefont {Damour}(2019)}]{Damour:2019lcq}%
  \BibitemOpen
  \bibfield  {author} {\bibinfo {author} {\bibfnamefont {T.}~\bibnamefont
  {Damour}},\ }\href@noop {} {\  (\bibinfo {year} {2019})},\ \Eprint
  {http://arxiv.org/abs/1912.02139} {arXiv:1912.02139 [gr-qc]} \BibitemShut
  {NoStop}%
\bibitem [{\citenamefont {Vines}\ \emph {et~al.}(2019)\citenamefont {Vines},
  \citenamefont {Steinhoff},\ and\ \citenamefont {Buonanno}}]{Vines:2018gqi}%
  \BibitemOpen
  \bibfield  {author} {\bibinfo {author} {\bibfnamefont {J.}~\bibnamefont
  {Vines}}, \bibinfo {author} {\bibfnamefont {J.}~\bibnamefont {Steinhoff}}, \
  and\ \bibinfo {author} {\bibfnamefont {A.}~\bibnamefont {Buonanno}},\ }\href
  {\doibase 10.1103/PhysRevD.99.064054} {\bibfield  {journal} {\bibinfo
  {journal} {Phys. Rev.}\ }\textbf {\bibinfo {volume} {D99}},\ \bibinfo {pages}
  {064054} (\bibinfo {year} {2019})},\ \Eprint
  {http://arxiv.org/abs/1812.00956} {arXiv:1812.00956 [gr-qc]} \BibitemShut
  {NoStop}%
\bibitem [{\citenamefont {Antonelli}\ \emph
  {et~al.}(2020{\natexlab{b}})\citenamefont {Antonelli}, \citenamefont
  {Kavanagh}, \citenamefont {Khalil}, \citenamefont {Steinhoff},\ and\
  \citenamefont {Vines}}]{Antonelli:2020aeb}%
  \BibitemOpen
  \bibfield  {author} {\bibinfo {author} {\bibfnamefont {A.}~\bibnamefont
  {Antonelli}}, \bibinfo {author} {\bibfnamefont {C.}~\bibnamefont {Kavanagh}},
  \bibinfo {author} {\bibfnamefont {M.}~\bibnamefont {Khalil}}, \bibinfo
  {author} {\bibfnamefont {J.}~\bibnamefont {Steinhoff}}, \ and\ \bibinfo
  {author} {\bibfnamefont {J.}~\bibnamefont {Vines}},\ }\href {\doibase
  10.1103/PhysRevLett.125.011103} {\bibfield  {journal} {\bibinfo  {journal}
  {Phys. Rev. Lett.}\ }\textbf {\bibinfo {volume} {125}},\ \bibinfo {pages}
  {011103} (\bibinfo {year} {2020}{\natexlab{b}})},\ \Eprint
  {http://arxiv.org/abs/2003.11391} {arXiv:2003.11391 [gr-qc]} \BibitemShut
  {NoStop}%
\bibitem [{\citenamefont {Blanchet}\ \emph {et~al.}(2013)\citenamefont
  {Blanchet}, \citenamefont {Buonanno},\ and\ \citenamefont
  {Le~Tiec}}]{Blanchet:2012at}%
  \BibitemOpen
  \bibfield  {author} {\bibinfo {author} {\bibfnamefont {L.}~\bibnamefont
  {Blanchet}}, \bibinfo {author} {\bibfnamefont {A.}~\bibnamefont {Buonanno}},
  \ and\ \bibinfo {author} {\bibfnamefont {A.}~\bibnamefont {Le~Tiec}},\ }\href
  {\doibase 10.1103/PhysRevD.87.024030} {\bibfield  {journal} {\bibinfo
  {journal} {Phys. Rev.}\ }\textbf {\bibinfo {volume} {D87}},\ \bibinfo {pages}
  {024030} (\bibinfo {year} {2013})},\ \Eprint {http://arxiv.org/abs/1211.1060}
  {arXiv:1211.1060 [gr-qc]} \BibitemShut {NoStop}%
\bibitem [{\citenamefont {Le~Tiec}(2015)}]{Tiec:2015cxa}%
  \BibitemOpen
  \bibfield  {author} {\bibinfo {author} {\bibfnamefont {A.}~\bibnamefont
  {Le~Tiec}},\ }\href {\doibase 10.1103/PhysRevD.92.084021} {\bibfield
  {journal} {\bibinfo  {journal} {Phys. Rev.}\ }\textbf {\bibinfo {volume}
  {D92}},\ \bibinfo {pages} {084021} (\bibinfo {year} {2015})},\ \Eprint
  {http://arxiv.org/abs/1506.05648} {arXiv:1506.05648 [gr-qc]} \BibitemShut
  {NoStop}%
\bibitem [{\citenamefont {Mathisson}(1937)}]{Mathisson:1937zz}%
  \BibitemOpen
  \bibfield  {author} {\bibinfo {author} {\bibfnamefont {M.}~\bibnamefont
  {Mathisson}},\ }\href@noop {} {\bibfield  {journal} {\bibinfo  {journal}
  {Acta Phys.\ Polon.}\ }\textbf {\bibinfo {volume} {6}},\ \bibinfo {pages}
  {163} (\bibinfo {year} {1937})}\BibitemShut {NoStop}%
\bibitem [{\citenamefont {Papapetrou}(1951)}]{Papapetrou:1951pa}%
  \BibitemOpen
  \bibfield  {author} {\bibinfo {author} {\bibfnamefont {A.}~\bibnamefont
  {Papapetrou}},\ }\href {\doibase 10.1098/rspa.1951.0200} {\bibfield
  {journal} {\bibinfo  {journal} {Proc.\ Roy.\ Soc.\ Lond.\ A}\ }\textbf
  {\bibinfo {volume} {A209}},\ \bibinfo {pages} {248} (\bibinfo {year}
  {1951})}\BibitemShut {NoStop}%
\bibitem [{\citenamefont {Dixon}(1970)}]{Dixon:1970zza}%
  \BibitemOpen
  \bibfield  {author} {\bibinfo {author} {\bibfnamefont {W.}~\bibnamefont
  {Dixon}},\ }\href {\doibase 10.1098/rspa.1970.0020} {\bibfield  {journal}
  {\bibinfo  {journal} {Proc. Roy. Soc. Lond. A}\ }\textbf {\bibinfo {volume}
  {314}},\ \bibinfo {pages} {499} (\bibinfo {year} {1970})}\BibitemShut
  {NoStop}%
\bibitem [{\citenamefont {{Dixon}}(1979)}]{Dixon:1979}%
  \BibitemOpen
  \bibfield  {author} {\bibinfo {author} {\bibfnamefont {W.~G.}\ \bibnamefont
  {{Dixon}}},\ }in\ \href@noop {} {\emph {\bibinfo {booktitle} {Isolated
  Gravitating Systems in General Relativity}}},\ \bibinfo {editor} {edited by\
  \bibinfo {editor} {\bibfnamefont {J.}~\bibnamefont {{Ehlers}}}}\ (\bibinfo
  {year} {1979})\ pp.\ \bibinfo {pages} {156--219}\BibitemShut {NoStop}%
\bibitem [{\citenamefont {Steinhoff}(2015)}]{Steinhoff:2014kwa}%
  \BibitemOpen
  \bibfield  {author} {\bibinfo {author} {\bibfnamefont {J.}~\bibnamefont
  {Steinhoff}},\ }\bibfield  {booktitle} {\emph {\bibinfo {booktitle}
  {{Proceedings, 524th WE-Heraeus-Seminar: Equations of Motion in Relativistic
  Gravity (EOM 2013)}}},\ }\href {\doibase 10.1007/978-3-319-18335-0_19}
  {\bibfield  {journal} {\bibinfo  {journal} {Fund. Theor. Phys.}\ }\textbf
  {\bibinfo {volume} {179}},\ \bibinfo {pages} {615} (\bibinfo {year}
  {2015})},\ \Eprint {http://arxiv.org/abs/1412.3251} {arXiv:1412.3251 [gr-qc]}
  \BibitemShut {NoStop}%
\bibitem [{\citenamefont {Tulczyjew}(1959)}]{Tulczyjew:1959}%
  \BibitemOpen
  \bibfield  {author} {\bibinfo {author} {\bibfnamefont {W.}~\bibnamefont
  {Tulczyjew}},\ }\href@noop {} {\bibfield  {journal} {\bibinfo  {journal}
  {Acta Phys. Polon.}\ }\textbf {\bibinfo {volume} {18}},\ \bibinfo {pages}
  {37} (\bibinfo {year} {1959})},\ \bibinfo {note} {[Erratum: Acta Phys. Polon.
  \textbf{18}, 534 (1959)]}\BibitemShut {NoStop}%
\bibitem [{\citenamefont {Fokker}(1929)}]{fokker1929relativiteitstheorie}%
  \BibitemOpen
  \bibfield  {author} {\bibinfo {author} {\bibfnamefont {A.~D.}\ \bibnamefont
  {Fokker}},\ }\href@noop {} {\emph {\bibinfo {title} {Relativiteitstheorie}}}\
  (\bibinfo  {publisher} {P. Noordhoff},\ \bibinfo {year} {1929})\BibitemShut
  {NoStop}%
\bibitem [{\citenamefont {Vines}(2018)}]{Vines:2017hyw}%
  \BibitemOpen
  \bibfield  {author} {\bibinfo {author} {\bibfnamefont {J.}~\bibnamefont
  {Vines}},\ }\href {\doibase 10.1088/1361-6382/aaa3a8} {\bibfield  {journal}
  {\bibinfo  {journal} {Class. Quant. Grav.}\ }\textbf {\bibinfo {volume}
  {35}},\ \bibinfo {pages} {084002} (\bibinfo {year} {2018})},\ \Eprint
  {http://arxiv.org/abs/1709.06016} {arXiv:1709.06016 [gr-qc]} \BibitemShut
  {NoStop}%
\bibitem [{\citenamefont {Maybee}\ \emph {et~al.}(2019)\citenamefont {Maybee},
  \citenamefont {O'Connell},\ and\ \citenamefont {Vines}}]{Maybee:2019jus}%
  \BibitemOpen
  \bibfield  {author} {\bibinfo {author} {\bibfnamefont {B.}~\bibnamefont
  {Maybee}}, \bibinfo {author} {\bibfnamefont {D.}~\bibnamefont {O'Connell}}, \
  and\ \bibinfo {author} {\bibfnamefont {J.}~\bibnamefont {Vines}},\ }\href
  {\doibase 10.1007/JHEP12(2019)156} {\bibfield  {journal} {\bibinfo  {journal}
  {JHEP}\ }\textbf {\bibinfo {volume} {12}},\ \bibinfo {pages} {156} (\bibinfo
  {year} {2019})},\ \Eprint {http://arxiv.org/abs/1906.09260} {arXiv:1906.09260
  [hep-th]} \BibitemShut {NoStop}%
\bibitem [{\citenamefont {Guevara}\ \emph
  {et~al.}(2019{\natexlab{a}})\citenamefont {Guevara}, \citenamefont
  {Ochirov},\ and\ \citenamefont {Vines}}]{Guevara:2019fsj}%
  \BibitemOpen
  \bibfield  {author} {\bibinfo {author} {\bibfnamefont {A.}~\bibnamefont
  {Guevara}}, \bibinfo {author} {\bibfnamefont {A.}~\bibnamefont {Ochirov}}, \
  and\ \bibinfo {author} {\bibfnamefont {J.}~\bibnamefont {Vines}},\ }\href
  {\doibase 10.1103/PhysRevD.100.104024} {\bibfield  {journal} {\bibinfo
  {journal} {Phys. Rev.}\ }\textbf {\bibinfo {volume} {D100}},\ \bibinfo
  {pages} {104024} (\bibinfo {year} {2019}{\natexlab{a}})},\ \Eprint
  {http://arxiv.org/abs/1906.10071} {arXiv:1906.10071 [hep-th]} \BibitemShut
  {NoStop}%
\bibitem [{\citenamefont {Bini}\ \emph {et~al.}(2017)\citenamefont {Bini},
  \citenamefont {Geralico},\ and\ \citenamefont {Vines}}]{Bini:2017pee}%
  \BibitemOpen
  \bibfield  {author} {\bibinfo {author} {\bibfnamefont {D.}~\bibnamefont
  {Bini}}, \bibinfo {author} {\bibfnamefont {A.}~\bibnamefont {Geralico}}, \
  and\ \bibinfo {author} {\bibfnamefont {J.}~\bibnamefont {Vines}},\ }\href
  {\doibase 10.1103/PhysRevD.96.084044} {\bibfield  {journal} {\bibinfo
  {journal} {Phys. Rev.}\ }\textbf {\bibinfo {volume} {D96}},\ \bibinfo {pages}
  {084044} (\bibinfo {year} {2017})},\ \Eprint
  {http://arxiv.org/abs/1707.09814} {arXiv:1707.09814 [gr-qc]} \BibitemShut
  {NoStop}%
\bibitem [{\citenamefont {Guevara}\ \emph
  {et~al.}(2019{\natexlab{b}})\citenamefont {Guevara}, \citenamefont
  {Ochirov},\ and\ \citenamefont {Vines}}]{Guevara:2018wpp}%
  \BibitemOpen
  \bibfield  {author} {\bibinfo {author} {\bibfnamefont {A.}~\bibnamefont
  {Guevara}}, \bibinfo {author} {\bibfnamefont {A.}~\bibnamefont {Ochirov}}, \
  and\ \bibinfo {author} {\bibfnamefont {J.}~\bibnamefont {Vines}},\ }\href
  {\doibase 10.1007/JHEP09(2019)056} {\bibfield  {journal} {\bibinfo  {journal}
  {JHEP}\ }\textbf {\bibinfo {volume} {09}},\ \bibinfo {pages} {056} (\bibinfo
  {year} {2019}{\natexlab{b}})},\ \Eprint {http://arxiv.org/abs/1812.06895}
  {arXiv:1812.06895 [hep-th]} \BibitemShut {NoStop}%
\bibitem [{\citenamefont {Pryce}(1948)}]{pryce1948mass}%
  \BibitemOpen
  \bibfield  {author} {\bibinfo {author} {\bibfnamefont {M.~H.~L.}\
  \bibnamefont {Pryce}},\ }\href@noop {} {\bibfield  {journal} {\bibinfo
  {journal} {Proceedings of the Royal Society of London. Series A. Mathematical
  and Physical Sciences}\ }\textbf {\bibinfo {volume} {195}},\ \bibinfo {pages}
  {62} (\bibinfo {year} {1948})}\BibitemShut {NoStop}%
\bibitem [{\citenamefont {Newton}\ and\ \citenamefont
  {Wigner}(1949)}]{newton1949localized}%
  \BibitemOpen
  \bibfield  {author} {\bibinfo {author} {\bibfnamefont {T.~D.}\ \bibnamefont
  {Newton}}\ and\ \bibinfo {author} {\bibfnamefont {E.~P.}\ \bibnamefont
  {Wigner}},\ }\href@noop {} {\bibfield  {journal} {\bibinfo  {journal}
  {Reviews of Modern Physics}\ }\textbf {\bibinfo {volume} {21}},\ \bibinfo
  {pages} {400} (\bibinfo {year} {1949})}\BibitemShut {NoStop}%
\bibitem [{\citenamefont {Barausse}\ \emph {et~al.}(2009)\citenamefont
  {Barausse}, \citenamefont {Racine},\ and\ \citenamefont
  {Buonanno}}]{Barausse:2009aa}%
  \BibitemOpen
  \bibfield  {author} {\bibinfo {author} {\bibfnamefont {E.}~\bibnamefont
  {Barausse}}, \bibinfo {author} {\bibfnamefont {E.}~\bibnamefont {Racine}}, \
  and\ \bibinfo {author} {\bibfnamefont {A.}~\bibnamefont {Buonanno}},\ }\href
  {\doibase 10.1103/PhysRevD.85.069904, 10.1103/PhysRevD.80.104025} {\bibfield
  {journal} {\bibinfo  {journal} {Phys. Rev.}\ }\textbf {\bibinfo {volume}
  {D80}},\ \bibinfo {pages} {104025} (\bibinfo {year} {2009})},\ \bibinfo
  {note} {[Erratum: Phys. Rev. \textbf{D85}, 069904 (2012)]},\ \Eprint
  {http://arxiv.org/abs/0907.4745} {arXiv:0907.4745 [gr-qc]} \BibitemShut
  {NoStop}%
\bibitem [{\citenamefont {Vines}\ \emph {et~al.}(2016)\citenamefont {Vines},
  \citenamefont {Kunst}, \citenamefont {Steinhoff},\ and\ \citenamefont
  {Hinderer}}]{Vines:2016unv}%
  \BibitemOpen
  \bibfield  {author} {\bibinfo {author} {\bibfnamefont {J.}~\bibnamefont
  {Vines}}, \bibinfo {author} {\bibfnamefont {D.}~\bibnamefont {Kunst}},
  \bibinfo {author} {\bibfnamefont {J.}~\bibnamefont {Steinhoff}}, \ and\
  \bibinfo {author} {\bibfnamefont {T.}~\bibnamefont {Hinderer}},\ }\href
  {\doibase 10.1103/PhysRevD.93.103008} {\bibfield  {journal} {\bibinfo
  {journal} {Phys. Rev.}\ }\textbf {\bibinfo {volume} {D93}},\ \bibinfo {pages}
  {103008} (\bibinfo {year} {2016})},\ \Eprint
  {http://arxiv.org/abs/1601.07529} {arXiv:1601.07529 [gr-qc]} \BibitemShut
  {NoStop}%
\bibitem [{\citenamefont {Bini}\ and\ \citenamefont
  {Damour}(2017{\natexlab{b}})}]{Bini:2017wfr}%
  \BibitemOpen
  \bibfield  {author} {\bibinfo {author} {\bibfnamefont {D.}~\bibnamefont
  {Bini}}\ and\ \bibinfo {author} {\bibfnamefont {T.}~\bibnamefont {Damour}},\
  }\href {\doibase 10.1103/PhysRevD.96.064021} {\bibfield  {journal} {\bibinfo
  {journal} {Phys. Rev. D}\ }\textbf {\bibinfo {volume} {96}},\ \bibinfo
  {pages} {064021} (\bibinfo {year} {2017}{\natexlab{b}})},\ \Eprint
  {http://arxiv.org/abs/1706.06877} {arXiv:1706.06877 [gr-qc]} \BibitemShut
  {NoStop}%
\bibitem [{\citenamefont {Siemonsen}\ \emph {et~al.}(2018)\citenamefont
  {Siemonsen}, \citenamefont {Steinhoff},\ and\ \citenamefont
  {Vines}}]{Siemonsen:2017yux}%
  \BibitemOpen
  \bibfield  {author} {\bibinfo {author} {\bibfnamefont {N.}~\bibnamefont
  {Siemonsen}}, \bibinfo {author} {\bibfnamefont {J.}~\bibnamefont
  {Steinhoff}}, \ and\ \bibinfo {author} {\bibfnamefont {J.}~\bibnamefont
  {Vines}},\ }\href {\doibase 10.1103/PhysRevD.97.124046} {\bibfield  {journal}
  {\bibinfo  {journal} {Phys. Rev. D}\ }\textbf {\bibinfo {volume} {97}},\
  \bibinfo {pages} {124046} (\bibinfo {year} {2018})},\ \Eprint
  {http://arxiv.org/abs/1712.08603} {arXiv:1712.08603 [gr-qc]} \BibitemShut
  {NoStop}%
\bibitem [{\citenamefont {Kabat}\ and\ \citenamefont
  {Ortiz}(1992)}]{Kabat:1992tb}%
  \BibitemOpen
  \bibfield  {author} {\bibinfo {author} {\bibfnamefont {D.~N.}\ \bibnamefont
  {Kabat}}\ and\ \bibinfo {author} {\bibfnamefont {M.}~\bibnamefont {Ortiz}},\
  }\href {\doibase 10.1016/0550-3213(92)90627-N} {\bibfield  {journal}
  {\bibinfo  {journal} {Nucl. Phys. B}\ }\textbf {\bibinfo {volume} {388}},\
  \bibinfo {pages} {570} (\bibinfo {year} {1992})},\ \Eprint
  {http://arxiv.org/abs/hep-th/9203082} {arXiv:hep-th/9203082} \BibitemShut
  {NoStop}%
\bibitem [{\citenamefont {Akhoury}\ \emph {et~al.}(2013)\citenamefont
  {Akhoury}, \citenamefont {Saotome},\ and\ \citenamefont
  {Sterman}}]{Akhoury:2013yua}%
  \BibitemOpen
  \bibfield  {author} {\bibinfo {author} {\bibfnamefont {R.}~\bibnamefont
  {Akhoury}}, \bibinfo {author} {\bibfnamefont {R.}~\bibnamefont {Saotome}}, \
  and\ \bibinfo {author} {\bibfnamefont {G.}~\bibnamefont {Sterman}},\
  }\href@noop {} {\  (\bibinfo {year} {2013})},\ \Eprint
  {http://arxiv.org/abs/1308.5204} {arXiv:1308.5204 [hep-th]} \BibitemShut
  {NoStop}%
\bibitem [{\citenamefont {Bjerrum-Bohr}\ \emph {et~al.}(2019)\citenamefont
  {Bjerrum-Bohr}, \citenamefont {Cristofoli},\ and\ \citenamefont
  {Damgaard}}]{Bjerrum-Bohr:2019kec}%
  \BibitemOpen
  \bibfield  {author} {\bibinfo {author} {\bibfnamefont {N.}~\bibnamefont
  {Bjerrum-Bohr}}, \bibinfo {author} {\bibfnamefont {A.}~\bibnamefont
  {Cristofoli}}, \ and\ \bibinfo {author} {\bibfnamefont {P.~H.}\ \bibnamefont
  {Damgaard}},\ }\href@noop {} {\  (\bibinfo {year} {2019})},\ \Eprint
  {http://arxiv.org/abs/1910.09366} {arXiv:1910.09366 [hep-th]} \BibitemShut
  {NoStop}%
\bibitem [{\citenamefont {Blanchet}\ and\ \citenamefont
  {Le~Tiec}(2017)}]{Blanchet:2017rcn}%
  \BibitemOpen
  \bibfield  {author} {\bibinfo {author} {\bibfnamefont {L.}~\bibnamefont
  {Blanchet}}\ and\ \bibinfo {author} {\bibfnamefont {A.}~\bibnamefont
  {Le~Tiec}},\ }\href {\doibase 10.1088/1361-6382/aa79d7} {\bibfield  {journal}
  {\bibinfo  {journal} {Class. Quant. Grav.}\ }\textbf {\bibinfo {volume}
  {34}},\ \bibinfo {pages} {164001} (\bibinfo {year} {2017})},\ \Eprint
  {http://arxiv.org/abs/1702.06839} {arXiv:1702.06839 [gr-qc]} \BibitemShut
  {NoStop}%
\bibitem [{\citenamefont {Fujita}\ \emph {et~al.}(2017)\citenamefont {Fujita},
  \citenamefont {Isoyama}, \citenamefont {Le~Tiec}, \citenamefont {Nakano},
  \citenamefont {Sago},\ and\ \citenamefont {Tanaka}}]{Fujita:2016igj}%
  \BibitemOpen
  \bibfield  {author} {\bibinfo {author} {\bibfnamefont {R.}~\bibnamefont
  {Fujita}}, \bibinfo {author} {\bibfnamefont {S.}~\bibnamefont {Isoyama}},
  \bibinfo {author} {\bibfnamefont {A.}~\bibnamefont {Le~Tiec}}, \bibinfo
  {author} {\bibfnamefont {H.}~\bibnamefont {Nakano}}, \bibinfo {author}
  {\bibfnamefont {N.}~\bibnamefont {Sago}}, \ and\ \bibinfo {author}
  {\bibfnamefont {T.}~\bibnamefont {Tanaka}},\ }\href {\doibase
  10.1088/1361-6382/aa7342} {\bibfield  {journal} {\bibinfo  {journal} {Class.
  Quant. Grav.}\ }\textbf {\bibinfo {volume} {34}},\ \bibinfo {pages} {134001}
  (\bibinfo {year} {2017})},\ \Eprint {http://arxiv.org/abs/1612.02504}
  {arXiv:1612.02504 [gr-qc]} \BibitemShut {NoStop}%
\bibitem [{\citenamefont {Hinderer}\ and\ \citenamefont
  {Flanagan}(2008)}]{Hinderer:2008dm}%
  \BibitemOpen
  \bibfield  {author} {\bibinfo {author} {\bibfnamefont {T.}~\bibnamefont
  {Hinderer}}\ and\ \bibinfo {author} {\bibfnamefont {E.~E.}\ \bibnamefont
  {Flanagan}},\ }\href {\doibase 10.1103/PhysRevD.78.064028} {\bibfield
  {journal} {\bibinfo  {journal} {Phys. Rev.}\ }\textbf {\bibinfo {volume}
  {D78}},\ \bibinfo {pages} {064028} (\bibinfo {year} {2008})},\ \Eprint
  {http://arxiv.org/abs/0805.3337} {arXiv:0805.3337 [gr-qc]} \BibitemShut
  {NoStop}%
\bibitem [{\citenamefont {Bini}\ and\ \citenamefont
  {Geralico}(2019{\natexlab{b}})}]{Bini:2019lkm}%
  \BibitemOpen
  \bibfield  {author} {\bibinfo {author} {\bibfnamefont {D.}~\bibnamefont
  {Bini}}\ and\ \bibinfo {author} {\bibfnamefont {A.}~\bibnamefont
  {Geralico}},\ }\href {\doibase 10.1103/PhysRevD.100.104003} {\bibfield
  {journal} {\bibinfo  {journal} {Phys. Rev.}\ }\textbf {\bibinfo {volume}
  {D100}},\ \bibinfo {pages} {104003} (\bibinfo {year} {2019}{\natexlab{b}})},\
  \Eprint {http://arxiv.org/abs/1907.11082} {arXiv:1907.11082 [gr-qc]}
  \BibitemShut {NoStop}%
\bibitem [{\citenamefont {Goldstein}\ \emph {et~al.}(2000)\citenamefont
  {Goldstein}, \citenamefont {Poole},\ and\ \citenamefont
  {Safko}}]{Goldstein:2000}%
  \BibitemOpen
  \bibfield  {author} {\bibinfo {author} {\bibfnamefont {H.}~\bibnamefont
  {Goldstein}}, \bibinfo {author} {\bibfnamefont {C.}~\bibnamefont {Poole}}, \
  and\ \bibinfo {author} {\bibfnamefont {J.}~\bibnamefont {Safko}},\
  }\href@noop {} {\emph {\bibinfo {title} {Classical Mechanics}}},\ \bibinfo
  {edition} {3rd}\ ed.\ (\bibinfo  {publisher} {Addison-Wesley},\ \bibinfo
  {address} {Boston},\ \bibinfo {year} {2000})\BibitemShut {NoStop}%
\bibitem [{\citenamefont {{Carter}}(2010)}]{Carter:2010}%
  \BibitemOpen
  \bibfield  {author} {\bibinfo {author} {\bibfnamefont {B.}~\bibnamefont
  {{Carter}}},\ }\href {\doibase 10.1007/s10714-009-0920-9} {\bibfield
  {journal} {\bibinfo  {journal} {General Relativity and Gravitation}\ }\textbf
  {\bibinfo {volume} {42}},\ \bibinfo {pages} {653} (\bibinfo {year}
  {2010})}\BibitemShut {NoStop}%
\bibitem [{\citenamefont {Barack}\ and\ \citenamefont
  {Sago}(2011)}]{Barack:2011ed}%
  \BibitemOpen
  \bibfield  {author} {\bibinfo {author} {\bibfnamefont {L.}~\bibnamefont
  {Barack}}\ and\ \bibinfo {author} {\bibfnamefont {N.}~\bibnamefont {Sago}},\
  }\href {\doibase 10.1103/PhysRevD.83.084023} {\bibfield  {journal} {\bibinfo
  {journal} {Phys. Rev.}\ }\textbf {\bibinfo {volume} {D83}},\ \bibinfo {pages}
  {084023} (\bibinfo {year} {2011})},\ \Eprint {http://arxiv.org/abs/1101.3331}
  {arXiv:1101.3331 [gr-qc]} \BibitemShut {NoStop}%
\bibitem [{anc()}]{ancmaterial}%
  \BibitemOpen
  \href@noop {} {\ }\bibinfo {note} {For the arXiv version, download the
  ancillary file \texttt{redshift\_precessfreq.m}, which contains the redshift
  and spin precession invariants for arbitrary mass ratios.}\BibitemShut
  {Stop}%
\bibitem [{\citenamefont {Bini}\ \emph
  {et~al.}(2016{\natexlab{b}})\citenamefont {Bini}, \citenamefont {Damour},\
  and\ \citenamefont {Geralico}}]{Bini:2016dvs}%
  \BibitemOpen
  \bibfield  {author} {\bibinfo {author} {\bibfnamefont {D.}~\bibnamefont
  {Bini}}, \bibinfo {author} {\bibfnamefont {T.}~\bibnamefont {Damour}}, \ and\
  \bibinfo {author} {\bibfnamefont {A.}~\bibnamefont {Geralico}},\ }\href
  {\doibase 10.1103/PhysRevD.93.124058} {\bibfield  {journal} {\bibinfo
  {journal} {Phys. Rev.}\ }\textbf {\bibinfo {volume} {D93}},\ \bibinfo {pages}
  {124058} (\bibinfo {year} {2016}{\natexlab{b}})},\ \Eprint
  {http://arxiv.org/abs/1602.08282} {arXiv:1602.08282 [gr-qc]} \BibitemShut
  {NoStop}%
\bibitem [{\citenamefont {Damour}\ \emph {et~al.}(2008)\citenamefont {Damour},
  \citenamefont {Jaranowski},\ and\ \citenamefont {Schäfer}}]{Damour:2008qf}%
  \BibitemOpen
  \bibfield  {author} {\bibinfo {author} {\bibfnamefont {T.}~\bibnamefont
  {Damour}}, \bibinfo {author} {\bibfnamefont {P.}~\bibnamefont {Jaranowski}},
  \ and\ \bibinfo {author} {\bibfnamefont {G.}~\bibnamefont {Schäfer}},\
  }\href {\doibase 10.1103/PhysRevD.78.024009} {\bibfield  {journal} {\bibinfo
  {journal} {Phys. Rev.}\ }\textbf {\bibinfo {volume} {D78}},\ \bibinfo {pages}
  {024009} (\bibinfo {year} {2008})},\ \Eprint {http://arxiv.org/abs/0803.0915}
  {arXiv:0803.0915 [gr-qc]} \BibitemShut {NoStop}%
\bibitem [{\citenamefont {Damour}\ \emph {et~al.}(2012)\citenamefont {Damour},
  \citenamefont {Nagar}, \citenamefont {Pollney},\ and\ \citenamefont
  {Reisswig}}]{Damour:2011fu}%
  \BibitemOpen
  \bibfield  {author} {\bibinfo {author} {\bibfnamefont {T.}~\bibnamefont
  {Damour}}, \bibinfo {author} {\bibfnamefont {A.}~\bibnamefont {Nagar}},
  \bibinfo {author} {\bibfnamefont {D.}~\bibnamefont {Pollney}}, \ and\
  \bibinfo {author} {\bibfnamefont {C.}~\bibnamefont {Reisswig}},\ }\href
  {\doibase 10.1103/PhysRevLett.108.131101} {\bibfield  {journal} {\bibinfo
  {journal} {Phys. Rev. Lett.}\ }\textbf {\bibinfo {volume} {108}},\ \bibinfo
  {pages} {131101} (\bibinfo {year} {2012})},\ \Eprint
  {http://arxiv.org/abs/1110.2938} {arXiv:1110.2938 [gr-qc]} \BibitemShut
  {NoStop}%
\bibitem [{\citenamefont {Nagar}\ \emph {et~al.}(2016)\citenamefont {Nagar},
  \citenamefont {Damour}, \citenamefont {Reisswig},\ and\ \citenamefont
  {Pollney}}]{Nagar:2015xqa}%
  \BibitemOpen
  \bibfield  {author} {\bibinfo {author} {\bibfnamefont {A.}~\bibnamefont
  {Nagar}}, \bibinfo {author} {\bibfnamefont {T.}~\bibnamefont {Damour}},
  \bibinfo {author} {\bibfnamefont {C.}~\bibnamefont {Reisswig}}, \ and\
  \bibinfo {author} {\bibfnamefont {D.}~\bibnamefont {Pollney}},\ }\href
  {\doibase 10.1103/PhysRevD.93.044046} {\bibfield  {journal} {\bibinfo
  {journal} {Phys. Rev.}\ }\textbf {\bibinfo {volume} {D93}},\ \bibinfo {pages}
  {044046} (\bibinfo {year} {2016})},\ \Eprint
  {http://arxiv.org/abs/1506.08457} {arXiv:1506.08457 [gr-qc]} \BibitemShut
  {NoStop}%
\bibitem [{\citenamefont {Ossokine}\ \emph {et~al.}(2018)\citenamefont
  {Ossokine}, \citenamefont {Dietrich}, \citenamefont {Foley}, \citenamefont
  {Katebi},\ and\ \citenamefont {Lovelace}}]{Ossokine:2017dge}%
  \BibitemOpen
  \bibfield  {author} {\bibinfo {author} {\bibfnamefont {S.}~\bibnamefont
  {Ossokine}}, \bibinfo {author} {\bibfnamefont {T.}~\bibnamefont {Dietrich}},
  \bibinfo {author} {\bibfnamefont {E.}~\bibnamefont {Foley}}, \bibinfo
  {author} {\bibfnamefont {R.}~\bibnamefont {Katebi}}, \ and\ \bibinfo {author}
  {\bibfnamefont {G.}~\bibnamefont {Lovelace}},\ }\href {\doibase
  10.1103/PhysRevD.98.104057} {\bibfield  {journal} {\bibinfo  {journal} {Phys.
  Rev.}\ }\textbf {\bibinfo {volume} {D98}},\ \bibinfo {pages} {104057}
  (\bibinfo {year} {2018})},\ \Eprint {http://arxiv.org/abs/1712.06533}
  {arXiv:1712.06533 [gr-qc]} \BibitemShut {NoStop}%
\bibitem [{SXS()}]{SXS}%
  \BibitemOpen
  \href@noop {} {}\bibinfo {howpublished}
  {\url{https://data.black-holes.org/waveforms}}\BibitemShut {NoStop}%
\bibitem [{\citenamefont {Dietrich}\ \emph {et~al.}(2017)\citenamefont
  {Dietrich}, \citenamefont {Bernuzzi}, \citenamefont {Ujevic},\ and\
  \citenamefont {Tichy}}]{Dietrich:2016lyp}%
  \BibitemOpen
  \bibfield  {author} {\bibinfo {author} {\bibfnamefont {T.}~\bibnamefont
  {Dietrich}}, \bibinfo {author} {\bibfnamefont {S.}~\bibnamefont {Bernuzzi}},
  \bibinfo {author} {\bibfnamefont {M.}~\bibnamefont {Ujevic}}, \ and\ \bibinfo
  {author} {\bibfnamefont {W.}~\bibnamefont {Tichy}},\ }\href {\doibase
  10.1103/PhysRevD.95.044045} {\bibfield  {journal} {\bibinfo  {journal} {Phys.
  Rev.}\ }\textbf {\bibinfo {volume} {D95}},\ \bibinfo {pages} {044045}
  (\bibinfo {year} {2017})},\ \Eprint {http://arxiv.org/abs/1611.07367}
  {arXiv:1611.07367 [gr-qc]} \BibitemShut {NoStop}%
\end{thebibliography}%

\end{document}